\documentclass[aps,prd,superscriptaddress,nofootinbib,floats,preprint,floats,floatfix]{revtex4-1}
\usepackage{CJK}
\usepackage{bm}
\usepackage{indentfirst}
\usepackage{amsmath}
\usepackage{graphicx}
\usepackage{float}
\usepackage{amssymb}
\usepackage{subfigure}
\usepackage{amssymb}
\usepackage{psfrag}
\usepackage{hyperref}
\usepackage{tensor}
\usepackage{braket}
\usepackage{multirow}
\usepackage{verbatim}
\usepackage[dvipsnames, svgnames, x11names]{xcolor}
\hypersetup{
	colorlinks=true,
	linkcolor=red,
	citecolor=blue,
} 

\allowdisplaybreaks[1]

\begin{document}
\begin{CJK*}{UTF8}{gbsn}
\title{Double-peaked inflation model: Scalar induced gravitational waves and primordial-black-hole suppression from primordial non-Gaussianity }
\author{Fengge Zhang(张丰阁)}
\email{fenggezhang@hust.edu.cn}
\affiliation{School of Physics, Huazhong University of Science and Technology,
Wuhan, Hubei 430074, China}

\author{Jiong Lin(林炯)}
\email{jionglin@hust.edu.cn}
\affiliation{School of Physics, Huazhong University of Science and Technology,
Wuhan, Hubei 430074, China}
\author{Yizhou Lu(卢一洲)}
\email{Corresponding author. louischou@hust.edu.cn}
\affiliation{School of Physics, Huazhong University of Science and Technology,
Wuhan, Hubei 430074, China}

\begin{abstract}
A significant abundance of primordial black hole (PBH) dark matter can be produced by curvature perturbations with power spectrum $\Delta_\zeta^2(k_{\mathrm{peak}})\sim \mathcal{O}(10^{-2})$ at small scales, associated with the generation of observable scalar induced gravitational waves (SIGWs). 
However, the primordial non-Gaussianity may play a non-negligible role, which is not usually considered.
We propose two inflation models that predict double peaks of order $\mathcal{O}(10^{-2})$ in the power spectrum and study the effects of primordial non-Gaussianity on PBHs and SIGWs. 
This model is driven by a power-law potential, and has a noncanonical kinetic term whose coupling function admits two peaks.
By field-redefinition, it can be recast into a canonical inflation model with two quasi-inflection points in the potential.
We find that the PBH abundance will be altered saliently if non-Gaussianity parameter satisfies $|f_{\mathrm{NL}}(k_{\text{peak}},k_{\text{peak}},k_{\text{peak}})|\gtrsim \Delta^2_{\zeta}(k_{\mathrm{peak}})/(23\delta^3_c) \sim \mathcal{O}(10^{-2})$. 
Whether the PBH abundance is suppressed or enhanced depends on the $f_{\mathrm{NL}}$ being positive or negative, respectively.
In our model, non-Gaussianity parameter $f_{\mathrm{NL}}(k_{\mathrm{peak}},k_{\mathrm{peak}},k_{\mathrm{peak}})\sim \mathcal{O}(1)$ takes positive sign, thus PBH abundance is suppressed dramatically. On the contrary, SIGWs are insensitive to primordial non-Gaussianity and hardly affected, so they are still within the sensitivities of space-based GWs observatories and Square Kilometer Array.
\end{abstract}
\maketitle
	
\section{introduction}
Ever since the detection of gravitational waves (GWs) by the Laser Interferometer Gravitational-Wave Observatory (LIGO) scientific collaboration and Virgo collaboration \cite{Abbott:2016nmj,Abbott:2016blz,Abbott:2017gyy,TheLIGOScientific:2017qsa,Abbott:2017oio,Abbott:2017vtc,LIGOScientific:2018mvr,Abbott:2020khf,Abbott:2020uma,LIGOScientific:2020stg}, primordial black holes (PBHs) \cite{Carr:1974nx,Hawking:1971ei} have been drawing much attention as PBHs might be the source of GW events \cite{Bird:2016dcv,Sasaki:2016jop,Takhistov:2020vxs,DeLuca:2020sae,Abbott:2020niy}. 
Recently, North American Nanohertz Observatory for Gravitational Waves (NANOGrav) also hints the existence of PBHs \cite{DeLuca:2020agl,Vaskonen:2020lbd,Kohri:2020qqd,Domenech:2020ers,Atal:2020yic,Yi:2021lxc}. 
In addition, it is attractive that PBHs could be the candidate of dark matter (DM).

PBHs are formed by gravitational collapse during a radiation-dominated era when large perturbations reenter the horizon. 
This mechanism requires the amplitude of the power spectrum of primordial Gaussian curvature perturbation $\mathcal{A}_{\zeta} \sim \mathcal{O}(10^{-2})$ at small scales \cite{Sato-Polito:2019hws}.
The constraints on power spectrum from the cosmic microwave background (CMB) is $\mathcal{A}_{\zeta}\sim \mathcal{O}(10^{-9})$ at the pivot scale $k_*=0.05~\mathrm{Mpc}^{-1}$ \cite{Akrami:2018odb}, so the power spectrum has to be enhanced at least seven orders at small scales during inflation to produce significant abundance of PBH DM \cite{Gong:2017qlj,Garcia-Bellido:2017mdw,Germani:2017bcs,Lu:2019sti,Motohashi:2017kbs,Espinosa:2017sgp,Belotsky:2018wph,Dalianis:2019vit,Passaglia:2019ueo,Fu:2019ttf,Xu:2019bdp,Lin:2020goi,Yi:2020kmq,Yi:2020cut,Gao:2020tsa,Fumagalli:2020adf,Gundhi:2020kzm,Ballesteros:2020qam,Ragavendra:2020sop,Palma:2020ejf,Braglia:2020eai}. 
Besides PBHs, large curvature perturbations also induce the scalar induced gravitational waves (SIGWs) that contribute to stochastic gravitational-wave background (SGWB) \cite{Baumann:2007zm,Saito:2008jc,Orlofsky:2016vbd,Nakama:2016gzw,Inomata:2016rbd,Cai:2018dig,Bartolo:2018evs,Kohri:2018awv,Espinosa:2018eve,Kuroyanagi:2018csn,Cai:2019elf,Drees:2019xpp,Inomata:2019ivs,Inomata:2019zqy,Fumagalli:2020nvq,Domenech:2020kqm,Braglia:2020taf,Fumagalli:2021cel}.
	
Generally, the power spectrum predicted by the slow-roll inflation is nearly scale-invariant and agrees with the observational constraints on large scales with $\mathcal{A}_\zeta \sim\mathcal{O}(10^{-9})$.
To amplify the power spectrum on small scales, it is necessary to consider the violation of slow-roll \cite{Motohashi:2017kbs}.
The canonical inflation model with a flat plateau, namely the so-called quasi-inflection point in potential is usually used to violate the slow-roll conditions and amplify the power spectrum \cite{Germani:2017bcs,Gong:2017qlj,Garcia-Bellido:2017mdw,Xu:2019bdp}.
However, it is difficult to find such a suitable potential with quasi-inflection points during inflation while keeping $e$-folds $50- 70$.
We will show that a canonical inflation model with a quasi-inflection point in the potential is equivalent to a noncanonical inflation model where the coupling function $G(\phi)$ of the kinetic term has a peak, up to a field-redefinition. 
We can have as many peaks in $G(\phi)$ as we want. 
This amounts to produce corresponding peaks in the power spectrum.

Usually, the power spectrum with a large peak only produces PBHs in a single mass range. 
We expect that the inflation models with multipeaked power spectrum produce PBHs at different mass ranges.
This can explain more DMs and different observational phenomena simultaneously.
For example, the canonical model proposed in Ref. \cite{Gao:2021dfi} predicts double peaks in the power spectrum which leads to a double-peaked PBH abundance and also the energy density of SIGWs. In Ref. \cite{Zheng:2021vda}, the power spectrum with triple peaks is produced by a triple-bumpy potential. 
However, the effect of primordial non-Gaussianity is not considered.
Based on this motivation, in this paper, with generalized G-inflation models \cite{Lin:2020goi,Yi:2020cut,Yi:2020kmq}, we use a noncanonical kinetic term with double-peaked coupling function $G(\phi)$ to produce a power spectrum with two culminations with magnitude $\mathcal{A}_{\zeta} \sim \mathcal{O}(10^{-2})$ at small scales while satisfying CMB constraints at large scales.
Moreover, the $e$-folds are reasonably $N\simeq 69$. 
With different choices of $G(\phi)$, both $\lambda\phi^{2/5}$ potential and Higgs potential $\lambda \phi^4$ can produce such power spectrum. 
In the period that the inflaton departs from the slow-roll, the non-Gaussianity may differ from that in slow-roll inflations, which predict negligible non-Gaussianity \cite{Maldacena:2002vr}.
We compute the primordial non-Gaussianity numerically in the squeezed and equilateral limits, and both have their maxima of order $f_{\mathrm{NL}}\sim\mathcal{O}(10)$.

Intuitively, it seems there are plenty of PBHs formed in two different mass ranges because of the double peaks in the power spectrum. 
However,  this is just an illusion owing to our ignorance of primordial non-Gaussianity.
We show that the PBH abundance is sensitive to primordial non-Gaussianity of curvature perturbation and will be strongly altered if the non-Gaussianity parameter satisfies $|f_{\mathrm{NL}}(k_{\text{peak}},k_{\text{peak}},k_{\text{peak}})|\gtrsim \Delta^2_{\zeta}(k_{\mathrm{peak}})/(23\delta^3_c) \sim \mathcal{O}(10^{-2})$, whether the PBH abundance is suppressed or enhanced depends on non-Gaussianity parameter being positive or negative, respectively.
Nevertheless, the magnitude of the fractional energy density of SIGWs is hardly affected by the primordial non-Gaussianity. 
Although this model predicts double-peaked SIGWs that can be detected by the space-based GW observatories such as Laser Interferometer Space Antenna (LISA) \cite{Audley:2017drz,Danzmann:1997hm}, TianQin \cite{Luo:2015ght}, Taiji\cite{Hu:2017mde} and etc, and the Square Kilometer Array (SKA) \cite{Moore:2014lga}, there is no significant abundance of PBHs produced with $f_{\mathrm{NL}}(k_{\text{peak}},k_{\text{peak}},k_{\text{peak}}) \sim \mathcal{O}(1)$ that is positive in the two models.

This paper is organized as follows.
In Sec. \ref{sec2}, we first explain the equivalence between a noncanonical inflation model with a peak coupling function of noncanonical kinetic term and a canonical model with a flat plateau in the potential and then give the power spectrum produced by noncanonical inflation models.
In Sec. \ref{NG}, we present the primordial non-Gaussianities. 
In Sec. \ref{PBHs}, we discuss the PBH abundance from the present models, where we also take consideration of non-Gaussianity. 
In Sec. \ref{SIGWs}, we compute the SIGWs during radiation dominant. Our conclusions are drew in Sec. \ref{conclusion}.
	
\section{Inflation models}\label{sec2}
Considering a kind of generalized G-inflation with a noncanonical kinetic term \cite{Lin:2020goi,Yi:2020kmq,Yi:2020cut}
\begin{equation}\label{act:kgmodel}
S=\int d^4 x \sqrt{-g}\left[\frac{R}{2}+[1+G(\phi)]X(\phi)-V(\phi)\right],
\end{equation}
where $X(x)=-\nabla_{\mu}x\nabla^{\mu}x/2$, $V(\phi)$ is the inflaton potential, and $G(\phi)$ is a function of $\phi$. Here we choose $8\pi G=1$. Performing a field-redefinition, 
\begin{equation}\label{redefinition}
d\varphi=\sqrt{1+G(\phi)}d\phi,
\end{equation}
we get $S\rightarrow \mathcal{S}$,
\begin{equation}\label{canonical}
\mathcal{S}=\int d^4 x \sqrt{-g}\left[\frac{R}{2}+X(\varphi)-\mathrm{U}(\varphi)\right],
\end{equation}
where $\mathrm{U}(\varphi)=V[\phi(\varphi)]$ is the potential of canonical field $\varphi=\varphi(\phi)$. 

The power spectrum of curvature perturbations produced by canonical model Eq. \eqref{canonical} under slow roll is
\begin{equation}\label{srps}
\Delta^2_{\zeta}\simeq \frac{\mathrm{U}}{24\pi^2\epsilon_\mathrm{U}},
\end{equation}
with
\begin{equation}
\epsilon_\mathrm{U}=\frac{\mathrm{U}^2_{\varphi}}{2\mathrm{U}^2},
\end{equation}
where $\mathrm{U}_{\varphi}=d\mathrm{U}/d\varphi$. From the expression of power spectrum Eq. \eqref{srps}, qualitatively, a flat potential, $\mathrm{U}_{\varphi}\ll 1$, will result in an enhancement of power spectrum. For a canonical inflation model, a flat plateau in potential is often required to amplify the power spectrum. In the following, we will see that a peak in $G(\phi)$ results in a quasi-inflection point in $\mathrm{U}(\varphi)$ of canonical field. Recalling the field-redefinition Eq. \eqref{redefinition}, we have
\begin{equation}\label{inflection1}
\epsilon_{\mathrm{U}}=\frac{V^2_\phi}{2V^2}\frac{1}{1+G(\phi)}, 
\end{equation}
\begin{equation}\label{inflection2}
\mathrm{U}_{\varphi \varphi}=V_{\phi \phi}\frac{1}{1+G(\phi)}-\frac{1}{2}\frac{V_\phi G_\phi}{(1+G(\phi))^2},
\end{equation}
where $\mathrm{U}_{\varphi \varphi}=d^2\mathrm{U}/d\varphi^2$, $V_{\phi}=d V/d\phi$, $V_{\phi \phi}=d^2V/d\phi^2$ and $G_\phi=dG/d\phi$. 
From Eqs.\eqref{inflection1} and \eqref{inflection2}, we see that $\epsilon_{\mathrm{U}},\mathrm{U}_{\varphi\varphi}\ll 1$ if $G(\phi)\gg 1$.
That is, a large peak in $G(\phi)$ leads to a quasi-inflection point in $\mathrm{U}(\varphi)$.
We can amplify the power spectrum by choosing a peaked $G(\phi)$.
We demonstrate this with two concrete examples. 
We generalize the choice of $G(\phi)$ in Ref. \cite{Lin:2020goi} and Refs. \cite{Yi:2020cut,Yi:2020kmq,Gao:2020tsa} to produce multiple peaks in power spectrum of curvature perturbation,
\begin{gather}
\label{G1}
G_1(\phi)=\sum_{i=1,2}\frac{d_i c_i^{q_i}}{c_i^{q_i}+|\phi-\phi_{p_i}|^{q_i}},\quad V_1(\phi)=\lambda_1\phi^{2/5},\\
\label{G2}
G_2(\phi)=\phi^{22}+\sum_{i=1,2}\frac{d_i c_i^{q_i}}{c_i^{q_i}+|\phi-\phi_{p_i}|^{q_i}},\quad V_2(\phi)=\frac{\lambda_2}{4}\phi^4,
\end{gather}
with $d_i c_i^{q_i}\sim \mathcal{O}(1)$. 
The choice of the function is inspired by Brans-Dicke theory with a noncanonical kinetic term $X/\phi$ performing a shift of $\phi$. 
In order to avoid singularity, we add a small parameter $c_i^{q_i}\ll 1$. 
These parameters $d_i$, $c_i$ and $q_i$ control the height, the width and the shape of the peak in function $G(\phi)$, separately.

\begin{figure}[htp]
\centering
\subfigure{\includegraphics[width=0.45\linewidth]{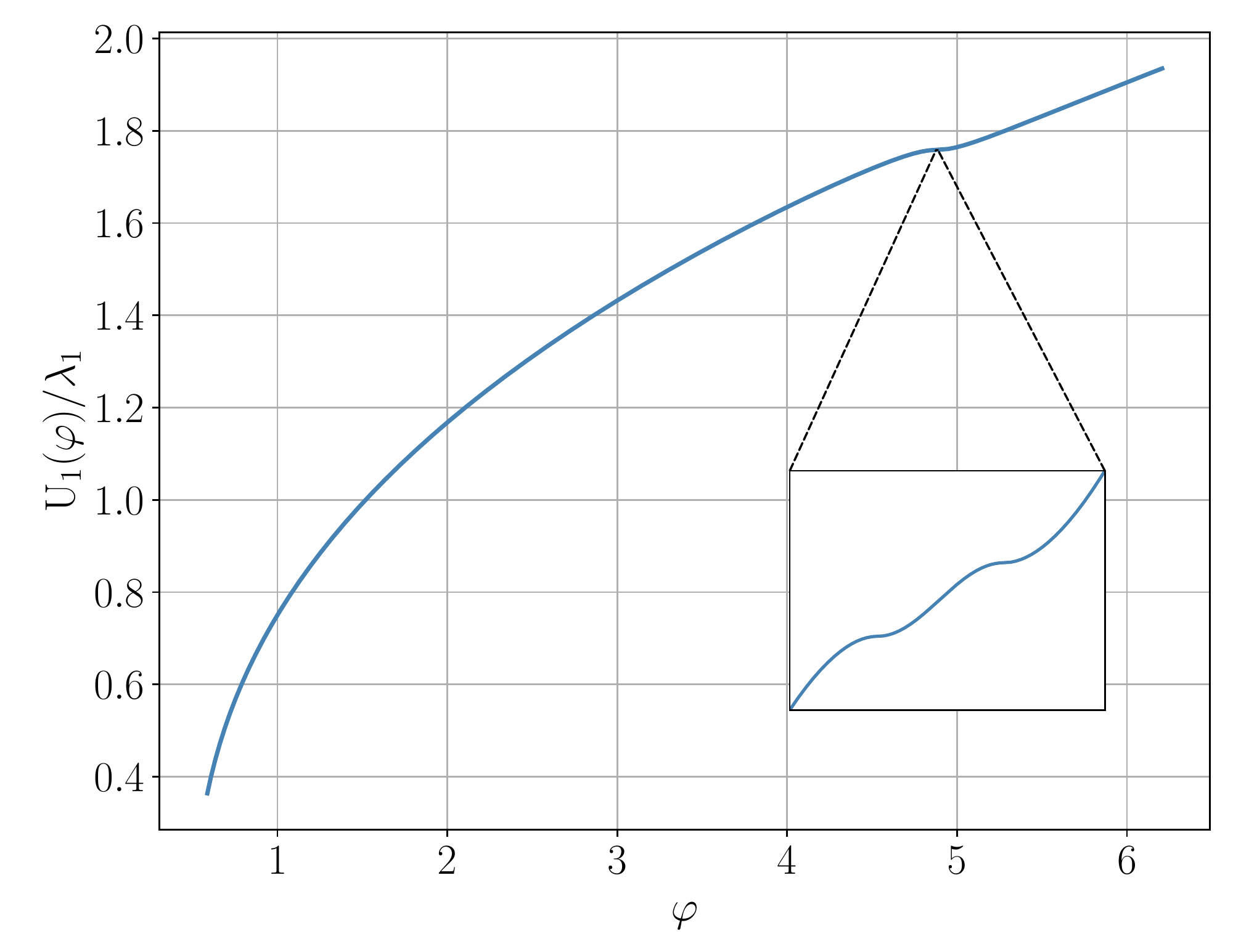}}
\subfigure{\includegraphics[width=0.45\linewidth]{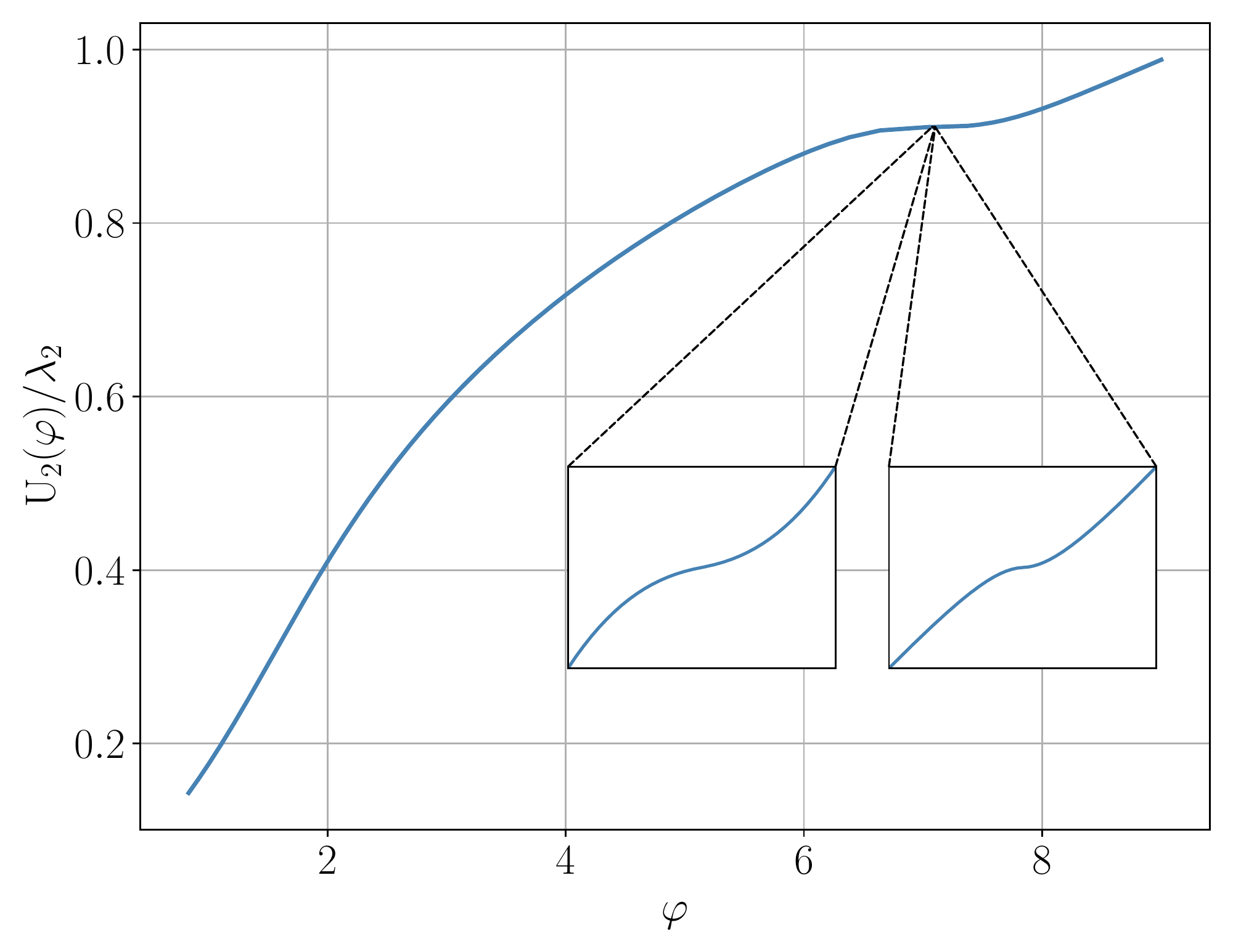}}
\caption{The potentials $\mathrm{U}_1(\varphi)=V_1(\phi)$ (left panel) $\mathrm{U}_2(\varphi)=V_2(\phi)$ (right panel) as the functions of canonical field $\varphi$.}
\label{fig:U_PHI}
\end{figure}

We denote the models with potential $V_1(\phi)=\lambda_1\phi^{2/5}$ and $V_2(\phi)=\lambda_2\phi^4/4$ by M1 and M2, respectively. 
$G_1(\phi)$ and $G_2(\phi)$ both contain two peaks and will lead to two flat plateaus in $\mathrm{U}(\varphi)$.
With the parameters listed in table \ref{para:tab}, the potentials $\mathrm{U}_1(\varphi)$ and $\mathrm{U}_2(\varphi)$ of canonical field $\varphi$ are shown in Fig. \ref{fig:U_PHI}. 
Here we choose $\varphi(\phi_*)=6.21$ and $\varphi(\phi_*)=9$ for models M1 and M2, where $\phi_*$ is the value of inflaton when modes $k_*=0.05~\mathrm{Mpc}^{-1}$ leave the horizon.
Note that we can always shift $\varphi$ by a constant because of Eq. \eqref{redefinition}.
From Fig. \ref{fig:U_PHI}, we can see that the potentials $\mathrm{U}_1(\varphi)$ and $\mathrm{U}_2(\varphi)$ of the canonical field $\varphi$ both have two flat plateaus, which could result in two large peaks in the power spectrum. 

Since the analytic expressions of $\mathrm{U_1}(\varphi)$ and $\mathrm{U}_2(\varphi)$ are hard to obtain, it is more convenient to work in the noncanonical models, as we will do in the followings.
	
\begin{table}[htp]
\centering
\renewcommand\tabcolsep{3.0pt}
\begin{tabular}{cccccccccc}
\hline
\hline
Model & $d_1$ & $d_2$ & $\phi_{p_1}$ & $\phi_{p_2}$ & $ \phi_*$ &  $c_1$ &  $c_2$ &$q_1$ & $q_2$\\
\hline
M1 &  $2.05\times 10^{9}$&  $1.003\times 10^{9}$ & $4.1$ & $4.09996$ & $5.21$& $2.22\times 10^{-11}$ &  $4.26\times 10^{-11}$ &$1$ &$1$\\
\hline
M2 &  $3.9\times 10^{11}$&  $5\times 10^{10}$ & $1.3815105$ & $1.3815102$ & $1.41$ & $4\times 10^{-12}$ &  $7\times 10^{-11}$ &$1$ &$5/4$ \\
\hline
\hline
\end{tabular}
\caption{The parameters for the peak function $G(\phi)$ in \eqref{G1} and \eqref{G2}. The parameters of potential for $\lambda_1$ and $\lambda_2$ are $7.20\times 10^{-10}$ and $9.72 \times 10^{-10}$, respectively. And $\phi_*$ is the value of the inflaton when the pivot scale $k_*=0.05\  \text{Mpc}^{-1}$ leaves the horizon.}
\label{para:tab}
\end{table}

\begin{figure}[htp]
\centering
\subfigure{\includegraphics[width=0.4\linewidth]{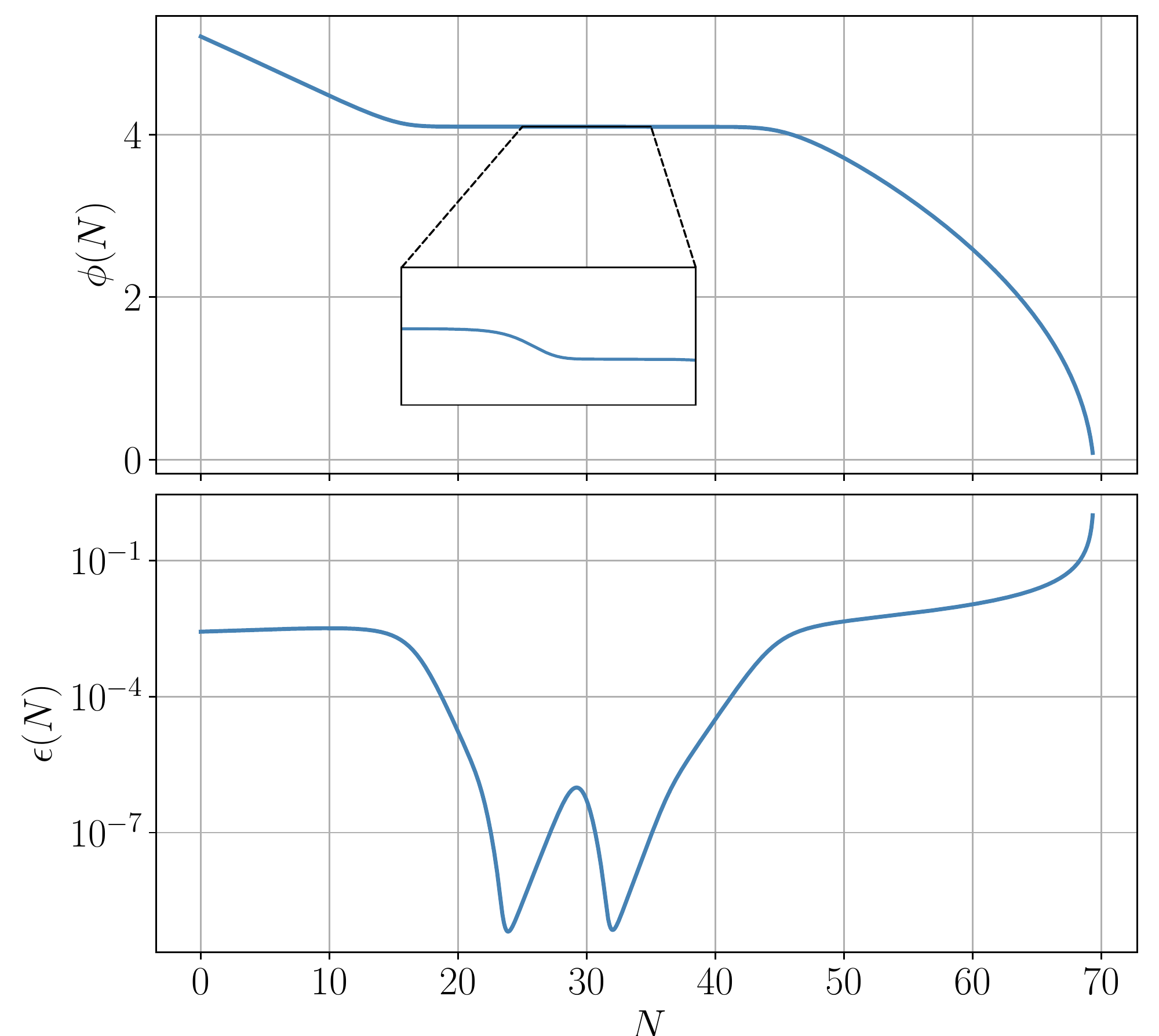}}
\subfigure{\includegraphics[width=0.4\linewidth]{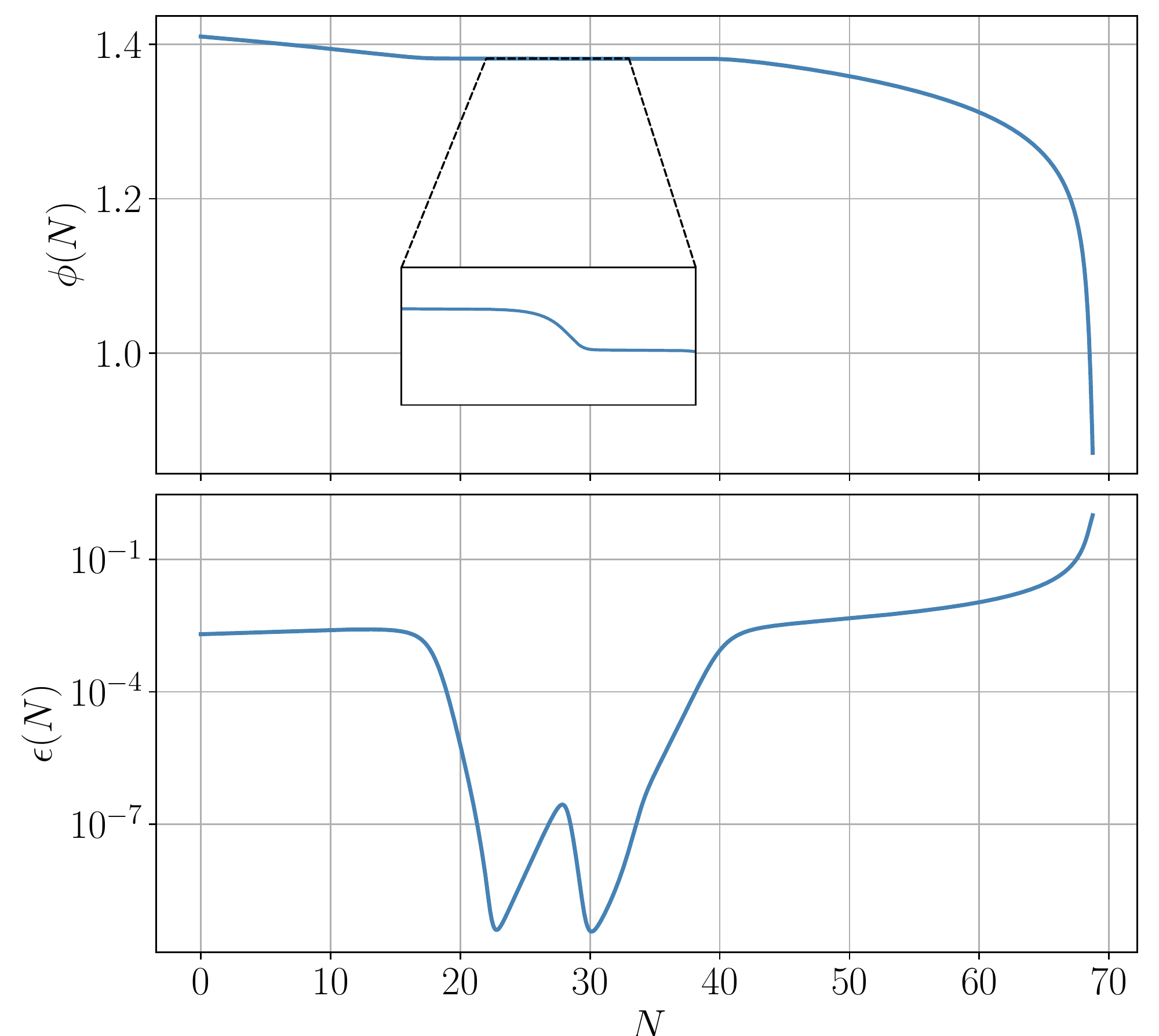}}
\caption{The evolution of inflaton $\phi$ and the slow-roll parameter $\epsilon$ with respect to $e$-folds $N,$ for models M1, the left panel, and M2, the right panel.}
\label{fig:bg}
\end{figure}

\begin{figure}
\centering
\includegraphics[width=0.7\linewidth]{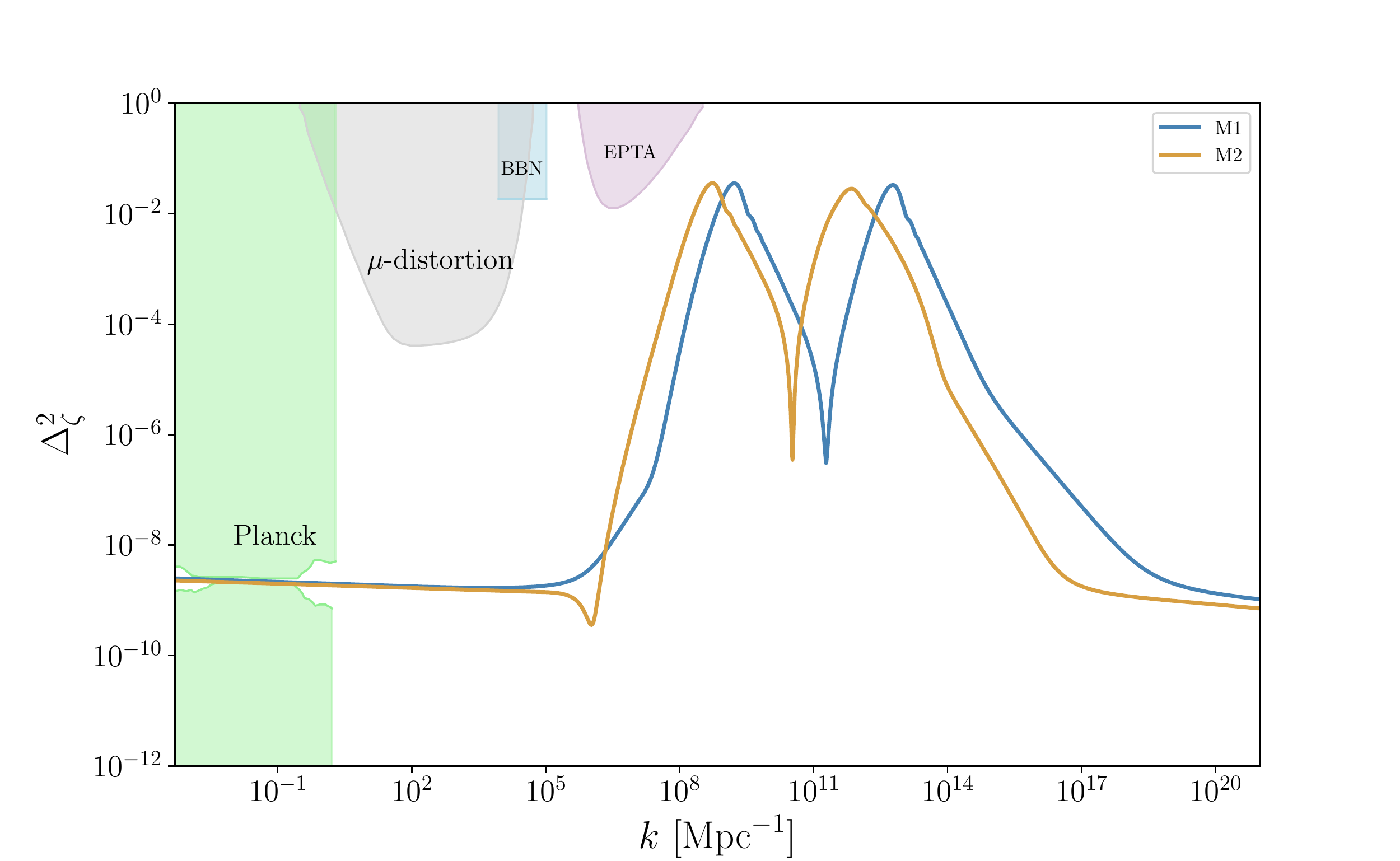}
\caption{The power spectra produce by models M1 and M2. They are enhanced to $\mathcal{O}(0.01)$ on small scales.
The shaded regions are constraints from observations \cite{Akrami:2018odb,Inomata:2018epa,Inomata:2016uip,Fixsen:1996nj}}.
\label{fig:spectrum}
\end{figure}

Working in the spatially flat Friedmann-Robertson-Walker (FRW) metric, we obtain the background equations from action \eqref{act:kgmodel}
\begin{gather}
\label{Eq:eom1}
3H^2=\frac{1}{2}\dot{\phi}^2+V(\phi)+\frac{1}{2}\dot{\phi}^2G(\phi),\\
\label{Eq:eom2}
\dot{H}=-\frac{1}{2}[1+G(\phi)]\dot{\phi}^2,\\
\label{Eq:eom3}
\ddot{\phi}+3H\dot{\phi}+\frac{V_{\phi}+\dot{\phi}^2G_{\phi}/2}{1+G(\phi)}=0,
\end{gather}
where the Hubble parameter $H=\dot{a}/a$, and the dot denotes the derivative with respect to the cosmological time $t$. 
To the second order of comoving curvature perturbation $\zeta$, the action reads 
\begin{equation}\label{S2}
S_2=\frac{1}{2} \int d \tau d^{3} x z^2\left[\left(\zeta^{\prime}\right)^{2}-(\partial \zeta)^{2}\right],
\end{equation}
where $z^2=2a^2\epsilon$, and the prime stands for the derivative with respect to the conformal time ${\tau}$, where $d\tau=dt/a(t)$. 
The slow-roll parameters $\epsilon$ and $\eta$ are defined as
\begin{equation}\label{SRPA}
\epsilon=-\frac{\dot{H}}{H^{2}},\ \eta=\frac{\dot \epsilon}{H \epsilon}.
\end{equation}
The leading order action Eq. \eqref{S2} predicts a Gaussian spectrum, while non-Gaussian features, which we will discuss in Sec. \ref{NG}, arise from cubic action.
Varying the quadratic action with respect to ${\zeta}$, we obtain the Mukhanov-Sasaki equation \cite{Mukhanov:1985rz,Sasaki:1986hm}
\begin{equation}
\label{E1}
\zeta_{k}^{\prime \prime}+2\frac{z'}{z}\zeta_{k}^{\prime}+k^2\zeta_k=0,
\end{equation}
with
\begin{equation}
\frac{z'}{z}=aH\left(1+\frac{\eta}{2}\right).
\end{equation}
The two-point correlation function and the power spectrum are given as follows
\begin{equation}
\begin{split}
\langle \hat{\zeta}_{\bm{k}}\hat{\zeta}_{\bm{k}'}\rangle&=
(2\pi)^3\delta^3\left(\bm{k}+\bm{k}'\right)|\zeta_k|^2\\&=
(2\pi)^3\delta^3\left(\bm{k}+\bm{k}'\right)P_{\zeta}\left(k\right),
\end{split}
\end{equation}	where the mode function $\zeta_k$ is the solution to Eq. \eqref{E1}, and $\hat{\zeta}_{\bm{k}}$ is the quantum field of the curvature perturbation that starts evolution from Bunch-Davies vacuum.
The dimensionless scalar power spectrum and its spectral index are defined by
\begin{equation}
\Delta^2_{\zeta}=\frac{k^3}{2\pi^2}P_\zeta\left(k\right),
\end{equation}
\begin{equation}
n_{\mathrm{s}}(k)-1=\frac{d\mathrm{ln}\Delta^2_{\zeta}}{d\mathrm{ln}k}.
\end{equation}
With the parameter sets in Table \ref{para:tab}, we solve the Eqs. \eqref{Eq:eom1}-\eqref{Eq:eom3} and \eqref{E1} numerically. 
We show in  Fig. \ref{fig:bg} the evolution of the inflaton $\phi$ and the slow-roll parameter $\epsilon$ against $e$-folds $N$, and in Fig. \ref{fig:spectrum} the power spectrum. 
In Tables. \ref{nsr:tab} and \ref{results:tab}, we list the power index, tensor to scalar ratio, the $e$-folds $N$, the peak scales, and the amplitude of the power spectra in models M1 and M2.

\begin{table}[htp]
\centering
\renewcommand\tabcolsep{4.0pt}
\begin{tabular}{llll}
\hline
\hline
Model &$n_s$& $r$ & $N$ \\
\hline
M1 & $0.971$ & $0.0426$& $69.3$  \\
\hline
M2 & $0.969$ & $0.0318$& $68.7$  \\
\hline
\hline
\end{tabular}
\caption{The power index, tensor to scalar ratio, and the $e$-folds $N$ in M1 and M2.}
\label{nsr:tab}
\end{table}

\begin{table}[htp]
\centering
\renewcommand\tabcolsep{3.0pt}
\begin{tabular}{|c|c|c|c|c|c|c|c|c|} 
\hline
Model & $k_{\mathrm{peak}}/\mathrm{Mpc}^{-1}$ & $\Delta^2_{\zeta}(k_{\mathrm{peak}})$ & $Y^G_{\mathrm{PBH}}$ & $M/M_{\odot}$ &$f_c/\mathrm{Hz}$ &$\Delta^{\mathrm{n}}_3(k_\mathrm{peak})$&$f_{\mathrm{NL}}$ &$\Delta^{\mathrm{a}}_3(k_\mathrm{peak})$\\
\hline \multirow{2}{*} {M1} & $1.66\times10^9$ & $0.0358$ & $0.003$ & $1.56\times 10^{-6}$ & $2.9\times 10^{-6}$ &$-44.1$&$0.76 $ &$-31.2$\\
\cline { 2-9} & $5.97\times 10^{12}$ & $0.0333$ & $0.959$ & $1.24\times 10^{-13}$ & $9.8\times 10^{-3}$&$-56.6$&$ 0.82$ &$-36.2$\\
\hline \multirow{2}{*} {M2} & $5.48\times10^8$ & $0.0360$ & $0.002$ & $1.51\times 10^{-5}$ & $9.2\times 10^{-7}$ &$-40.6$ &$0.71 $ &$-28.9$\\
\cline { 2-9} & $7.05\times 10^{11}$ & $0.0285$ &$0.012$ & $7.89\times 10^{-12}$ & $1.2\times 10^{-3}$&$-21.2$ & $0.34 $ &$-17.5$\\
\hline
\end{tabular}
\caption{The peak amplitude of the power spectrum, PBH DM abundance for Gaussian approximation, the 3rd cumulant $\Delta_3$ at peak scales, non-Gaussianity parameter $f_{\mathrm{NL}}(k_{\text{peak}},k_{\text{peak}},k_{\text{peak}})$ and the critical frequency of SIGWs for models M1 and M2. In particular, $\Delta_3^\mathrm{n}$ is the numerical result, and $\Delta_3^\mathrm{a}$ is approximation.}
\label{results:tab}
\end{table}
	
From the upper panels of Fig. \ref{fig:bg}, there are two successive plateaus for $\phi(N)$, where the velocity of inflaton $\partial_N\phi$ dramatically decreases.
In other words, there are two transitory phases where the inflaton behaves like in ultra slow-roll (USR) inflation \cite{Kinney:2005vj,Dimopoulos:2017ged} which correspond to the two valleys of $\epsilon(N)$ in the lower panels and lead to the double peaks in the power spectrum, as shown in Fig. \ref{fig:spectrum}. 
The second USR phase takes over {before the end of the first one}, which keeps $e$-folds $N$ within a reasonable range.
From Fig. \ref{fig:spectrum}, both the power spectra produced by M1 and M2 are the order of $\mathcal{O}(10^{-9})$ at large scales, satisfying the constraints from CMB. 
At small scales, the power spectra are enhanced to the order of $\mathcal{O}(10^{-2})$, which serves to produce PBHs after the horizon reentry.
The power spectra also satisfy the constraints from CMB $\mu$-distortion, big bang nucleosynthesis (BBN) and pulsar timing array (PTA) observations \cite{Inomata:2018epa,Inomata:2016uip,Fixsen:1996nj}.

\section{primordial non-Gaussianity}\label{NG}
The inflaton in both models experiences a departure from slow roll during several $e$-folds, and the primordial non-Gaussianity may be very different from slow-roll inflation models which predict negligible non-Gaussianity. 
In this section, we compute the non-Gaussianity of the two models.

\begin{figure}[htp]
\centering
\includegraphics[width=0.7\linewidth]{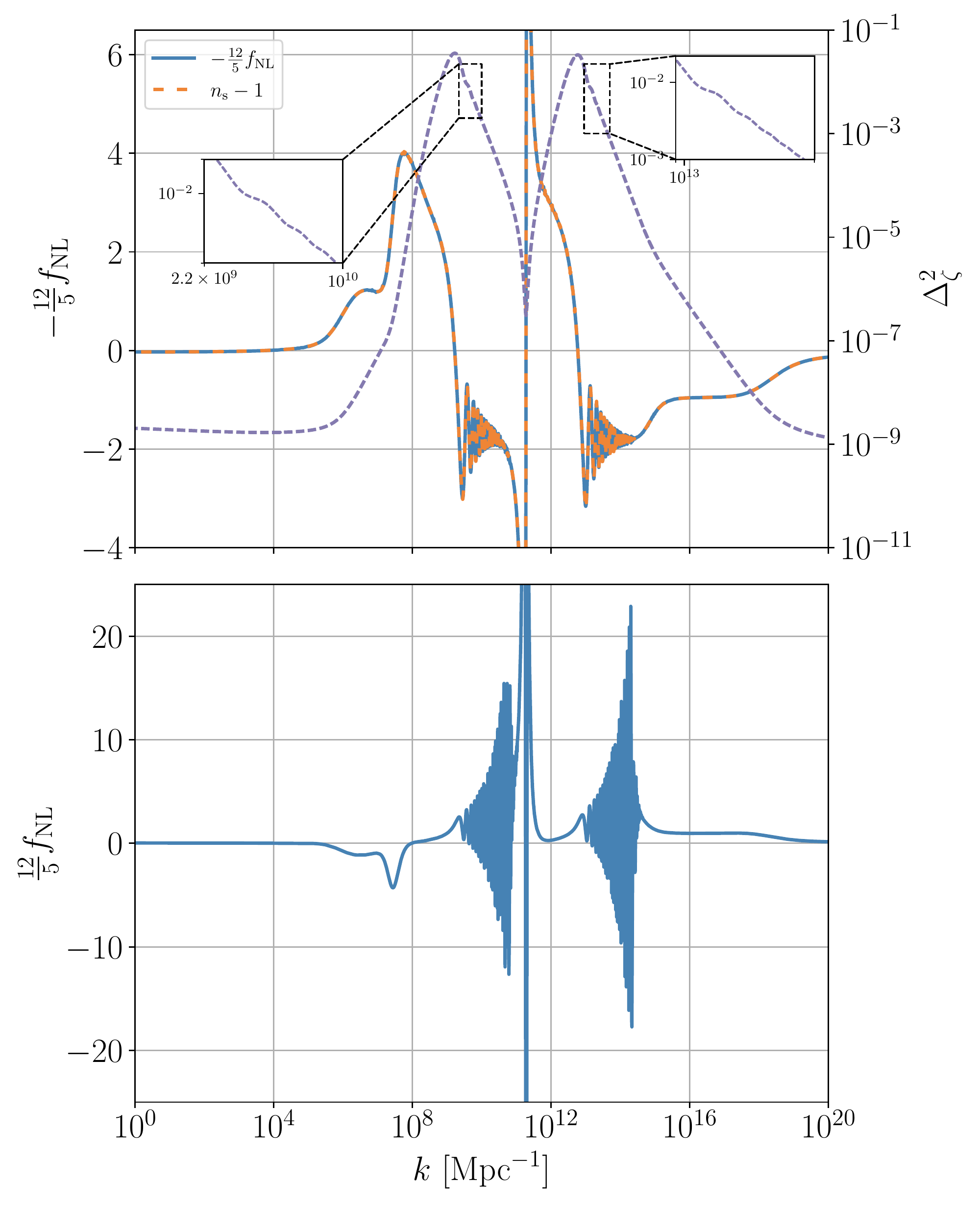}
\caption{The primordial scalar power spectrum $\Delta^2_\zeta$ and the non-Gaussianity parameter $f_{\mathrm{NL}}$ for model M1. 
In the upper panel, $-\frac{12}{5}f_{\mathrm{NL}}$ in squeezed limit coincides with the spectral tilt $n_\mathrm{s}-1$, and the power spectrum has two peaks. 
We have set $k_1=k_2=10^6 k_3=k$ for squeezed limit.
The insets show the wiggles in $\Delta^2_\zeta$ which leads to oscillations in $f_{\mathrm{NL}}$. 
The lower panel shows $\frac{12}{5}f_{\mathrm{NL}}$ in the equilateral limit for the modes $k_1=k_2=k_3=k$.}
\label{fig:kgfnlps}
\end{figure}

\begin{figure}[htp]
\centering
\includegraphics[width=0.7\linewidth]{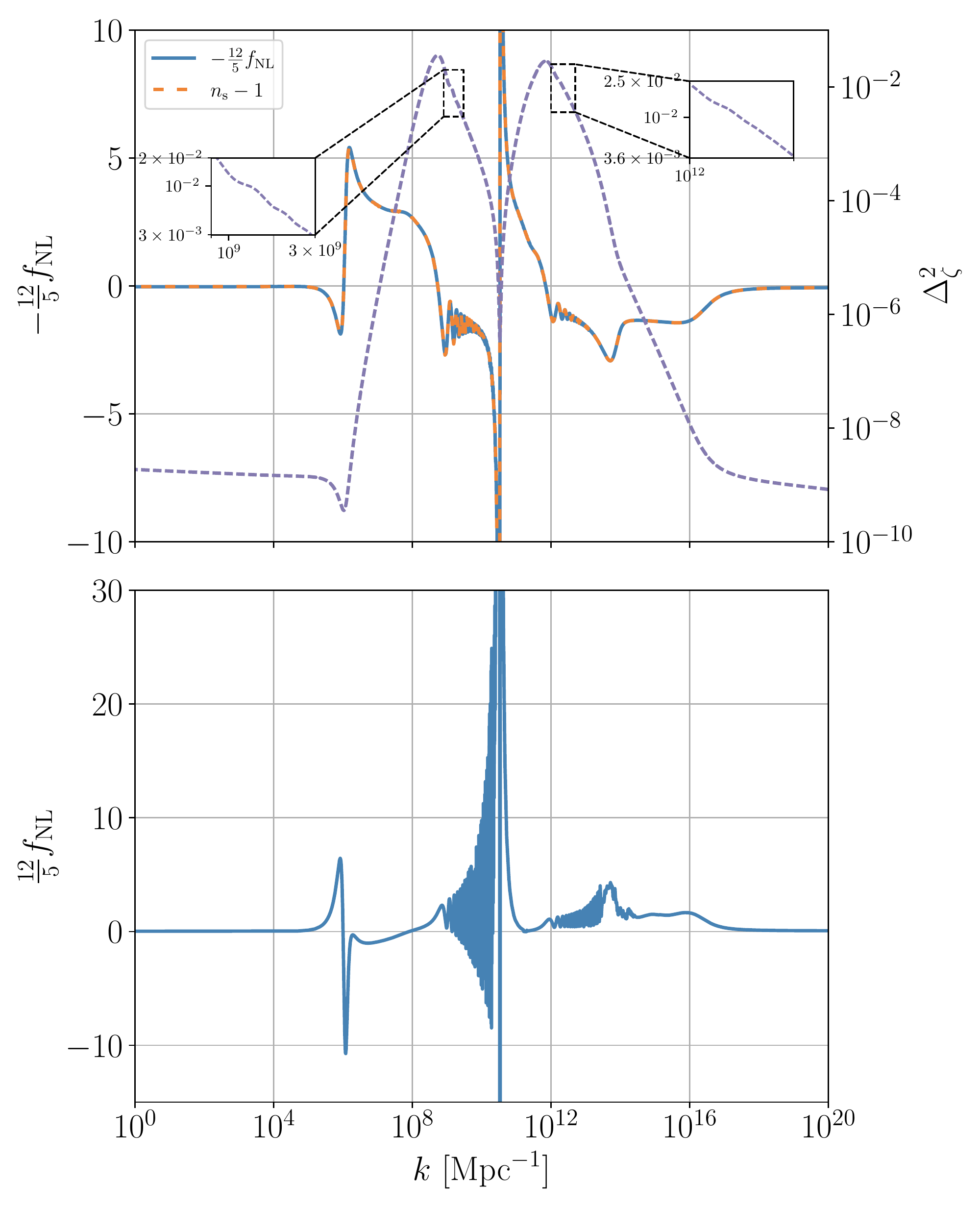}
\caption{The primordial scalar power spectrum $\Delta^2_\zeta$ and the non-Gaussianity parameter $f_{\mathrm{NL}}$ for the model M2.
In the upper panel, the power spectrum admits two peaks and the squeezed limit $-\frac{12}{5}f_{\mathrm{NL}}$ with $k_1=k_2=10^6 k_3=k$ coincides the scalar spectral tilt $n_\mathrm{s}-1$. 
The insets show the wiggles in $\Delta_\zeta^2$ which leads to oscillations in $f_{\mathrm{NL}}$.
The lower panel shows $\frac{12}{5}f_{\mathrm{NL}}$ in the equilateral limit.}
\label{fig:higgsfnlps}
\end{figure}

The bispectrum $B_{\zeta}$ is related to the three-point function as \cite{Byrnes:2010ft,Ade:2015ava} 
\begin{equation}\label{Bi}
\left\langle\hat{\zeta}_{\bm{k}_{1}}\hat{\zeta}_{\bm{k}_{2}}\hat{\zeta}_{\bm{k}_{3}}\right\rangle=(2 \pi)^{3} \delta^{3}\left(\bm{k}_{1}+\bm{k}_{2}+\bm{k}_{3}\right) B_{\zeta}\left(k_{1}, k_{2}, k_{3}\right),
\end{equation}
and the explicit lengthy expression of bispectrum $B_\zeta(k_1,k_2,k_3)$ can be found in Refs. \cite{Hazra:2012yn,Arroja:2011yj,Zhang:2020uek}.
The non-Gaussianity parameter $f_\text{NL}$ is defined as \cite{Creminelli:2006rz,Byrnes:2010ft}
\begin{equation}\label{Fnl}
f_{\text{NL}}(k_1,k_2,k_3)=\frac{5}{6}\frac{B_{\zeta}(k_1,k_2,k_3)}{P_{\zeta}(k_1)
P_{\zeta}(k_2)+P_{\zeta}(k_2)P_{\zeta}(k_3)+P_{\zeta}(k_3)P_{\zeta}(k_1)}.
\end{equation}
We numerically compute the non-Gaussianity parameter $f_{\mathrm{NL}}$ in the squeezed limit and the equilateral limit.
The results are shown in Figs. \ref{fig:kgfnlps} and \ref{fig:higgsfnlps} for models M1 and M2, respectively.
In squeezed limit, the non-Gaussianity parameter $f_{\mathrm{NL}}$ is related to the power index $n_\mathrm{s}-1$ by the consistency relation
\begin{equation}\label{consistency}
\lim_{k_3\rightarrow 0} f_{\mathrm{NL}}(k_1,k_2,k_3)=\frac{5}{12}(1-n_{\mathrm s}),\quad \text{for}~ k_1=k_2.
\end{equation}
The consistency relation was derived originally in the canonical single-field inflation with slow-roll conditions \cite{Maldacena:2002vr}. 
Then it was proved to be true for any inflationary model as long as the inflaton is the only dynamical field in addition to the gravitational field during inflation \cite{Creminelli:2004yq}. 
From the upper panels in Figs. \ref{fig:kgfnlps} and \ref{fig:higgsfnlps}, we can see that the consistency relation \eqref{consistency} holds for the two models.
Besides, the non-Gaussianity parameter $f_{\text{NL}}$ can reach as large as order $10$ at certain scales. 
Specifically, $\left|12f_{\mathrm{NL}}\right|/5\simeq 4$ in the squeezed limit and $\left|12f_{\mathrm{NL}}\right|/5\simeq 25$ in the equilateral limit for model M1.
For model M2, $\left|12f_{\mathrm{NL}}\right|/5\simeq 5$ in the squeezed limit and $\left|12f_{\mathrm{NL}}\right|/5\simeq 10$ in the equilateral limit.
For both models M1 and M2, $f_{\text{NL}}$ has a pretty large value due to the steep growth and decrease of the power spectrum. 
Moreover, the oscillation nature of $f_{\text{NL}}$ in both squeezed and equilateral limits originates from the small waggles in the power spectrum, 
as can be seen from the insets of Figs. \ref{fig:kgfnlps} and \ref{fig:higgsfnlps}.
 
\section{PBHs}\label{PBHs}
When the primordial curvature perturbation reenters the horizon during radiation dominated era, if the density contrast exceeds the threshold, it may cause gravitational collapse to form PBHs. 
The mass of PBH is $\gamma M_{\mathrm{hor}}$, where $M_{\mathrm{hor}}$ is the horizon mass and we choose the factor
$\gamma= 0.2$ \cite{Carr:1975qj}. 
The current fractional energy density of PBHs with mass $M$ in DM is \cite{Carr:2016drx,Gong:2017qlj}
\begin{equation}
\label{fpbheq1}
\begin{split}
Y_{\text{PBH}}(M)=&\frac{\beta(M)}{3.94\times10^{-9}}\left(\frac{\gamma}{0.2}\right)^{1/2}
\left(\frac{g_*}{10.75}\right)^{-1/4}\times \left(\frac{0.12}{\Omega_{\text{DM}}h^2}\right)
\left(\frac{M}{M_\odot}\right)^{-1/2},
\end{split}
\end{equation}
where $M_{\odot}$ is the solar mass, $g_*$ is the effective degrees of freedom at the formation time, $\Omega_{\text{DM}}$ is the current	energy density parameter of DM, 
and $\beta$ is the fractional energy density of PBHs at the formation. 

\begin{figure}[htp]
\centering
\includegraphics[width=0.9\linewidth]{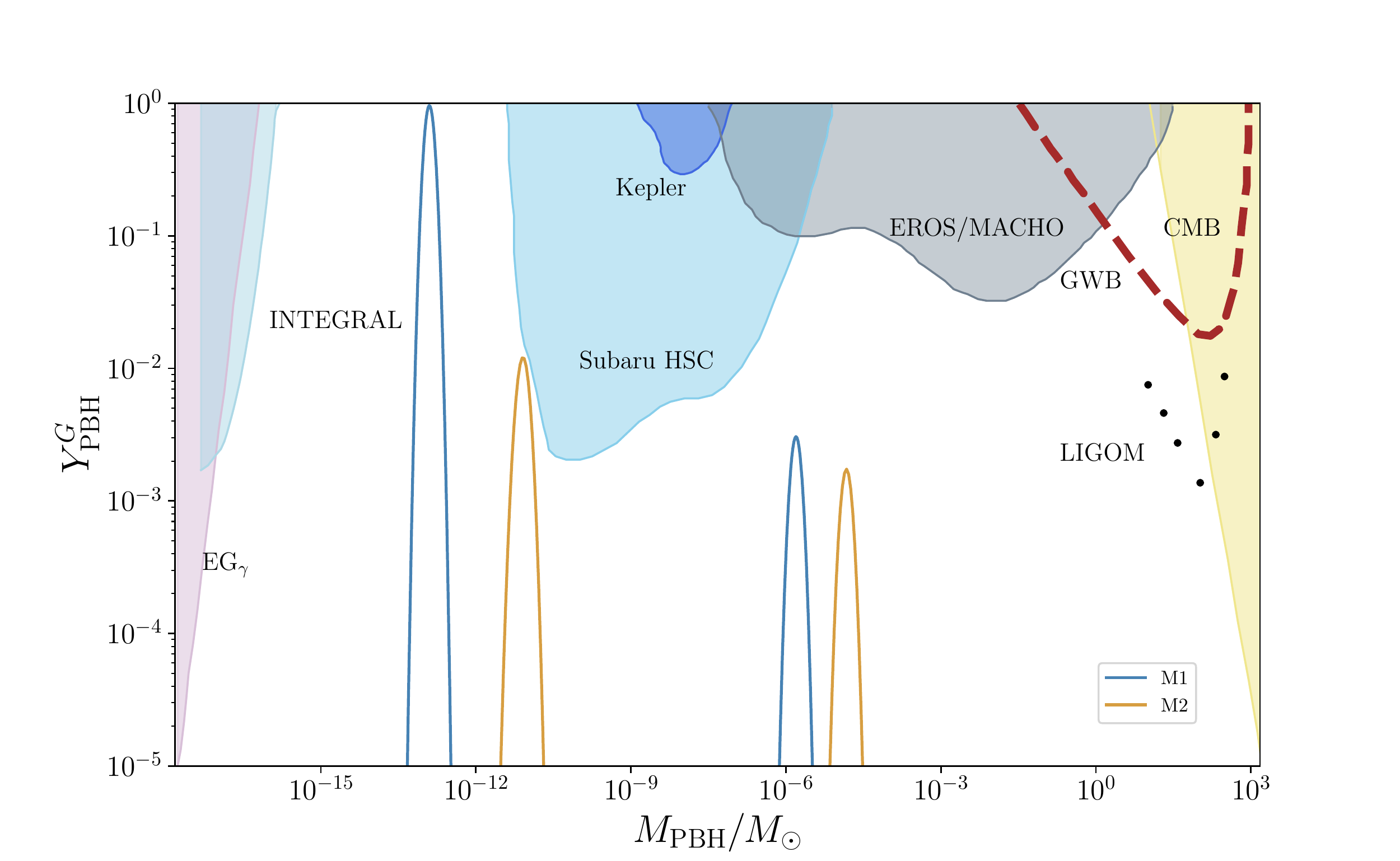}
\caption{The PBH DM abundance $Y^G_{\mathrm{PBH}}$ produced by model M1 and model M2 with Gaussian statistic perturbation. The shaded regions, the dashed line and the dotted line are the constraints on PBHs abundance from various observations, please refer to \cite{Sato-Polito:2019hws,Carr:2009jm,Laha:2019ssq,Dasgupta:2019cae,Niikura:2017zjd,Griest:2013esa,Tisserand:2006zx,Ali-Haimoud:2017rtz,Raidal:2017mfl,Ali-Haimoud:2016mbv,Poulin:2017bwe,Wang:2019kaf} and references therein for the details.
}
\label{fig:pbhs}
\end{figure}

For Gaussian statistic comoving curvature perturbation $\zeta$ \cite{Ozsoy:2018flq,Tada:2019amh}
\begin{equation}
\label{eq:beta}
\beta^G(M)\approx\sqrt{\frac{2}{\pi}}\frac{\sigma_R(M)}{\delta_c}
\exp\left(-\frac{\delta_c^2}{2\sigma^2_R(M)}\right),
\end{equation}
where $\delta_c$ is the critical density contrast for the PBHs formation, and $\sigma^2_R$ is the variance of $\delta_R$, the density contrast $\delta$ smoothed on the horizon scales $R=1/(aH)$
\begin{equation}
\delta_R(\bm{x})=\int d^3y \text{W}(|\bm{x}-\bm{y}|,R)\delta(\bm{y}),
\end{equation}
where $\text{W}$ is a window function. During radiation domination, the comoving curvature perturbation is related to density contrast as \cite{Musco:2018rwt,DeLuca:2019qsy},
\begin{equation}
  \delta(\bm{x})=\frac{4}{9}\left(\frac{1}{aH}\right)^2\nabla^2\zeta(\bm{x}).  
\end{equation}
Then
\begin{equation}
\label{sigmaeq1}
\sigma^2_R=\left(\frac{4}{9}\right)^2\int \frac{dk}{k} \widetilde{\text{W}}^2(kR)(kR)^4\Delta^2_{\zeta}(k),
\end{equation}
$\widetilde{\text{W}}$ is the Fourier transform of window function. 
We will use a Gaussian window function
 $\widetilde{\mathrm{W}}(x)=\mathrm{e}^{-x^2/2}$. The effective degrees of freedom $g_*=107.5$ for $T>300$ GeV
and $g_*=10.75$ for $0.5\text{~MeV}<T<300\text{~GeV}$.
We take the observational value $\Omega_{\text{DM}}h^2=0.12$ \cite{Aghanim:2018eyx}
and $\delta_c=0.4$ \cite{Harada:2013epa,Tada:2019amh,Escriva:2019phb}
in the calculation of PBH abundance \footnote{For the dependence of $\delta_c$ on the shape of the power spectrum, please refer to \cite{Musco:2018rwt,Musco:2020jjb} }.
The relation between the PBH mass $M$ and the scale $k$ is \cite{Gong:2017qlj}
\begin{equation}
\label{mkeq1}
M(k)=3.68\left(\frac{\gamma}{0.2}\right)\left(\frac{g_*}{10.75}\right)^{-1/6}
\left(\frac{k}{10^6\ \text{Mpc}^{-1}}\right)^{-2} M_{\odot}.
\end{equation}
With Eqs. \eqref{fpbheq1}, \eqref{eq:beta}, \eqref{mkeq1} and the power spectrum obtained in section \ref{sec2}, we compute the PBH DM abundance with Gaussian perturbation, $Y^G_{\mathrm{PBH}}$, the results are shown in Table \ref{results:tab} and Fig. \ref{fig:pbhs}.
As excepted, there are two mass ranges of PBHs. 
From Fig. \ref{fig:pbhs}, for the mass range $M_{\mathrm{PBH}}\sim 10^{-13}-10^{-12}M_{\odot}$, the PBHs can be the dominance of DM in model M1.
The PBH abundance at the peak is about $Y^G_{\mathrm{PBH}}\simeq 0.959$, then almost all DM is PBHs. 
For M2, $Y^G_{\mathrm{PBH}}\simeq 0.012$ is about one percent. 
In the mass range $M_{\mathrm{PBH}}\sim 10^{-6}-10^{-5}M_{\odot}$, the peak of $Y^G_\mathrm{PBH}$ is about $0.003$ and $0.002$ for M1 and M2, respectively. 
In this mass range, PBHs can be used to explain the origin of Planet 9.
	
As shown in Sec. \ref{NG}, large non-Gaussianities are produced in M1 and M2, so it is necessary to consider the effect of non-Gaussianity on PBH abundance. 
The non-Gaussianity-corrected $\beta$ is related to $\beta^{G}$ from Gaussian spectrum by \cite{Franciolini:2018vbk,Kehagias:2019eil,Atal:2018neu,Riccardi:2021rlf}
\begin{equation}\label{betaNG}
 \beta=\text{e}^{\Delta_3}\beta^{G},
\end{equation}
where $\Delta_3$ is called the 3rd cumulant,
\begin{equation}\label{delta3}
\Delta_3=\frac{1}{3!}\left(\frac{\delta_c}{\sigma_R}\right)^2 S_3 \delta_c,
\end{equation}
and
\begin{equation}
S_3=\frac{\left\langle\delta_R(\bm{x}) \delta_R(\bm{x}) \delta_R(\bm{x})\right\rangle}{\sigma^4_R}.
\end{equation}
With Gaussian window function, after some calculation, we get
\begin{equation}\label{d3}
\begin{split}
\left\langle\delta_R(\bm{x}) \delta_R(\bm{x}) \delta_R(\bm{x})\right\rangle &=
-64\left(\frac{4}{9}\right)^3\frac{2}{\left(2\pi\right)^4}k^6 \\&  \quad
\times \int^{\infty}_0d u\int^{\infty}_0 d v\int^{u+v}_{|u-v|}d w u^3 v^3 w^3 \text{e}^{-u^2}\text{e}^{-v^2}\text{e}^{-w^2}B_{\zeta}(\sqrt{2}uk,\sqrt{2}vk,\sqrt{2}wk).
\end{split}
\end{equation}
Since the mass of PBHs is almost monochromatic, we can consider only the correction on peak scales.
This triple integral is too complicated to perform analytically.
We approximately replace $B_\zeta(\sqrt{2}uk,\sqrt{2}vk,\sqrt{2}wk)$ by $B_\zeta(\sqrt{2}k_{\mathrm{peak}},\sqrt{2}k_{\mathrm{peak}},\sqrt{2}k_{\mathrm{peak}})$ at peaks.
This will be justified by numerical computation in our models.

Around the peak, the integral is approximately
\begin{equation}
\left\langle\delta_R(\bm{x}) \delta_R(\bm{x}) \delta_R(\bm{x})\right\rangle_{\mathrm{peak}} \simeq -\frac{800}{19683\sqrt{3}\pi^3}k_{\text{peak}}^6
B_{\zeta}(\sqrt{2}k_{\text{peak}},\sqrt{2}k_{\text{peak}},\sqrt{2}k_{\text{peak}}).
\end{equation}
Considering that $B_{\zeta}$ scaling as $k^{-6}$, then
\begin{equation}\label{apd3}
\left\langle\delta_R(\bm{x}) \delta_R(\bm{x}) \delta_R(\bm{x})\right\rangle_{\mathrm{peak}} \simeq -\frac{1}{8}\frac{800}{19683\sqrt{3}\pi^3}k_{\text{peak}}^6
B_{\zeta}(k_{\text{peak}},k_{\text{peak}},k_{\text{peak}}).
\end{equation}
Substituting Eqs. \eqref{sigmaeq1} and \eqref{apd3} into \eqref{delta3}, we get
\begin{equation}\label{effector}
\begin{split}
\Delta_3(k_{\mathrm{peak}}) &\simeq -\frac{80\sqrt{3}\pi \delta^3_c}{19683}\frac{\left(\Delta^2_{\zeta}(k_{\text{peak}})\right)^2}{\sigma^6_R}f_{\mathrm{NL}}(k_{\text{peak}},k_{\text{peak}},k_{\text{peak}}) \\ &
\sim -\frac{135\sqrt{3}\pi}{32}\frac{\delta^3_c}{\Delta^2_{\zeta}(k_{\text{peak}})}f_{\mathrm{NL}}(k_{\text{peak}},k_{\text{peak}},k_{\text{peak}}) \\ & \approx  -23\frac{\delta^3_c}{\Delta^2_{\zeta}(k_{\text{peak}})}f_{\mathrm{NL}}(k_{\text{peak}},k_{\text{peak}},k_{\text{peak}}),
\end{split}
\end{equation}
where we used the approximation $\sigma^2_R \sim 8\Delta^2_\zeta/81$. Note that the expression of $\Delta_3(k_{\mathrm{peak}})$ \eqref{effector} is underestimated because $\sigma^2_R$ is actually smaller than $8\Delta^2_\zeta/81$. 
The effect of non-Gaussianity of curvature perturbation $\zeta$ on PBH abundance is significant unless $|\Delta_3(k_{\mathrm{peak}})|\lesssim 1$, namely $|f_{\mathrm{NL}}(k_{\text{peak}},k_{\text{peak}},k_{\text{peak}})|\lesssim \Delta^2_{\zeta}(k_{\text{peak}})/(23\delta^3_c)\sim \mathcal{O}(10^{-2})$.
The numerical result $\Delta^{\mathrm{n}}_3(k_{\mathrm{peak}})$ and approximate result $\Delta^{\mathrm{a}}_3(k_\mathrm{peak})$ given by Eq. \eqref{effector} are shown in Table \ref{results:tab}. We see that the analytic and numerical results are of the same order. The factor $\text{e}^{\Delta_3}$ can be extremely small as $\Delta_3\lesssim -10$, and thus $\beta$ is rather infinitesimal. 
It seems that even though the power spectrum of $\zeta$ is amplified to order $\mathcal{O}(10^{-2})$, the production of PBHs may be violently suppressed by the effect from non-Gaussianities. 
But honestly, enhancement is also possible as long as $\Delta_3>0$, that is, $f_{\mathrm{NL}}(k_{\text{peak}},k_{\text{peak}},k_{\text{peak}})<0$ for some models.

The above discussion implies that the abundance of PBH DM could be highly overestimated (or underestimated) without considering non-Gaussianity of $\zeta$. In our model, $f_{\mathrm{NL}}(k_{\text{peak}},k_{\text{peak}},k_{\text{peak}})\sim \mathcal{O}(1)$, which leads to a suppression of PBH formation.
	
\section{SIGWs}\label{SIGWs}
In Sec. \ref{sec2}, we got the power spectrum enhanced at small scales, and we expect that the SIGWs are also amplified. 
In this section, we compute the energy density fraction of SIGWs.
	 
In Newtonian gauge\footnote{For a discussion on the gauge issue of SIGWs, see \cite{Lu:2020diy,Ali:2020sfw,Chang:2020tji,Chang:2020iji,Chang:2020mky,Domenech:2020xin,Inomata:2019yww,Tomikawa:2019tvi,Yuan:2019fwv,DeLuca:2019ufz}}, the perturbed metric is
\begin{equation}
ds^2=a^2(\tau)\left[-(1+2\Phi)d\tau^2+\left\{(1-2\Phi)\delta_{ij}+\frac{1}{2}h_{ij}\right\}dx^i x^j\right],
\end{equation}
where ${\Phi}$ is the Bardeen potential, and the second-order tensor perturbation ${h_{ij}}$ is transverse and traceless, ${\partial_{i}h_{ij}=h_{ii}=0}$. In Fourier space, the tensor perturbation is
\begin{equation}
h_{ij}\left(\bm{x},\tau\right)=\int\frac{d^3\bm{k}}{\left(2\pi\right)^{3/2}}\text{e}^{i\bm{k}\cdot\bm{x}}\left[h^{+}_{\bm{k}}\left(\tau\right)e^{+}_{ij}\left(\bm{k}\right)
+h^{\times}_{\bm{k}}\left(\tau\right)e^{\times}_{ij}\left(\bm{k}\right)\right],
\end{equation}
where the polarization tensors ${e^{+}_{ij}\left(\bm{k}\right)}$ and ${e^{\times}_{ij}\left(\bm{k}\right)}$ are
\begin{equation}
\begin{split}
&e^{+}_{ij}\left(\bm{k}\right)=\frac{1}{\sqrt{2}}\left[e_i\left(\bm{k}\right)e_j\left(\bm{k}\right)
-\overline{e}_i\left(\bm{k}\right)\overline{e}_j\left(\bm{k}\right)\right],\\& 
e^{\times}_{ij}\left(\bm{k}\right)=\frac{1}{\sqrt{2}}\left[e_i\left(\bm{k}\right)
\overline{e}_j\left(\bm{k}\right)+\overline{e}_i\left(\bm{k}\right)e_j\left(\bm{k}\right)\right],
\end{split}
\end{equation}
with ${e_{i}\left(\bm{k}\right)}$ and ${\overline{e}_{i}\left(\bm{k}\right)}$ are two orthonormal basis vectors which are orthogonal to the wave vector ${\bm{k}}$.
Neglecting the anisotropic stress, the equation of motion for the tensor perturbation $h_{\bm k}$ of either polarization is
\begin{equation}\label{E4}
h^{''}_{\bm{k}}\left(\tau\right)+2\mathcal{H}h^{'}_{\bm{k}}\left(\tau\right)+k^2h^{}_{\bm{k}}\left(\tau\right)=4S_{\bm{k}}\left(\tau\right),
\end{equation}
where ${\mathcal{H}=a'/a}$ and $S_{\bm{k}}\left(\tau\right)$ is 
\begin{equation}\label{ES}
S_{\bm{k}}\left(\tau\right)=\int\frac{d^3\tilde{\bm{k}}}{\left(2\pi\right)^{3/2}}
e^{ij}\left(\bm{k}\right)\tilde{k}_i\tilde{k}_j
f(\bm{k},\tilde{\bm{k}},\tau)\phi_{\tilde{\bm{k}}}\phi_{\bm{k}-\tilde{\bm{k}}},
\end{equation}
in which
\begin{equation}
\begin{split}
f(\bm{k},\tilde{\bm{k}},\tau)=&
2T_{\phi}(\tilde{k}\tau)T_{\phi}(|\bm{k}-\tilde{\bm{k}}|\tau)\\
&+\frac{4}{3\left(1+w\right)\mathcal{H}^2}
\left(T'_{\phi}(\tilde{k}\tau)+\mathcal{H}T_{\phi}(\tilde{k}\tau)\right)\left(T'_{\phi}(|\bm{k}-\tilde{\bm{k}}|\tau)+\mathcal{H} T_{\phi}(|\bm{k}-\tilde{\bm{k}}|\tau)\right),
\end{split}
\end{equation}
The transfer function $T_\phi$ for the Bardeen potential $\Phi$ is defined with its primordial value $\phi_{\bm{k}}$ as
\begin{equation}
\Phi_{\bm{k}}\left(\tau\right)=\phi_{\bm{k}}T_{\phi}\left(k\tau\right).
\end{equation}
The primordial value ${\phi_{\bm{k}}}$ is related to the comoving curvature perturbation as
\begin{equation}\label{Eps}
\langle\phi_{\bm{k}}\phi_{\bm{p}}\rangle=\delta^3\left(\bm{k}+\bm{p}\right)\frac{2\pi^2}{k^3}\left(\frac{3+3w}{5+3w}\right)^2
\mathcal{P}_{\zeta}\left(k\right),
\end{equation}
here $\mathcal{P}_\zeta$ is defined by
\begin{equation}
\begin{split}
\langle \hat{\zeta}_{\bm{k}}\hat{\zeta}_{\bm{k}'}\rangle&=
(2\pi)^3\delta^3\left(\bm{k}+\bm{k}'\right)|\zeta_k|^2\\&=
(2\pi)^3\delta^3\left(\bm{k}+\bm{k}'\right)\frac{2\pi^2}{k^3}\mathcal{P}_{\zeta}\left(k\right).
\end{split}
\end{equation}	
It is worth mentioning that we do not assume anything about the statistical nature of the primordial curvature perturbation, so $\mathcal{P}_\zeta$ is affected by the self-interactions of $\zeta$, in which the leading is the primordial non-Gaussianity, and there will be the difference between $\mathcal{P}_\zeta$ and $\Delta^2_\zeta$ obtained in Sec. \ref{sec2}.
	
The solution to Eq. \eqref{E4} is given by the Green function
\begin{equation}\label{E7}
h_{\bm{k}}\left(\tau\right)=\frac{4}{a\left(\tau\right)}\int^{\tau}d\overline{\tau}G_{k}\left(\tau,\overline{\tau}\right)
a\left(\overline{\tau}\right)
S_{\bm{k}}\left(\overline{\tau}\right),
\end{equation}
where the Green's function satisfies the equation
\begin{equation}
G^{''}_{\bm{k}}(\tau,\overline{\tau})+\left(k^2-\frac{a^{''}\left(\tau\right)}{a\left(\tau\right)}\right)G_{\bm{k}}\left(\tau,\overline{\tau}\right)=
\delta\left(\tau-\overline{\tau}\right).
\end{equation}
During radiation domination, ${w=1/3}$, the Green's function is 
\begin{equation}
G_{\bm{k}}\left(\tau,\overline{\tau}\right)=\frac{\sin\left[k\left(\tau-\overline{\tau}\right)\right]}{k},
\end{equation}
and the transfer function is 
\begin{equation}
T_{\phi}\left(x\right)=\frac{9}{x^2}\left(\frac{\sin\left(x/\sqrt{3}\right)}{x/\sqrt{3}}-\cos\left(x/\sqrt{3}\right)\right),
\end{equation}
where ${x=k\tau}$.
The power spectrum of tensor perturbation is defined in a similar way to $\mathcal{P}_\zeta$ as
\begin{equation}\label{Ep}
\langle h_{\bm{k}}\left(\tau\right)h_{\bm{p}}\left(\tau\right)\rangle
=\delta^3\left(\bm{k}+\bm{p}\right) \frac{2\pi^2}{k^3}\mathcal{P}_h\left(k,\tau\right).
\end{equation}
Combing the Eqs. \eqref{ES}, \eqref{E7} and \eqref{Ep},	we get the semianalytic expression for $\mathcal{P}_h$ \cite{Kohri:2018awv,Espinosa:2018eve}
\begin{equation}\label{Eph2}
\mathcal{P}_{h}\left(k,\tau\right)=4\int^{\infty}_{0}dv\int^{1+v}_{\left|1-v\right|}du \left(\frac{4v^2-\left(1+v^2-u^2\right)^2}{4vu}\right)^2
I^{2}_{RD}\left(u,v,x\right)\mathcal{P}_\zeta\left(vk\right)\mathcal{P}_\zeta\left(uk\right),
\end{equation}
where ${u=|\bm{k}-\tilde{\bm{k}}|/k, v=\tilde{|\bm{k}|}/k}$.
At late time, ${k\tau \gg 1\Leftrightarrow x\rightarrow \infty}$, the time average of ${I^{2}_{\mathrm{RD}}\left(u,v,x\rightarrow\infty \right)}$ is
\begin{equation}\label{IRD}
\begin{split}
\overline{I_{\mathrm{RD}}^2(v,u,x\rightarrow \infty)} =& \frac{1}{2x^2} \left( \frac{3(u^2+v^2-3)}{4 u^3 v^3 } \right)^2 \left\{ \left( -4uv+(u^2+v^2-3) \log \left| \frac{3-(u+v)^2}{3-(u-v)^2} \right| \right)^2  \right. \\
&  \left.\vphantom{\log \left| \frac{3-(u+v)^2}{3-(u-v)^2} \right|^2} \qquad \qquad \qquad \qquad \qquad   + \pi^2 (u^2+v^2-3)^2 \Theta \left( v+u-\sqrt{3}\right)\right\}. 
\end{split}
\end{equation}
The energy density parameter of SIGWs per logarithmic interval of ${k}$ is
\begin{equation}\label{EGW0}
\Omega_{\mathrm{GW}}\left(k\right)=\frac{1}{24}\left(\frac{k}{\mathcal{H\left(\tau\right)}}\right)^2\overline{\mathcal{P}_h\left(k,\tau\right)}.
\end{equation}
Notice that at late time, $\Omega_{\mathrm{GW}}$ is a constant since GWs behave like free radiation. With this featrue, we can also express the fractional energy density of SIGWs today $\Omega_{\mathrm{GW},0}$ as
\cite{Espinosa:2018eve}
\begin{equation}\label{EGW}
\Omega_{\mathrm{GW},0}\left(k\right)=\Omega_{\mathrm{GW}}\left(k\right)\frac{\Omega_{r,0}}{\Omega_{r}\left(\tau\right)},
\end{equation}
where we choose $\Omega_{r}=1$ because we are considering the radiation domination.
	
As mentioned above, the perturbations that induce SIGWs can be non-Gaussian, in order to estimate the contribution of non-Gaussianity to SIGWs, using the nonlinear coupling constant $f_{\text{NL}}$ defined as \cite{Verde:1999ij,Komatsu:2001rj}
\begin{equation}\label{local}
\zeta(\bm{x})=\zeta^G(\bm{x})+\frac{3}{5}f_\text{NL}(\zeta^G(\bm{x})^2-\langle \zeta^G(\bm{x})^2 \rangle),
\end{equation}
where $\zeta^G$ is the linear Gaussian part of the curvature perturbation. Then the power spectrum of $\zeta$ taking into account of the non-Gaussianity can be expressed as
\begin{equation}\label{sumps}
\mathcal{P}_{\zeta}(k)=\mathcal{P}^G_{\zeta}(k)+\mathcal{P}^{NG}_{\zeta}(k),
\end{equation}
where $\mathcal{P}^G_{\zeta}(k)=\Delta^2_{\zeta}(k)$, and 
\begin{equation}
\mathcal{P}^{NG}_{\zeta}(k)
=\left(\frac{3}{5}\right)^2\frac{k^3}{2\pi}f^2_{\mathrm{NL}}
\int d^3\bm{p}\frac{\mathcal{P}^G_{\zeta}(p)}{p^3}
\frac{\mathcal{P}^G_{\zeta}(|\bm{k}-\bm{p}|)}{|\bm{k}-\bm{p}|^3}.
\end{equation}    
It was shown that the non-Gaussian contribution to SIGWs exceeds the Gaussian part if $ \left(\frac{3}{5}\right)^2f^2_{\text{NL}} \mathcal{P}^G_\zeta\gtrsim 1$ \cite{Cai:2018dig}.
Taking $f_\mathrm{NL}$ as the estimator to parameterize the magnitude of non-Gaussianity \cite{Chen:2006nt}, we find the contribution from the non-Gaussian curvature perturbations to SIGWs is negligible, because $f_{\text{NL}}\sim \mathcal{O}(1)$ is not very large at the peak where power spectrum $\mathcal{P}^G_\zeta$ culminates, as shown in Figs. \ref{fig:kgfnlps} and \ref{fig:higgsfnlps}. In the following, we safely use $\Delta^2_\zeta$ we get in Sec. \ref{sec2} instead of $\mathcal{P}_\zeta$ to compute SIGWs.

\begin{figure}[htp]
\centering
\includegraphics[width=0.7\linewidth]{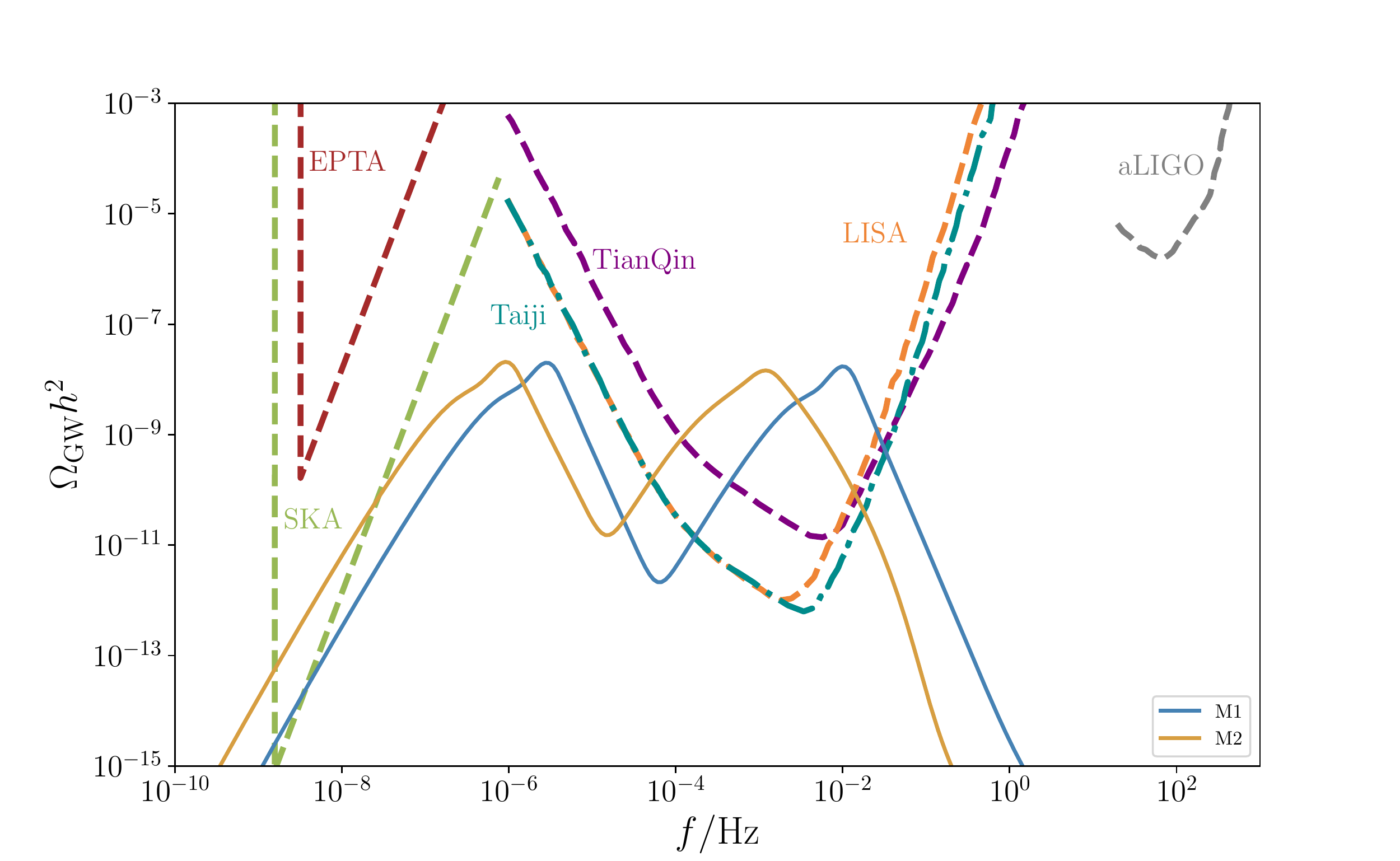}
\caption{The SIGWs generate by model M1 and model M2.
The fuchsia dashed curve denotes the EPTA limit \cite{Ferdman:2010xq,Hobbs:2009yy,McLaughlin:2013ira,Hobbs:2013aka} ,
the lime dashed curve denotes the SKA limit \cite{Moore:2014lga},
the purple dashed curve in the middle denotes the TianQin limit \cite{Luo:2015ght},
the dark cyan dotted-dashed curve shows the Taiji limit \cite{Hu:2017mde},
the orange dashed curve shows the LISA limit \cite{Audley:2017drz},
and the gray dashed curve denotes the aLIGO limit \cite{Harry:2010zz,TheLIGOScientific:2014jea}.}
\label{fig:sigws}
\end{figure}

We compute the $\Omega_{\mathrm{GW}}$ and the results are shown in Table \ref{results:tab} and Fig. \ref{fig:sigws}. As expected, there are also double peaks in SIGWs at the corresponding scales, and both the peaks are observable in the future. The milli-Hertz SIGWs can be tested by TianQin, Taiji, and LISA. 
The $10^{-6}$ Hz SIGWs leave a blue-tilted spectrum on the SKA region.

\section{conclusion}\label{conclusion}
For PBHs to be all (or most of) the DM, the violation of slow-roll conditions in single-field inflation is necessary as the slow-roll inflation predicts nearly scale-invariant power spectrum of order $\mathcal{O}(10^{-9})$ which hardly generates a significant abundance of PBHs and observable SIGWs. 
On the other hand, the violation of slow-roll conditions may lead to considerable primordial non-Gaussianity of curvature perturbation. 
We find that the abundance of PBHs is sensitive to the primordial non-Gaussianity of curvature perturbation and  will be altered significantly if the non-Gaussianity parameter $|f_{\mathrm{NL}}(k_{\text{peak}},k_{\text{peak}},k_{\text{peak}})|\gtrsim \Delta^2_{\zeta}(k_{\mathrm{peak}})/(23\delta^3_c) \sim \mathcal{O}(10^{-2})$.
Whether the PBH abundance is suppressed or enhanced depends on $f_{\mathrm{NL}}(k_{\text{peak}},k_{\text{peak}},k_{\text{peak}})$ being positive or negative, respectively. 

We propose a noncanonical inflation model with a double-peak coupling function in the noncanonical kinetic term, which is equivalent to a canonical model with double quasi-inflection points in the potential by field-redefinition.
The model driven by both $\lambda\phi^{2/5}$ potential and Higgs potential can amplify the power spectrum to $\mathcal{O}(10^{-2})$ at two different small scales while satisfying the constraints from CMB at large scales and keeping $e$-folds $N\sim 69$. 
We also numerically calculate the primordial non-Gaussianity in squeezed and equilateral limits.
We find that the consistency relation is valid even when the slow-roll conditions are violated.
Regrettably, our model predicts large non-Gaussianity with $f_{\mathrm{NL}}(k_{\text{peak}},k_{\text{peak}},k_{\text{peak}}) \sim \mathcal{O}(1)$ that is positive and the PBH abundance is suppressed saliently, even though the power spectrum is about $\mathcal{O}(0.01)$.
However, the energy density of SIGWs is insensitive to primordial non-Gaussianity and thus observable. 
In this regard, SIGWs are better at characterizing the small-scale power spectrum.
In our models, SIGWs will leave a blue-tilted spectrum in SKA sensitivity, and a peak in the sensitivities of LISA, Taiji and TianQin.
In addition, it is possible that we observe SIGWs in SGWB, but PBHs hardly exist in the corresponding mass ranges, taking into account the primordial non-Gaussianity.

\begin{acknowledgments}
The author F.Z. would like to thank Prof. Yungui Gong for useful discussion. This research was supported in part by the National Natural Science Foundation of China under Grant No. 11875136; 
and the Major Program of the National Natural Science Foundation of China under Grant No. 11690021.
\end{acknowledgments}


\begin{thebibliography}{130}%
\makeatletter
\providecommand \@ifxundefined [1]{%
 \@ifx{#1\undefined}
}%
\providecommand \@ifnum [1]{%
 \ifnum #1\expandafter \@firstoftwo
 \else \expandafter \@secondoftwo
 \fi
}%
\providecommand \@ifx [1]{%
 \ifx #1\expandafter \@firstoftwo
 \else \expandafter \@secondoftwo
 \fi
}%
\providecommand \natexlab [1]{#1}%
\providecommand \enquote  [1]{``#1''}%
\providecommand \bibnamefont  [1]{#1}%
\providecommand \bibfnamefont [1]{#1}%
\providecommand \citenamefont [1]{#1}%
\providecommand \href@noop [0]{\@secondoftwo}%
\providecommand \href [0]{\begingroup \@sanitize@url \@href}%
\providecommand \@href[1]{\@@startlink{#1}\@@href}%
\providecommand \@@href[1]{\endgroup#1\@@endlink}%
\providecommand \@sanitize@url [0]{\catcode `\\12\catcode `\$12\catcode
  `\&12\catcode `\#12\catcode `\^12\catcode `\_12\catcode `\%12\relax}%
\providecommand \@@startlink[1]{}%
\providecommand \@@endlink[0]{}%
\providecommand \url  [0]{\begingroup\@sanitize@url \@url }%
\providecommand \@url [1]{\endgroup\@href {#1}{\urlprefix }}%
\providecommand \urlprefix  [0]{URL }%
\providecommand \Eprint [0]{\href }%
\providecommand \doibase [0]{https://doi.org/}%
\providecommand \selectlanguage [0]{\@gobble}%
\providecommand \bibinfo  [0]{\@secondoftwo}%
\providecommand \bibfield  [0]{\@secondoftwo}%
\providecommand \translation [1]{[#1]}%
\providecommand \BibitemOpen [0]{}%
\providecommand \bibitemStop [0]{}%
\providecommand \bibitemNoStop [0]{.\EOS\space}%
\providecommand \EOS [0]{\spacefactor3000\relax}%
\providecommand \BibitemShut  [1]{\csname bibitem#1\endcsname}%
\let\auto@bib@innerbib\@empty
\bibitem [{\citenamefont {Abbott}\ {\it
  et~al.}(2016{\natexlab{a}})\citenamefont {Abbott} {\it
  et~al.}}]{Abbott:2016nmj}%
  \BibitemOpen
  \bibfield  {author} {\bibinfo {author} {\bibfnamefont {B.~P.}\ \bibnamefont
  {Abbott}} {\it et~al.} (\bibinfo {collaboration} {LIGO Scientific, Virgo}),\
  }\bibinfo {title} {{GW151226: Observation of Gravitational Waves from a
  22-Solar-Mass Binary Black Hole Coalescence}},\ \href
  {https://doi.org/10.1103/PhysRevLett.116.241103} {\bibfield  {journal}
  {\bibinfo  {journal} {Phys. Rev. Lett.}\ }\textbf {\bibinfo {volume} {116}},\
  \bibinfo {pages} {241103} (\bibinfo {year} {2016}{\natexlab{a}})},\ \Eprint
  {https://arxiv.org/abs/1606.04855} {arXiv:1606.04855} \BibitemShut {NoStop}%
\bibitem [{\citenamefont {Abbott}\ {\it
  et~al.}(2016{\natexlab{b}})\citenamefont {Abbott} {\it
  et~al.}}]{Abbott:2016blz}%
  \BibitemOpen
  \bibfield  {author} {\bibinfo {author} {\bibfnamefont {B.~P.}\ \bibnamefont
  {Abbott}} {\it et~al.} (\bibinfo {collaboration} {LIGO Scientific, Virgo}),\
  }\bibinfo {title} {{Observation of Gravitational Waves from a Binary Black
  Hole Merger}},\ \href {https://doi.org/10.1103/PhysRevLett.116.061102}
  {\bibfield  {journal} {\bibinfo  {journal} {Phys. Rev. Lett.}\ }\textbf
  {\bibinfo {volume} {116}},\ \bibinfo {pages} {061102} (\bibinfo {year}
  {2016}{\natexlab{b}})},\ \Eprint {https://arxiv.org/abs/1602.03837}
  {arXiv:1602.03837} \BibitemShut {NoStop}%
\bibitem [{\citenamefont {Abbott}\ {\it
  et~al.}(2017{\natexlab{a}})\citenamefont {Abbott} {\it
  et~al.}}]{Abbott:2017gyy}%
  \BibitemOpen
  \bibfield  {author} {\bibinfo {author} {\bibfnamefont {B.~. P.~.}\
  \bibnamefont {Abbott}} {\it et~al.} (\bibinfo {collaboration} {LIGO
  Scientific, Virgo}),\ }\bibinfo {title} {{GW170608: Observation of a
  19-solar-mass Binary Black Hole Coalescence}},\ \href
  {https://doi.org/10.3847/2041-8213/aa9f0c} {\bibfield  {journal} {\bibinfo
  {journal} {Astrophys. J. Lett.}\ }\textbf {\bibinfo {volume} {851}},\
  \bibinfo {pages} {L35} (\bibinfo {year} {2017}{\natexlab{a}})},\ \Eprint
  {https://arxiv.org/abs/1711.05578} {arXiv:1711.05578} \BibitemShut {NoStop}%
\bibitem [{\citenamefont {Abbott}\ {\it
  et~al.}(2017{\natexlab{b}})\citenamefont {Abbott} {\it
  et~al.}}]{TheLIGOScientific:2017qsa}%
  \BibitemOpen
  \bibfield  {author} {\bibinfo {author} {\bibfnamefont {B.~P.}\ \bibnamefont
  {Abbott}} {\it et~al.} (\bibinfo {collaboration} {LIGO Scientific, Virgo}),\
  }\bibinfo {title} {{GW170817: Observation of Gravitational Waves from a
  Binary Neutron Star Inspiral}},\ \href
  {https://doi.org/10.1103/PhysRevLett.119.161101} {\bibfield  {journal}
  {\bibinfo  {journal} {Phys. Rev. Lett.}\ }\textbf {\bibinfo {volume} {119}},\
  \bibinfo {pages} {161101} (\bibinfo {year} {2017}{\natexlab{b}})},\ \Eprint
  {https://arxiv.org/abs/1710.05832} {arXiv:1710.05832} \BibitemShut {NoStop}%
\bibitem [{\citenamefont {Abbott}\ {\it
  et~al.}(2017{\natexlab{c}})\citenamefont {Abbott} {\it
  et~al.}}]{Abbott:2017oio}%
  \BibitemOpen
  \bibfield  {author} {\bibinfo {author} {\bibfnamefont {B.~P.}\ \bibnamefont
  {Abbott}} {\it et~al.} (\bibinfo {collaboration} {LIGO Scientific, Virgo}),\
  }\bibinfo {title} {{GW170814: A Three-Detector Observation of Gravitational
  Waves from a Binary Black Hole Coalescence}},\ \href
  {https://doi.org/10.1103/PhysRevLett.119.141101} {\bibfield  {journal}
  {\bibinfo  {journal} {Phys. Rev. Lett.}\ }\textbf {\bibinfo {volume} {119}},\
  \bibinfo {pages} {141101} (\bibinfo {year} {2017}{\natexlab{c}})},\ \Eprint
  {https://arxiv.org/abs/1709.09660} {arXiv:1709.09660} \BibitemShut {NoStop}%
\bibitem [{\citenamefont {Abbott}\ {\it
  et~al.}(2017{\natexlab{d}})\citenamefont {Abbott} {\it
  et~al.}}]{Abbott:2017vtc}%
  \BibitemOpen
  \bibfield  {author} {\bibinfo {author} {\bibfnamefont {B.~P.}\ \bibnamefont
  {Abbott}} {\it et~al.} (\bibinfo {collaboration} {LIGO Scientific, VIRGO}),\
  }\bibinfo {title} {{GW170104: Observation of a 50-Solar-Mass Binary Black
  Hole Coalescence at Redshift 0.2}},\ \href
  {https://doi.org/10.1103/PhysRevLett.118.221101} {\bibfield  {journal}
  {\bibinfo  {journal} {Phys. Rev. Lett.}\ }\textbf {\bibinfo {volume} {118}},\
  \bibinfo {pages} {221101} (\bibinfo {year} {2017}{\natexlab{d}})},\ \bibinfo
  {note} {[Erratum: Phys.Rev.Lett. 121, 129901 (2018)]},\ \Eprint
  {https://arxiv.org/abs/1706.01812} {arXiv:1706.01812} \BibitemShut {NoStop}%
\bibitem [{\citenamefont {Abbott}\ {\it et~al.}(2019)\citenamefont {Abbott}
  {\it et~al.}}]{LIGOScientific:2018mvr}%
  \BibitemOpen
  \bibfield  {author} {\bibinfo {author} {\bibfnamefont {B.~P.}\ \bibnamefont
  {Abbott}} {\it et~al.} (\bibinfo {collaboration} {LIGO Scientific, Virgo}),\
  }\bibinfo {title} {{GWTC-1: A Gravitational-Wave Transient Catalog of Compact
  Binary Mergers Observed by LIGO and Virgo during the First and Second
  Observing Runs}},\ \href {https://doi.org/10.1103/PhysRevX.9.031040}
  {\bibfield  {journal} {\bibinfo  {journal} {Phys. Rev. X}\ }\textbf {\bibinfo
  {volume} {9}},\ \bibinfo {pages} {031040} (\bibinfo {year} {2019})},\ \Eprint
  {https://arxiv.org/abs/1811.12907} {arXiv:1811.12907} \BibitemShut {NoStop}%
\bibitem [{\citenamefont {Abbott}\ {\it
  et~al.}(2020{\natexlab{a}})\citenamefont {Abbott} {\it
  et~al.}}]{Abbott:2020khf}%
  \BibitemOpen
  \bibfield  {author} {\bibinfo {author} {\bibfnamefont {R.}~\bibnamefont
  {Abbott}} {\it et~al.} (\bibinfo {collaboration} {LIGO Scientific, Virgo}),\
  }\bibinfo {title} {{GW190814: Gravitational Waves from the Coalescence of a
  23 Solar Mass Black Hole with a 2.6 Solar Mass Compact Object}},\ \href
  {https://doi.org/10.3847/2041-8213/ab960f} {\bibfield  {journal} {\bibinfo
  {journal} {Astrophys. J. Lett.}\ }\textbf {\bibinfo {volume} {896}},\
  \bibinfo {pages} {L44} (\bibinfo {year} {2020}{\natexlab{a}})},\ \Eprint
  {https://arxiv.org/abs/2006.12611} {arXiv:2006.12611} \BibitemShut {NoStop}%
\bibitem [{\citenamefont {Abbott}\ {\it
  et~al.}(2020{\natexlab{b}})\citenamefont {Abbott} {\it
  et~al.}}]{Abbott:2020uma}%
  \BibitemOpen
  \bibfield  {author} {\bibinfo {author} {\bibfnamefont {B.~P.}\ \bibnamefont
  {Abbott}} {\it et~al.} (\bibinfo {collaboration} {LIGO Scientific, Virgo}),\
  }\bibinfo {title} {{GW190425: Observation of a Compact Binary Coalescence
  with Total Mass $\sim 3.4 M_{\odot}$}},\ \href
  {https://doi.org/10.3847/2041-8213/ab75f5} {\bibfield  {journal} {\bibinfo
  {journal} {Astrophys. J. Lett.}\ }\textbf {\bibinfo {volume} {892}},\
  \bibinfo {pages} {L3} (\bibinfo {year} {2020}{\natexlab{b}})},\ \Eprint
  {https://arxiv.org/abs/2001.01761} {arXiv:2001.01761} \BibitemShut {NoStop}%
\bibitem [{\citenamefont {Abbott}\ {\it
  et~al.}(2020{\natexlab{c}})\citenamefont {Abbott} {\it
  et~al.}}]{LIGOScientific:2020stg}%
  \BibitemOpen
  \bibfield  {author} {\bibinfo {author} {\bibfnamefont {R.}~\bibnamefont
  {Abbott}} {\it et~al.} (\bibinfo {collaboration} {LIGO Scientific, Virgo}),\
  }\bibinfo {title} {{GW190412: Observation of a Binary-Black-Hole Coalescence
  with Asymmetric Masses}},\ \href
  {https://doi.org/10.1103/PhysRevD.102.043015} {\bibfield  {journal} {\bibinfo
   {journal} {Phys. Rev. D}\ }\textbf {\bibinfo {volume} {102}},\ \bibinfo
  {pages} {043015} (\bibinfo {year} {2020}{\natexlab{c}})},\ \Eprint
  {https://arxiv.org/abs/2004.08342} {arXiv:2004.08342} \BibitemShut {NoStop}%
\bibitem [{\citenamefont {Carr}\ and\ \citenamefont
  {Hawking}(1974)}]{Carr:1974nx}%
  \BibitemOpen
  \bibfield  {author} {\bibinfo {author} {\bibfnamefont {B.~J.}\ \bibnamefont
  {Carr}}\ and\ \bibinfo {author} {\bibfnamefont {S.~W.}\ \bibnamefont
  {Hawking}},\ }\bibinfo {title} {{Black holes in the early Universe}},\
  \href@noop {} {\bibfield  {journal} {\bibinfo  {journal} {Mon. Not. Roy.
  Astron. Soc.}\ }\textbf {\bibinfo {volume} {168}},\ \bibinfo {pages} {399}
  (\bibinfo {year} {1974})}\BibitemShut {NoStop}%
\bibitem [{\citenamefont {Hawking}(1971)}]{Hawking:1971ei}%
  \BibitemOpen
  \bibfield  {author} {\bibinfo {author} {\bibfnamefont {S.}~\bibnamefont
  {Hawking}},\ }\bibinfo {title} {{Gravitationally collapsed objects of very
  low mass}},\ \href@noop {} {\bibfield  {journal} {\bibinfo  {journal} {Mon.
  Not. Roy. Astron. Soc.}\ }\textbf {\bibinfo {volume} {152}},\ \bibinfo
  {pages} {75} (\bibinfo {year} {1971})}\BibitemShut {NoStop}%
\bibitem [{\citenamefont {Bird}\ {\it et~al.}(2016)\citenamefont {Bird},
  \citenamefont {Cholis}, \citenamefont {Mu\~noz}, \citenamefont
  {Ali-Ha\"\i{}moud}, \citenamefont {Kamionkowski}, \citenamefont {Kovetz},
  \citenamefont {Raccanelli},\ and\ \citenamefont {Riess}}]{Bird:2016dcv}%
  \BibitemOpen
  \bibfield  {author} {\bibinfo {author} {\bibfnamefont {S.}~\bibnamefont
  {Bird}}, \bibinfo {author} {\bibfnamefont {I.}~\bibnamefont {Cholis}},
  \bibinfo {author} {\bibfnamefont {J.~B.}\ \bibnamefont {Mu\~noz}}, \bibinfo
  {author} {\bibfnamefont {Y.}~\bibnamefont {Ali-Ha\"\i{}moud}}, \bibinfo
  {author} {\bibfnamefont {M.}~\bibnamefont {Kamionkowski}}, \bibinfo {author}
  {\bibfnamefont {E.~D.}\ \bibnamefont {Kovetz}}, \bibinfo {author}
  {\bibfnamefont {A.}~\bibnamefont {Raccanelli}},\ and\ \bibinfo {author}
  {\bibfnamefont {A.~G.}\ \bibnamefont {Riess}},\ }\bibinfo {title} {{Did LIGO
  detect dark matter?}},\ \href
  {https://doi.org/10.1103/PhysRevLett.116.201301} {\bibfield  {journal}
  {\bibinfo  {journal} {Phys. Rev. Lett.}\ }\textbf {\bibinfo {volume} {116}},\
  \bibinfo {pages} {201301} (\bibinfo {year} {2016})},\ \Eprint
  {https://arxiv.org/abs/1603.00464} {arXiv:1603.00464} \BibitemShut {NoStop}%
\bibitem [{\citenamefont {Sasaki}\ {\it et~al.}(2016)\citenamefont {Sasaki},
  \citenamefont {Suyama}, \citenamefont {Tanaka},\ and\ \citenamefont
  {Yokoyama}}]{Sasaki:2016jop}%
  \BibitemOpen
  \bibfield  {author} {\bibinfo {author} {\bibfnamefont {M.}~\bibnamefont
  {Sasaki}}, \bibinfo {author} {\bibfnamefont {T.}~\bibnamefont {Suyama}},
  \bibinfo {author} {\bibfnamefont {T.}~\bibnamefont {Tanaka}},\ and\ \bibinfo
  {author} {\bibfnamefont {S.}~\bibnamefont {Yokoyama}},\ }\bibinfo {title}
  {{Primordial Black Hole Scenario for the Gravitational-Wave Event
  GW150914}},\ \href {https://doi.org/10.1103/PhysRevLett.117.061101}
  {\bibfield  {journal} {\bibinfo  {journal} {Phys. Rev. Lett.}\ }\textbf
  {\bibinfo {volume} {117}},\ \bibinfo {pages} {061101} (\bibinfo {year}
  {2016})},\ \bibinfo {note} {[Erratum: Phys.Rev.Lett. 121, 059901 (2018)]},\
  \Eprint {https://arxiv.org/abs/1603.08338} {arXiv:1603.08338} \BibitemShut
  {NoStop}%
\bibitem [{\citenamefont {Takhistov}\ {\it et~al.}(2021)\citenamefont
  {Takhistov}, \citenamefont {Fuller},\ and\ \citenamefont
  {Kusenko}}]{Takhistov:2020vxs}%
  \BibitemOpen
  \bibfield  {author} {\bibinfo {author} {\bibfnamefont {V.}~\bibnamefont
  {Takhistov}}, \bibinfo {author} {\bibfnamefont {G.~M.}\ \bibnamefont
  {Fuller}},\ and\ \bibinfo {author} {\bibfnamefont {A.}~\bibnamefont
  {Kusenko}},\ }\bibinfo {title} {{Test for the Origin of Solar Mass Black
  Holes}},\ \href {https://doi.org/10.1103/PhysRevLett.126.071101} {\bibfield
  {journal} {\bibinfo  {journal} {Phys. Rev. Lett.}\ }\textbf {\bibinfo
  {volume} {126}},\ \bibinfo {pages} {071101} (\bibinfo {year} {2021})},\
  \Eprint {https://arxiv.org/abs/2008.12780} {arXiv:2008.12780} \BibitemShut
  {NoStop}%
\bibitem [{\citenamefont {De~Luca}\ {\it
  et~al.}(2021{\natexlab{a}})\citenamefont {De~Luca}, \citenamefont
  {Desjacques}, \citenamefont {Franciolini}, \citenamefont {Pani},\ and\
  \citenamefont {Riotto}}]{DeLuca:2020sae}%
  \BibitemOpen
  \bibfield  {author} {\bibinfo {author} {\bibfnamefont {V.}~\bibnamefont
  {De~Luca}}, \bibinfo {author} {\bibfnamefont {V.}~\bibnamefont {Desjacques}},
  \bibinfo {author} {\bibfnamefont {G.}~\bibnamefont {Franciolini}}, \bibinfo
  {author} {\bibfnamefont {P.}~\bibnamefont {Pani}},\ and\ \bibinfo {author}
  {\bibfnamefont {A.}~\bibnamefont {Riotto}},\ }\bibinfo {title} {{GW190521
  Mass Gap Event and the Primordial Black Hole Scenario}},\ \href
  {https://doi.org/10.1103/PhysRevLett.126.051101} {\bibfield  {journal}
  {\bibinfo  {journal} {Phys. Rev. Lett.}\ }\textbf {\bibinfo {volume} {126}},\
  \bibinfo {pages} {051101} (\bibinfo {year} {2021}{\natexlab{a}})},\ \Eprint
  {https://arxiv.org/abs/2009.01728} {arXiv:2009.01728} \BibitemShut {NoStop}%
\bibitem [{\citenamefont {Abbott}\ {\it et~al.}(2021)\citenamefont {Abbott}
  {\it et~al.}}]{Abbott:2020niy}%
  \BibitemOpen
  \bibfield  {author} {\bibinfo {author} {\bibfnamefont {R.}~\bibnamefont
  {Abbott}} {\it et~al.} (\bibinfo {collaboration} {LIGO Scientific, Virgo}),\
  }\bibinfo {title} {{GWTC-2: Compact Binary Coalescences Observed by LIGO and
  Virgo During the First Half of the Third Observing Run}},\ \href
  {https://doi.org/10.1103/PhysRevX.11.021053} {\bibfield  {journal} {\bibinfo
  {journal} {Phys. Rev. X}\ }\textbf {\bibinfo {volume} {11}},\ \bibinfo
  {pages} {021053} (\bibinfo {year} {2021})},\ \Eprint
  {https://arxiv.org/abs/2010.14527} {arXiv:2010.14527} \BibitemShut {NoStop}%
\bibitem [{\citenamefont {De~Luca}\ {\it
  et~al.}(2021{\natexlab{b}})\citenamefont {De~Luca}, \citenamefont
  {Franciolini},\ and\ \citenamefont {Riotto}}]{DeLuca:2020agl}%
  \BibitemOpen
  \bibfield  {author} {\bibinfo {author} {\bibfnamefont {V.}~\bibnamefont
  {De~Luca}}, \bibinfo {author} {\bibfnamefont {G.}~\bibnamefont
  {Franciolini}},\ and\ \bibinfo {author} {\bibfnamefont {A.}~\bibnamefont
  {Riotto}},\ }\bibinfo {title} {{NANOGrav Data Hints at Primordial Black Holes
  as Dark Matter}},\ \href {https://doi.org/10.1103/PhysRevLett.126.041303}
  {\bibfield  {journal} {\bibinfo  {journal} {Phys. Rev. Lett.}\ }\textbf
  {\bibinfo {volume} {126}},\ \bibinfo {pages} {041303} (\bibinfo {year}
  {2021}{\natexlab{b}})},\ \Eprint {https://arxiv.org/abs/2009.08268}
  {arXiv:2009.08268} \BibitemShut {NoStop}%
\bibitem [{\citenamefont {Vaskonen}\ and\ \citenamefont
  {Veerm\"ae}(2021)}]{Vaskonen:2020lbd}%
  \BibitemOpen
  \bibfield  {author} {\bibinfo {author} {\bibfnamefont {V.}~\bibnamefont
  {Vaskonen}}\ and\ \bibinfo {author} {\bibfnamefont {H.}~\bibnamefont
  {Veerm\"ae}},\ }\bibinfo {title} {{Did NANOGrav see a signal from primordial
  black hole formation?}},\ \href
  {https://doi.org/10.1103/PhysRevLett.126.051303} {\bibfield  {journal}
  {\bibinfo  {journal} {Phys. Rev. Lett.}\ }\textbf {\bibinfo {volume} {126}},\
  \bibinfo {pages} {051303} (\bibinfo {year} {2021})},\ \Eprint
  {https://arxiv.org/abs/2009.07832} {arXiv:2009.07832} \BibitemShut {NoStop}%
\bibitem [{\citenamefont {Kohri}\ and\ \citenamefont
  {Terada}(2021)}]{Kohri:2020qqd}%
  \BibitemOpen
  \bibfield  {author} {\bibinfo {author} {\bibfnamefont {K.}~\bibnamefont
  {Kohri}}\ and\ \bibinfo {author} {\bibfnamefont {T.}~\bibnamefont {Terada}},\
  }\bibinfo {title} {{Solar-Mass Primordial Black Holes Explain NANOGrav Hint
  of Gravitational Waves}},\ \href
  {https://doi.org/10.1016/j.physletb.2020.136040} {\bibfield  {journal}
  {\bibinfo  {journal} {Phys. Lett. B}\ }\textbf {\bibinfo {volume} {813}},\
  \bibinfo {pages} {136040} (\bibinfo {year} {2021})},\ \Eprint
  {https://arxiv.org/abs/2009.11853} {arXiv:2009.11853} \BibitemShut {NoStop}%
\bibitem [{\citenamefont {Dom\`enech}\ and\ \citenamefont
  {Pi}(2020)}]{Domenech:2020ers}%
  \BibitemOpen
  \bibfield  {author} {\bibinfo {author} {\bibfnamefont {G.}~\bibnamefont
  {Dom\`enech}}\ and\ \bibinfo {author} {\bibfnamefont {S.}~\bibnamefont
  {Pi}},\ }\bibinfo {title} {{NANOGrav Hints on Planet-Mass Primordial Black
  Holes}},\ \href@noop {} {\  (\bibinfo {year} {2020})},\ \Eprint
  {https://arxiv.org/abs/2010.03976} {arXiv:2010.03976} \BibitemShut {NoStop}%
\bibitem [{\citenamefont {Atal}\ {\it et~al.}(2021)\citenamefont {Atal},
  \citenamefont {Sanglas},\ and\ \citenamefont {Triantafyllou}}]{Atal:2020yic}%
  \BibitemOpen
  \bibfield  {author} {\bibinfo {author} {\bibfnamefont {V.}~\bibnamefont
  {Atal}}, \bibinfo {author} {\bibfnamefont {A.}~\bibnamefont {Sanglas}},\ and\
  \bibinfo {author} {\bibfnamefont {N.}~\bibnamefont {Triantafyllou}},\
  }\bibinfo {title} {{NANOGrav signal as mergers of Stupendously Large
  Primordial Black Holes}},\ \href
  {https://doi.org/10.1088/1475-7516/2021/06/022} {J. Cosmol. Astropart. Phys.\
  \bibinfo {volume} {06}\bibfield  {year} {\bibinfo  {year} { (\textbf
  {2021})}\ }\bibfield  {pages} {\bibinfo  {pages} {022}},\ }\Eprint
  {https://arxiv.org/abs/2012.14721} {arXiv:2012.14721} \BibitemShut {NoStop}%
\bibitem [{\citenamefont {Yi}\ and\ \citenamefont {Zhu}(2021)}]{Yi:2021lxc}%
  \BibitemOpen
  \bibfield  {author} {\bibinfo {author} {\bibfnamefont {Z.}~\bibnamefont
  {Yi}}\ and\ \bibinfo {author} {\bibfnamefont {Z.-H.}\ \bibnamefont {Zhu}},\
  }\bibinfo {title} {{NANOGrav signal and LIGO-Virgo Primordial Black Holes
  from Higgs inflation}},\ \href@noop {} {\  (\bibinfo {year} {2021})},\
  \Eprint {https://arxiv.org/abs/2105.01943} {arXiv:2105.01943} \BibitemShut
  {NoStop}%
\bibitem [{\citenamefont {Sato-Polito}\ {\it et~al.}(2019)\citenamefont
  {Sato-Polito}, \citenamefont {Kovetz},\ and\ \citenamefont
  {Kamionkowski}}]{Sato-Polito:2019hws}%
  \BibitemOpen
  \bibfield  {author} {\bibinfo {author} {\bibfnamefont {G.}~\bibnamefont
  {Sato-Polito}}, \bibinfo {author} {\bibfnamefont {E.~D.}\ \bibnamefont
  {Kovetz}},\ and\ \bibinfo {author} {\bibfnamefont {M.}~\bibnamefont
  {Kamionkowski}},\ }\bibinfo {title} {{Constraints on the primordial curvature
  power spectrum from primordial black holes}},\ \href
  {https://doi.org/10.1103/PhysRevD.100.063521} {\bibfield  {journal} {\bibinfo
   {journal} {Phys. Rev. D}\ }\textbf {\bibinfo {volume} {100}},\ \bibinfo
  {pages} {063521} (\bibinfo {year} {2019})},\ \Eprint
  {https://arxiv.org/abs/1904.10971} {arXiv:1904.10971} \BibitemShut {NoStop}%
\bibitem [{\citenamefont {Akrami}\ {\it et~al.}(2020)\citenamefont {Akrami}
  {\it et~al.}}]{Akrami:2018odb}%
  \BibitemOpen
  \bibfield  {author} {\bibinfo {author} {\bibfnamefont {Y.}~\bibnamefont
  {Akrami}} {\it et~al.} (\bibinfo {collaboration} {Planck}),\ }\bibinfo
  {title} {{Planck 2018 results. X. Constraints on inflation}},\ \href
  {https://doi.org/10.1051/0004-6361/201833887} {\bibfield  {journal} {\bibinfo
   {journal} {Astron. Astrophys.}\ }\textbf {\bibinfo {volume} {641}},\
  \bibinfo {pages} {A10} (\bibinfo {year} {2020})},\ \Eprint
  {https://arxiv.org/abs/1807.06211} {arXiv:1807.06211} \BibitemShut {NoStop}%
\bibitem [{\citenamefont {Di}\ and\ \citenamefont {Gong}(2018)}]{Gong:2017qlj}%
  \BibitemOpen
  \bibfield  {author} {\bibinfo {author} {\bibfnamefont {H.}~\bibnamefont
  {Di}}\ and\ \bibinfo {author} {\bibfnamefont {Y.}~\bibnamefont {Gong}},\
  }\bibinfo {title} {{Primordial black holes and second order gravitational
  waves from ultra-slow-roll inflation}},\ \href
  {https://doi.org/10.1088/1475-7516/2018/07/007} {J. Cosmol. Astropart. Phys.\
  \bibinfo {volume} {07}\bibfield  {year} {\bibinfo  {year} { (\textbf
  {2018})}\ }\bibfield  {pages} {\bibinfo  {pages} {007}},\ }\Eprint
  {https://arxiv.org/abs/1707.09578} {arXiv:1707.09578} \BibitemShut {NoStop}%
\bibitem [{\citenamefont {Garcia-Bellido}\ and\ \citenamefont
  {Ruiz~Morales}(2017)}]{Garcia-Bellido:2017mdw}%
  \BibitemOpen
  \bibfield  {author} {\bibinfo {author} {\bibfnamefont {J.}~\bibnamefont
  {Garcia-Bellido}}\ and\ \bibinfo {author} {\bibfnamefont {E.}~\bibnamefont
  {Ruiz~Morales}},\ }\bibinfo {title} {{Primordial black holes from single
  field models of inflation}},\ \href
  {https://doi.org/10.1016/j.dark.2017.09.007} {\bibfield  {journal} {\bibinfo
  {journal} {Phys. Dark Univ.}\ }\textbf {\bibinfo {volume} {18}},\ \bibinfo
  {pages} {47} (\bibinfo {year} {2017})},\ \Eprint
  {https://arxiv.org/abs/1702.03901} {arXiv:1702.03901} \BibitemShut {NoStop}%
\bibitem [{\citenamefont {Germani}\ and\ \citenamefont
  {Prokopec}(2017)}]{Germani:2017bcs}%
  \BibitemOpen
  \bibfield  {author} {\bibinfo {author} {\bibfnamefont {C.}~\bibnamefont
  {Germani}}\ and\ \bibinfo {author} {\bibfnamefont {T.}~\bibnamefont
  {Prokopec}},\ }\bibinfo {title} {{On primordial black holes from an
  inflection point}},\ \href {https://doi.org/10.1016/j.dark.2017.09.001}
  {\bibfield  {journal} {\bibinfo  {journal} {Phys. Dark Univ.}\ }\textbf
  {\bibinfo {volume} {18}},\ \bibinfo {pages} {6} (\bibinfo {year} {2017})},\
  \Eprint {https://arxiv.org/abs/1706.04226} {arXiv:1706.04226} \BibitemShut
  {NoStop}%
\bibitem [{\citenamefont {Lu}\ {\it et~al.}(2019)\citenamefont {Lu},
  \citenamefont {Gong}, \citenamefont {Yi},\ and\ \citenamefont
  {Zhang}}]{Lu:2019sti}%
  \BibitemOpen
  \bibfield  {author} {\bibinfo {author} {\bibfnamefont {Y.}~\bibnamefont
  {Lu}}, \bibinfo {author} {\bibfnamefont {Y.}~\bibnamefont {Gong}}, \bibinfo
  {author} {\bibfnamefont {Z.}~\bibnamefont {Yi}},\ and\ \bibinfo {author}
  {\bibfnamefont {F.}~\bibnamefont {Zhang}},\ }\bibinfo {title} {{Constraints
  on primordial curvature perturbations from primordial black hole dark matter
  and secondary gravitational waves}},\ \href
  {https://doi.org/10.1088/1475-7516/2019/12/031} {J. Cosmol. Astropart. Phys.\
  \bibinfo {volume} {12}\bibfield  {year} {\bibinfo  {year} { (\textbf
  {2019})}\ }\bibfield  {pages} {\bibinfo  {pages} {031}},\ }\Eprint
  {https://arxiv.org/abs/1907.11896} {arXiv:1907.11896} \BibitemShut {NoStop}%
\bibitem [{\citenamefont {Motohashi}\ and\ \citenamefont
  {Hu}(2017)}]{Motohashi:2017kbs}%
  \BibitemOpen
  \bibfield  {author} {\bibinfo {author} {\bibfnamefont {H.}~\bibnamefont
  {Motohashi}}\ and\ \bibinfo {author} {\bibfnamefont {W.}~\bibnamefont {Hu}},\
  }\bibinfo {title} {{Primordial Black Holes and Slow-Roll Violation}},\ \href
  {https://doi.org/10.1103/PhysRevD.96.063503} {\bibfield  {journal} {\bibinfo
  {journal} {Phys. Rev. D}\ }\textbf {\bibinfo {volume} {96}},\ \bibinfo
  {pages} {063503} (\bibinfo {year} {2017})},\ \Eprint
  {https://arxiv.org/abs/1706.06784} {arXiv:1706.06784} \BibitemShut {NoStop}%
\bibitem [{\citenamefont {Espinosa}\ {\it
  et~al.}(2018{\natexlab{a}})\citenamefont {Espinosa}, \citenamefont {Racco},\
  and\ \citenamefont {Riotto}}]{Espinosa:2017sgp}%
  \BibitemOpen
  \bibfield  {author} {\bibinfo {author} {\bibfnamefont {J.~R.}\ \bibnamefont
  {Espinosa}}, \bibinfo {author} {\bibfnamefont {D.}~\bibnamefont {Racco}},\
  and\ \bibinfo {author} {\bibfnamefont {A.}~\bibnamefont {Riotto}},\ }\bibinfo
  {title} {{Cosmological Signature of the Standard Model Higgs Vacuum
  Instability: Primordial Black Holes as Dark Matter}},\ \href
  {https://doi.org/10.1103/PhysRevLett.120.121301} {\bibfield  {journal}
  {\bibinfo  {journal} {Phys. Rev. Lett.}\ }\textbf {\bibinfo {volume} {120}},\
  \bibinfo {pages} {121301} (\bibinfo {year} {2018}{\natexlab{a}})},\ \Eprint
  {https://arxiv.org/abs/1710.11196} {arXiv:1710.11196} \BibitemShut {NoStop}%
\bibitem [{\citenamefont {Belotsky}\ {\it et~al.}(2019)\citenamefont
  {Belotsky}, \citenamefont {Dokuchaev}, \citenamefont {Eroshenko},
  \citenamefont {Esipova}, \citenamefont {Khlopov}, \citenamefont {Khromykh},
  \citenamefont {Kirillov}, \citenamefont {Nikulin}, \citenamefont {Rubin},\
  and\ \citenamefont {Svadkovsky}}]{Belotsky:2018wph}%
  \BibitemOpen
  \bibfield  {author} {\bibinfo {author} {\bibfnamefont {K.~M.}\ \bibnamefont
  {Belotsky}}, \bibinfo {author} {\bibfnamefont {V.~I.}\ \bibnamefont
  {Dokuchaev}}, \bibinfo {author} {\bibfnamefont {Y.~N.}\ \bibnamefont
  {Eroshenko}}, \bibinfo {author} {\bibfnamefont {E.~A.}\ \bibnamefont
  {Esipova}}, \bibinfo {author} {\bibfnamefont {M.~Y.}\ \bibnamefont
  {Khlopov}}, \bibinfo {author} {\bibfnamefont {L.~A.}\ \bibnamefont
  {Khromykh}}, \bibinfo {author} {\bibfnamefont {A.~A.}\ \bibnamefont
  {Kirillov}}, \bibinfo {author} {\bibfnamefont {V.~V.}\ \bibnamefont
  {Nikulin}}, \bibinfo {author} {\bibfnamefont {S.~G.}\ \bibnamefont {Rubin}},\
  and\ \bibinfo {author} {\bibfnamefont {I.~V.}\ \bibnamefont {Svadkovsky}},\
  }\bibinfo {title} {{Clusters of primordial black holes}},\ \href
  {https://doi.org/10.1140/epjc/s10052-019-6741-4} {\bibfield  {journal}
  {\bibinfo  {journal} {Eur. Phys. J. C}\ }\textbf {\bibinfo {volume} {79}},\
  \bibinfo {pages} {246} (\bibinfo {year} {2019})},\ \Eprint
  {https://arxiv.org/abs/1807.06590} {arXiv:1807.06590} \BibitemShut {NoStop}%
\bibitem [{\citenamefont {Dalianis}\ {\it et~al.}(2020)\citenamefont
  {Dalianis}, \citenamefont {Karydas},\ and\ \citenamefont
  {Papantonopoulos}}]{Dalianis:2019vit}%
  \BibitemOpen
  \bibfield  {author} {\bibinfo {author} {\bibfnamefont {I.}~\bibnamefont
  {Dalianis}}, \bibinfo {author} {\bibfnamefont {S.}~\bibnamefont {Karydas}},\
  and\ \bibinfo {author} {\bibfnamefont {E.}~\bibnamefont {Papantonopoulos}},\
  }\bibinfo {title} {{Generalized Non-Minimal Derivative Coupling: Application
  to Inflation and Primordial Black Hole Production}},\ \href
  {https://doi.org/10.1088/1475-7516/2020/06/040} {J. Cosmol. Astropart. Phys.\
  \bibinfo {volume} {06}\bibfield  {year} {\bibinfo  {year} { (\textbf
  {2020})}\ }\bibfield  {pages} {\bibinfo  {pages} {040}},\ }\Eprint
  {https://arxiv.org/abs/1910.00622} {arXiv:1910.00622} \BibitemShut {NoStop}%
\bibitem [{\citenamefont {Passaglia}\ {\it et~al.}(2020)\citenamefont
  {Passaglia}, \citenamefont {Hu},\ and\ \citenamefont
  {Motohashi}}]{Passaglia:2019ueo}%
  \BibitemOpen
  \bibfield  {author} {\bibinfo {author} {\bibfnamefont {S.}~\bibnamefont
  {Passaglia}}, \bibinfo {author} {\bibfnamefont {W.}~\bibnamefont {Hu}},\ and\
  \bibinfo {author} {\bibfnamefont {H.}~\bibnamefont {Motohashi}},\ }\bibinfo
  {title} {{Primordial black holes as dark matter through Higgs field
  criticality}},\ \href {https://doi.org/10.1103/PhysRevD.101.123523}
  {\bibfield  {journal} {\bibinfo  {journal} {Phys. Rev. D}\ }\textbf {\bibinfo
  {volume} {101}},\ \bibinfo {pages} {123523} (\bibinfo {year} {2020})},\
  \Eprint {https://arxiv.org/abs/1912.02682} {arXiv:1912.02682} \BibitemShut
  {NoStop}%
\bibitem [{\citenamefont {Fu}\ {\it et~al.}(2019)\citenamefont {Fu},
  \citenamefont {Wu},\ and\ \citenamefont {Yu}}]{Fu:2019ttf}%
  \BibitemOpen
  \bibfield  {author} {\bibinfo {author} {\bibfnamefont {C.}~\bibnamefont
  {Fu}}, \bibinfo {author} {\bibfnamefont {P.}~\bibnamefont {Wu}},\ and\
  \bibinfo {author} {\bibfnamefont {H.}~\bibnamefont {Yu}},\ }\bibinfo {title}
  {{Primordial Black Holes from Inflation with Nonminimal Derivative
  Coupling}},\ \href {https://doi.org/10.1103/PhysRevD.100.063532} {\bibfield
  {journal} {\bibinfo  {journal} {Phys. Rev. D}\ }\textbf {\bibinfo {volume}
  {100}},\ \bibinfo {pages} {063532} (\bibinfo {year} {2019})},\ \Eprint
  {https://arxiv.org/abs/1907.05042} {arXiv:1907.05042} \BibitemShut {NoStop}%
\bibitem [{\citenamefont {Xu}\ {\it et~al.}(2020)\citenamefont {Xu},
  \citenamefont {Liu}, \citenamefont {Gao},\ and\ \citenamefont
  {Guo}}]{Xu:2019bdp}%
  \BibitemOpen
  \bibfield  {author} {\bibinfo {author} {\bibfnamefont {W.-T.}\ \bibnamefont
  {Xu}}, \bibinfo {author} {\bibfnamefont {J.}~\bibnamefont {Liu}}, \bibinfo
  {author} {\bibfnamefont {T.-J.}\ \bibnamefont {Gao}},\ and\ \bibinfo {author}
  {\bibfnamefont {Z.-K.}\ \bibnamefont {Guo}},\ }\bibinfo {title}
  {{Gravitational waves from double-inflection-point inflation}},\ \href
  {https://doi.org/10.1103/PhysRevD.101.023505} {\bibfield  {journal} {\bibinfo
   {journal} {Phys. Rev. D}\ }\textbf {\bibinfo {volume} {101}},\ \bibinfo
  {pages} {023505} (\bibinfo {year} {2020})},\ \Eprint
  {https://arxiv.org/abs/1907.05213} {arXiv:1907.05213} \BibitemShut {NoStop}%
\bibitem [{\citenamefont {Lin}\ {\it et~al.}(2020)\citenamefont {Lin},
  \citenamefont {Gao}, \citenamefont {Gong}, \citenamefont {Lu}, \citenamefont
  {Zhang},\ and\ \citenamefont {Zhang}}]{Lin:2020goi}%
  \BibitemOpen
  \bibfield  {author} {\bibinfo {author} {\bibfnamefont {J.}~\bibnamefont
  {Lin}}, \bibinfo {author} {\bibfnamefont {Q.}~\bibnamefont {Gao}}, \bibinfo
  {author} {\bibfnamefont {Y.}~\bibnamefont {Gong}}, \bibinfo {author}
  {\bibfnamefont {Y.}~\bibnamefont {Lu}}, \bibinfo {author} {\bibfnamefont
  {C.}~\bibnamefont {Zhang}},\ and\ \bibinfo {author} {\bibfnamefont
  {F.}~\bibnamefont {Zhang}},\ }\bibinfo {title} {{Primordial black holes and
  secondary gravitational waves from $k$ and $G$ inflation}},\ \href
  {https://doi.org/10.1103/PhysRevD.101.103515} {\bibfield  {journal} {\bibinfo
   {journal} {Phys. Rev. D}\ }\textbf {\bibinfo {volume} {101}},\ \bibinfo
  {pages} {103515} (\bibinfo {year} {2020})},\ \Eprint
  {https://arxiv.org/abs/2001.05909} {arXiv:2001.05909} \BibitemShut {NoStop}%
\bibitem [{\citenamefont {Yi}\ {\it et~al.}(2021{\natexlab{a}})\citenamefont
  {Yi}, \citenamefont {Gong}, \citenamefont {Wang},\ and\ \citenamefont
  {Zhu}}]{Yi:2020kmq}%
  \BibitemOpen
  \bibfield  {author} {\bibinfo {author} {\bibfnamefont {Z.}~\bibnamefont
  {Yi}}, \bibinfo {author} {\bibfnamefont {Y.}~\bibnamefont {Gong}}, \bibinfo
  {author} {\bibfnamefont {B.}~\bibnamefont {Wang}},\ and\ \bibinfo {author}
  {\bibfnamefont {Z.-h.}\ \bibnamefont {Zhu}},\ }\bibinfo {title} {{Primordial
  black holes and secondary gravitational waves from the Higgs field}},\ \href
  {https://doi.org/10.1103/PhysRevD.103.063535} {\bibfield  {journal} {\bibinfo
   {journal} {Phys. Rev. D}\ }\textbf {\bibinfo {volume} {103}},\ \bibinfo
  {pages} {063535} (\bibinfo {year} {2021}{\natexlab{a}})},\ \Eprint
  {https://arxiv.org/abs/2007.09957} {arXiv:2007.09957} \BibitemShut {NoStop}%
\bibitem [{\citenamefont {Yi}\ {\it et~al.}(2021{\natexlab{b}})\citenamefont
  {Yi}, \citenamefont {Gao}, \citenamefont {Gong},\ and\ \citenamefont
  {Zhu}}]{Yi:2020cut}%
  \BibitemOpen
  \bibfield  {author} {\bibinfo {author} {\bibfnamefont {Z.}~\bibnamefont
  {Yi}}, \bibinfo {author} {\bibfnamefont {Q.}~\bibnamefont {Gao}}, \bibinfo
  {author} {\bibfnamefont {Y.}~\bibnamefont {Gong}},\ and\ \bibinfo {author}
  {\bibfnamefont {Z.-h.}\ \bibnamefont {Zhu}},\ }\bibinfo {title} {{Primordial
  black holes and scalar-induced secondary gravitational waves from
  inflationary models with a noncanonical kinetic term}},\ \href
  {https://doi.org/10.1103/PhysRevD.103.063534} {\bibfield  {journal} {\bibinfo
   {journal} {Phys. Rev. D}\ }\textbf {\bibinfo {volume} {103}},\ \bibinfo
  {pages} {063534} (\bibinfo {year} {2021}{\natexlab{b}})},\ \Eprint
  {https://arxiv.org/abs/2011.10606} {arXiv:2011.10606} \BibitemShut {NoStop}%
\bibitem [{\citenamefont {Gao}\ {\it et~al.}(2021)\citenamefont {Gao},
  \citenamefont {Gong},\ and\ \citenamefont {Yi}}]{Gao:2020tsa}%
  \BibitemOpen
  \bibfield  {author} {\bibinfo {author} {\bibfnamefont {Q.}~\bibnamefont
  {Gao}}, \bibinfo {author} {\bibfnamefont {Y.}~\bibnamefont {Gong}},\ and\
  \bibinfo {author} {\bibfnamefont {Z.}~\bibnamefont {Yi}},\ }\bibinfo {title}
  {{Primordial black holes and secondary gravitational waves from natural
  inflation}},\ \href {https://doi.org/10.1016/j.nuclphysb.2021.115480}
  {\bibfield  {journal} {\bibinfo  {journal} {Nucl. Phys. B}\ }\textbf
  {\bibinfo {volume} {969}},\ \bibinfo {pages} {115480} (\bibinfo {year}
  {2021})},\ \Eprint {https://arxiv.org/abs/2012.03856} {arXiv:2012.03856}
  \BibitemShut {NoStop}%
\bibitem [{\citenamefont {Fumagalli}\ {\it
  et~al.}(2020{\natexlab{a}})\citenamefont {Fumagalli}, \citenamefont
  {Renaux-Petel}, \citenamefont {Ronayne},\ and\ \citenamefont
  {Witkowski}}]{Fumagalli:2020adf}%
  \BibitemOpen
  \bibfield  {author} {\bibinfo {author} {\bibfnamefont {J.}~\bibnamefont
  {Fumagalli}}, \bibinfo {author} {\bibfnamefont {S.}~\bibnamefont
  {Renaux-Petel}}, \bibinfo {author} {\bibfnamefont {J.~W.}\ \bibnamefont
  {Ronayne}},\ and\ \bibinfo {author} {\bibfnamefont {L.~T.}\ \bibnamefont
  {Witkowski}},\ }\bibinfo {title} {{Turning in the landscape: a new mechanism
  for generating Primordial Black Holes}},\ \href@noop {} {\  (\bibinfo {year}
  {2020}{\natexlab{a}})},\ \Eprint {https://arxiv.org/abs/2004.08369}
  {arXiv:2004.08369} \BibitemShut {NoStop}%
\bibitem [{\citenamefont {Gundhi}\ {\it et~al.}(2021)\citenamefont {Gundhi},
  \citenamefont {Ketov},\ and\ \citenamefont {Steinwachs}}]{Gundhi:2020kzm}%
  \BibitemOpen
  \bibfield  {author} {\bibinfo {author} {\bibfnamefont {A.}~\bibnamefont
  {Gundhi}}, \bibinfo {author} {\bibfnamefont {S.~V.}\ \bibnamefont {Ketov}},\
  and\ \bibinfo {author} {\bibfnamefont {C.~F.}\ \bibnamefont {Steinwachs}},\
  }\bibinfo {title} {{Primordial black hole dark matter in dilaton-extended
  two-field Starobinsky inflation}},\ \href
  {https://doi.org/10.1103/PhysRevD.103.083518} {\bibfield  {journal} {\bibinfo
   {journal} {Phys. Rev. D}\ }\textbf {\bibinfo {volume} {103}},\ \bibinfo
  {pages} {083518} (\bibinfo {year} {2021})},\ \Eprint
  {https://arxiv.org/abs/2011.05999} {arXiv:2011.05999} \BibitemShut {NoStop}%
\bibitem [{\citenamefont {Ballesteros}\ {\it et~al.}(2020)\citenamefont
  {Ballesteros}, \citenamefont {Rey}, \citenamefont {Taoso},\ and\
  \citenamefont {Urbano}}]{Ballesteros:2020qam}%
  \BibitemOpen
  \bibfield  {author} {\bibinfo {author} {\bibfnamefont {G.}~\bibnamefont
  {Ballesteros}}, \bibinfo {author} {\bibfnamefont {J.}~\bibnamefont {Rey}},
  \bibinfo {author} {\bibfnamefont {M.}~\bibnamefont {Taoso}},\ and\ \bibinfo
  {author} {\bibfnamefont {A.}~\bibnamefont {Urbano}},\ }\bibinfo {title}
  {{Primordial black holes as dark matter and gravitational waves from
  single-field polynomial inflation}},\ \href
  {https://doi.org/10.1088/1475-7516/2020/07/025} {J. Cosmol. Astropart. Phys.\
  \bibinfo {volume} {07}\bibfield  {year} {\bibinfo  {year} { (\textbf
  {2020})}\ }\bibfield  {pages} {\bibinfo  {pages} {025}},\ }\Eprint
  {https://arxiv.org/abs/2001.08220} {arXiv:2001.08220} \BibitemShut {NoStop}%
\bibitem [{\citenamefont {Ragavendra}\ {\it et~al.}(2021)\citenamefont
  {Ragavendra}, \citenamefont {Saha}, \citenamefont {Sriramkumar},\ and\
  \citenamefont {Silk}}]{Ragavendra:2020sop}%
  \BibitemOpen
  \bibfield  {author} {\bibinfo {author} {\bibfnamefont {H.~V.}\ \bibnamefont
  {Ragavendra}}, \bibinfo {author} {\bibfnamefont {P.}~\bibnamefont {Saha}},
  \bibinfo {author} {\bibfnamefont {L.}~\bibnamefont {Sriramkumar}},\ and\
  \bibinfo {author} {\bibfnamefont {J.}~\bibnamefont {Silk}},\ }\bibinfo
  {title} {{Primordial black holes and secondary gravitational waves from
  ultraslow roll and punctuated inflation}},\ \href
  {https://doi.org/10.1103/PhysRevD.103.083510} {\bibfield  {journal} {\bibinfo
   {journal} {Phys. Rev. D}\ }\textbf {\bibinfo {volume} {103}},\ \bibinfo
  {pages} {083510} (\bibinfo {year} {2021})},\ \Eprint
  {https://arxiv.org/abs/2008.12202} {arXiv:2008.12202} \BibitemShut {NoStop}%
\bibitem [{\citenamefont {Palma}\ {\it et~al.}(2020)\citenamefont {Palma},
  \citenamefont {Sypsas},\ and\ \citenamefont {Zenteno}}]{Palma:2020ejf}%
  \BibitemOpen
  \bibfield  {author} {\bibinfo {author} {\bibfnamefont {G.~A.}\ \bibnamefont
  {Palma}}, \bibinfo {author} {\bibfnamefont {S.}~\bibnamefont {Sypsas}},\ and\
  \bibinfo {author} {\bibfnamefont {C.}~\bibnamefont {Zenteno}},\ }\bibinfo
  {title} {{Seeding primordial black holes in multifield inflation}},\ \href
  {https://doi.org/10.1103/PhysRevLett.125.121301} {\bibfield  {journal}
  {\bibinfo  {journal} {Phys. Rev. Lett.}\ }\textbf {\bibinfo {volume} {125}},\
  \bibinfo {pages} {121301} (\bibinfo {year} {2020})},\ \Eprint
  {https://arxiv.org/abs/2004.06106} {arXiv:2004.06106} \BibitemShut {NoStop}%
\bibitem [{\citenamefont {Braglia}\ {\it et~al.}(2020)\citenamefont {Braglia},
  \citenamefont {Hazra}, \citenamefont {Finelli}, \citenamefont {Smoot},
  \citenamefont {Sriramkumar},\ and\ \citenamefont
  {Starobinsky}}]{Braglia:2020eai}%
  \BibitemOpen
  \bibfield  {author} {\bibinfo {author} {\bibfnamefont {M.}~\bibnamefont
  {Braglia}}, \bibinfo {author} {\bibfnamefont {D.~K.}\ \bibnamefont {Hazra}},
  \bibinfo {author} {\bibfnamefont {F.}~\bibnamefont {Finelli}}, \bibinfo
  {author} {\bibfnamefont {G.~F.}\ \bibnamefont {Smoot}}, \bibinfo {author}
  {\bibfnamefont {L.}~\bibnamefont {Sriramkumar}},\ and\ \bibinfo {author}
  {\bibfnamefont {A.~A.}\ \bibnamefont {Starobinsky}},\ }\bibinfo {title}
  {{Generating PBHs and small-scale GWs in two-field models of inflation}},\
  \href {https://doi.org/10.1088/1475-7516/2020/08/001} {J. Cosmol. Astropart.
  Phys.\ \bibinfo {volume} {08}\bibfield  {year} {\bibinfo  {year} { (\textbf
  {2020})}\ }\bibfield  {pages} {\bibinfo  {pages} {001}},\ }\Eprint
  {https://arxiv.org/abs/2005.02895} {arXiv:2005.02895} \BibitemShut {NoStop}%
\bibitem [{\citenamefont {Baumann}\ {\it et~al.}(2007)\citenamefont {Baumann},
  \citenamefont {Steinhardt}, \citenamefont {Takahashi},\ and\ \citenamefont
  {Ichiki}}]{Baumann:2007zm}%
  \BibitemOpen
  \bibfield  {author} {\bibinfo {author} {\bibfnamefont {D.}~\bibnamefont
  {Baumann}}, \bibinfo {author} {\bibfnamefont {P.~J.}\ \bibnamefont
  {Steinhardt}}, \bibinfo {author} {\bibfnamefont {K.}~\bibnamefont
  {Takahashi}},\ and\ \bibinfo {author} {\bibfnamefont {K.}~\bibnamefont
  {Ichiki}},\ }\bibinfo {title} {{Gravitational Wave Spectrum Induced by
  Primordial Scalar Perturbations}},\ \href
  {https://doi.org/10.1103/PhysRevD.76.084019} {\bibfield  {journal} {\bibinfo
  {journal} {Phys. Rev. D}\ }\textbf {\bibinfo {volume} {76}},\ \bibinfo
  {pages} {084019} (\bibinfo {year} {2007})},\ \Eprint
  {https://arxiv.org/abs/hep-th/0703290} {arXiv:hep-th/0703290} \BibitemShut
  {NoStop}%
\bibitem [{\citenamefont {Saito}\ and\ \citenamefont
  {Yokoyama}(2009)}]{Saito:2008jc}%
  \BibitemOpen
  \bibfield  {author} {\bibinfo {author} {\bibfnamefont {R.}~\bibnamefont
  {Saito}}\ and\ \bibinfo {author} {\bibfnamefont {J.}~\bibnamefont
  {Yokoyama}},\ }\bibinfo {title} {{Gravitational wave background as a probe of
  the primordial black hole abundance}},\ \href
  {https://doi.org/10.1103/PhysRevLett.102.161101} {\bibfield  {journal}
  {\bibinfo  {journal} {Phys. Rev. Lett.}\ }\textbf {\bibinfo {volume} {102}},\
  \bibinfo {pages} {161101} (\bibinfo {year} {2009})},\ \bibinfo {note}
  {[Erratum: Phys.Rev.Lett. 107, 069901 (2011)]},\ \Eprint
  {https://arxiv.org/abs/0812.4339} {arXiv:0812.4339} \BibitemShut {NoStop}%
\bibitem [{\citenamefont {Orlofsky}\ {\it et~al.}(2017)\citenamefont
  {Orlofsky}, \citenamefont {Pierce},\ and\ \citenamefont
  {Wells}}]{Orlofsky:2016vbd}%
  \BibitemOpen
  \bibfield  {author} {\bibinfo {author} {\bibfnamefont {N.}~\bibnamefont
  {Orlofsky}}, \bibinfo {author} {\bibfnamefont {A.}~\bibnamefont {Pierce}},\
  and\ \bibinfo {author} {\bibfnamefont {J.~D.}\ \bibnamefont {Wells}},\
  }\bibinfo {title} {{Inflationary theory and pulsar timing investigations of
  primordial black holes and gravitational waves}},\ \href
  {https://doi.org/10.1103/PhysRevD.95.063518} {\bibfield  {journal} {\bibinfo
  {journal} {Phys. Rev. D}\ }\textbf {\bibinfo {volume} {95}},\ \bibinfo
  {pages} {063518} (\bibinfo {year} {2017})},\ \Eprint
  {https://arxiv.org/abs/1612.05279} {arXiv:1612.05279} \BibitemShut {NoStop}%
\bibitem [{\citenamefont {Nakama}\ {\it et~al.}(2017)\citenamefont {Nakama},
  \citenamefont {Silk},\ and\ \citenamefont {Kamionkowski}}]{Nakama:2016gzw}%
  \BibitemOpen
  \bibfield  {author} {\bibinfo {author} {\bibfnamefont {T.}~\bibnamefont
  {Nakama}}, \bibinfo {author} {\bibfnamefont {J.}~\bibnamefont {Silk}},\ and\
  \bibinfo {author} {\bibfnamefont {M.}~\bibnamefont {Kamionkowski}},\
  }\bibinfo {title} {{Stochastic gravitational waves associated with the
  formation of primordial black holes}},\ \href
  {https://doi.org/10.1103/PhysRevD.95.043511} {\bibfield  {journal} {\bibinfo
  {journal} {Phys. Rev. D}\ }\textbf {\bibinfo {volume} {95}},\ \bibinfo
  {pages} {043511} (\bibinfo {year} {2017})},\ \Eprint
  {https://arxiv.org/abs/1612.06264} {arXiv:1612.06264} \BibitemShut {NoStop}%
\bibitem [{\citenamefont {Inomata}\ {\it et~al.}(2017)\citenamefont {Inomata},
  \citenamefont {Kawasaki}, \citenamefont {Mukaida}, \citenamefont {Tada},\
  and\ \citenamefont {Yanagida}}]{Inomata:2016rbd}%
  \BibitemOpen
  \bibfield  {author} {\bibinfo {author} {\bibfnamefont {K.}~\bibnamefont
  {Inomata}}, \bibinfo {author} {\bibfnamefont {M.}~\bibnamefont {Kawasaki}},
  \bibinfo {author} {\bibfnamefont {K.}~\bibnamefont {Mukaida}}, \bibinfo
  {author} {\bibfnamefont {Y.}~\bibnamefont {Tada}},\ and\ \bibinfo {author}
  {\bibfnamefont {T.~T.}\ \bibnamefont {Yanagida}},\ }\bibinfo {title}
  {{Inflationary primordial black holes for the LIGO gravitational wave events
  and pulsar timing array experiments}},\ \href
  {https://doi.org/10.1103/PhysRevD.95.123510} {\bibfield  {journal} {\bibinfo
  {journal} {Phys. Rev. D}\ }\textbf {\bibinfo {volume} {95}},\ \bibinfo
  {pages} {123510} (\bibinfo {year} {2017})},\ \Eprint
  {https://arxiv.org/abs/1611.06130} {arXiv:1611.06130} \BibitemShut {NoStop}%
\bibitem [{\citenamefont {Cai}\ {\it et~al.}(2019{\natexlab{a}})\citenamefont
  {Cai}, \citenamefont {Pi},\ and\ \citenamefont {Sasaki}}]{Cai:2018dig}%
  \BibitemOpen
  \bibfield  {author} {\bibinfo {author} {\bibfnamefont {R.-g.}\ \bibnamefont
  {Cai}}, \bibinfo {author} {\bibfnamefont {S.}~\bibnamefont {Pi}},\ and\
  \bibinfo {author} {\bibfnamefont {M.}~\bibnamefont {Sasaki}},\ }\bibinfo
  {title} {{Gravitational Waves Induced by non-Gaussian Scalar
  Perturbations}},\ \href {https://doi.org/10.1103/PhysRevLett.122.201101}
  {\bibfield  {journal} {\bibinfo  {journal} {Phys. Rev. Lett.}\ }\textbf
  {\bibinfo {volume} {122}},\ \bibinfo {pages} {201101} (\bibinfo {year}
  {2019}{\natexlab{a}})},\ \Eprint {https://arxiv.org/abs/1810.11000}
  {arXiv:1810.11000} \BibitemShut {NoStop}%
\bibitem [{\citenamefont {Bartolo}\ {\it et~al.}(2019)\citenamefont {Bartolo},
  \citenamefont {De~Luca}, \citenamefont {Franciolini}, \citenamefont {Lewis},
  \citenamefont {Peloso},\ and\ \citenamefont {Riotto}}]{Bartolo:2018evs}%
  \BibitemOpen
  \bibfield  {author} {\bibinfo {author} {\bibfnamefont {N.}~\bibnamefont
  {Bartolo}}, \bibinfo {author} {\bibfnamefont {V.}~\bibnamefont {De~Luca}},
  \bibinfo {author} {\bibfnamefont {G.}~\bibnamefont {Franciolini}}, \bibinfo
  {author} {\bibfnamefont {A.}~\bibnamefont {Lewis}}, \bibinfo {author}
  {\bibfnamefont {M.}~\bibnamefont {Peloso}},\ and\ \bibinfo {author}
  {\bibfnamefont {A.}~\bibnamefont {Riotto}},\ }\bibinfo {title} {{Primordial
  Black Hole Dark Matter: LISA Serendipity}},\ \href
  {https://doi.org/10.1103/PhysRevLett.122.211301} {\bibfield  {journal}
  {\bibinfo  {journal} {Phys. Rev. Lett.}\ }\textbf {\bibinfo {volume} {122}},\
  \bibinfo {pages} {211301} (\bibinfo {year} {2019})},\ \Eprint
  {https://arxiv.org/abs/1810.12218} {arXiv:1810.12218} \BibitemShut {NoStop}%
\bibitem [{\citenamefont {Kohri}\ and\ \citenamefont
  {Terada}(2018)}]{Kohri:2018awv}%
  \BibitemOpen
  \bibfield  {author} {\bibinfo {author} {\bibfnamefont {K.}~\bibnamefont
  {Kohri}}\ and\ \bibinfo {author} {\bibfnamefont {T.}~\bibnamefont {Terada}},\
  }\bibinfo {title} {{Semianalytic calculation of gravitational wave spectrum
  nonlinearly induced from primordial curvature perturbations}},\ \href
  {https://doi.org/10.1103/PhysRevD.97.123532} {\bibfield  {journal} {\bibinfo
  {journal} {Phys. Rev. D}\ }\textbf {\bibinfo {volume} {97}},\ \bibinfo
  {pages} {123532} (\bibinfo {year} {2018})},\ \Eprint
  {https://arxiv.org/abs/1804.08577} {arXiv:1804.08577} \BibitemShut {NoStop}%
\bibitem [{\citenamefont {Espinosa}\ {\it
  et~al.}(2018{\natexlab{b}})\citenamefont {Espinosa}, \citenamefont {Racco},\
  and\ \citenamefont {Riotto}}]{Espinosa:2018eve}%
  \BibitemOpen
  \bibfield  {author} {\bibinfo {author} {\bibfnamefont {J.~R.}\ \bibnamefont
  {Espinosa}}, \bibinfo {author} {\bibfnamefont {D.}~\bibnamefont {Racco}},\
  and\ \bibinfo {author} {\bibfnamefont {A.}~\bibnamefont {Riotto}},\ }\bibinfo
  {title} {{A Cosmological Signature of the SM Higgs Instability: Gravitational
  Waves}},\ \href {https://doi.org/10.1088/1475-7516/2018/09/012} {J. Cosmol.
  Astropart. Phys.\ \bibinfo {volume} {09}\bibfield  {year} {\bibinfo  {year} {
  (\textbf {2018})}\ }\bibfield  {pages} {\bibinfo  {pages} {012}},\ }\Eprint
  {https://arxiv.org/abs/1804.07732} {arXiv:1804.07732} \BibitemShut {NoStop}%
\bibitem [{\citenamefont {Kuroyanagi}\ {\it et~al.}(2018)\citenamefont
  {Kuroyanagi}, \citenamefont {Chiba},\ and\ \citenamefont
  {Takahashi}}]{Kuroyanagi:2018csn}%
  \BibitemOpen
  \bibfield  {author} {\bibinfo {author} {\bibfnamefont {S.}~\bibnamefont
  {Kuroyanagi}}, \bibinfo {author} {\bibfnamefont {T.}~\bibnamefont {Chiba}},\
  and\ \bibinfo {author} {\bibfnamefont {T.}~\bibnamefont {Takahashi}},\
  }\bibinfo {title} {{Probing the Universe through the Stochastic Gravitational
  Wave Background}},\ \href {https://doi.org/10.1088/1475-7516/2018/11/038} {J.
  Cosmol. Astropart. Phys.\ \bibinfo {volume} {11}\bibfield  {year} {\bibinfo
  {year} { (\textbf {2018})}\ }\bibfield  {pages} {\bibinfo  {pages} {038}},\
  }\Eprint {https://arxiv.org/abs/1807.00786} {arXiv:1807.00786} \BibitemShut
  {NoStop}%
\bibitem [{\citenamefont {Cai}\ {\it et~al.}(2019{\natexlab{b}})\citenamefont
  {Cai}, \citenamefont {Pi}, \citenamefont {Wang},\ and\ \citenamefont
  {Yang}}]{Cai:2019elf}%
  \BibitemOpen
  \bibfield  {author} {\bibinfo {author} {\bibfnamefont {R.-G.}\ \bibnamefont
  {Cai}}, \bibinfo {author} {\bibfnamefont {S.}~\bibnamefont {Pi}}, \bibinfo
  {author} {\bibfnamefont {S.-J.}\ \bibnamefont {Wang}},\ and\ \bibinfo
  {author} {\bibfnamefont {X.-Y.}\ \bibnamefont {Yang}},\ }\bibinfo {title}
  {{Pulsar Timing Array Constraints on the Induced Gravitational Waves}},\
  \href {https://doi.org/10.1088/1475-7516/2019/10/059} {J. Cosmol. Astropart.
  Phys.\ \bibinfo {volume} {10}\bibfield  {year} {\bibinfo  {year} { (\textbf
  {2019})}\ }\bibfield  {pages} {\bibinfo  {pages} {059}},\ }\Eprint
  {https://arxiv.org/abs/1907.06372} {arXiv:1907.06372} \BibitemShut {NoStop}%
\bibitem [{\citenamefont {Drees}\ and\ \citenamefont
  {Xu}(2021)}]{Drees:2019xpp}%
  \BibitemOpen
  \bibfield  {author} {\bibinfo {author} {\bibfnamefont {M.}~\bibnamefont
  {Drees}}\ and\ \bibinfo {author} {\bibfnamefont {Y.}~\bibnamefont {Xu}},\
  }\bibinfo {title} {{Overshooting, Critical Higgs Inflation and Second Order
  Gravitational Wave Signatures}},\ \href
  {https://doi.org/10.1140/epjc/s10052-021-08976-2} {\bibfield  {journal}
  {\bibinfo  {journal} {Eur. Phys. J. C}\ }\textbf {\bibinfo {volume} {81}},\
  \bibinfo {pages} {182} (\bibinfo {year} {2021})},\ \Eprint
  {https://arxiv.org/abs/1905.13581} {arXiv:1905.13581} \BibitemShut {NoStop}%
\bibitem [{\citenamefont {Inomata}\ {\it
  et~al.}(2019{\natexlab{a}})\citenamefont {Inomata}, \citenamefont {Kohri},
  \citenamefont {Nakama},\ and\ \citenamefont {Terada}}]{Inomata:2019ivs}%
  \BibitemOpen
  \bibfield  {author} {\bibinfo {author} {\bibfnamefont {K.}~\bibnamefont
  {Inomata}}, \bibinfo {author} {\bibfnamefont {K.}~\bibnamefont {Kohri}},
  \bibinfo {author} {\bibfnamefont {T.}~\bibnamefont {Nakama}},\ and\ \bibinfo
  {author} {\bibfnamefont {T.}~\bibnamefont {Terada}},\ }\bibinfo {title}
  {{Enhancement of Gravitational Waves Induced by Scalar Perturbations due to a
  Sudden Transition from an Early Matter Era to the Radiation Era}},\ \href
  {https://doi.org/10.1103/PhysRevD.100.043532} {\bibfield  {journal} {\bibinfo
   {journal} {Phys. Rev. D}\ }\textbf {\bibinfo {volume} {100}},\ \bibinfo
  {pages} {043532} (\bibinfo {year} {2019}{\natexlab{a}})},\ \Eprint
  {https://arxiv.org/abs/1904.12879} {arXiv:1904.12879} \BibitemShut {NoStop}%
\bibitem [{\citenamefont {Inomata}\ {\it
  et~al.}(2019{\natexlab{b}})\citenamefont {Inomata}, \citenamefont {Kohri},
  \citenamefont {Nakama},\ and\ \citenamefont {Terada}}]{Inomata:2019zqy}%
  \BibitemOpen
  \bibfield  {author} {\bibinfo {author} {\bibfnamefont {K.}~\bibnamefont
  {Inomata}}, \bibinfo {author} {\bibfnamefont {K.}~\bibnamefont {Kohri}},
  \bibinfo {author} {\bibfnamefont {T.}~\bibnamefont {Nakama}},\ and\ \bibinfo
  {author} {\bibfnamefont {T.}~\bibnamefont {Terada}},\ }\bibinfo {title}
  {{Gravitational Waves Induced by Scalar Perturbations during a Gradual
  Transition from an Early Matter Era to the Radiation Era}},\ \href
  {https://doi.org/10.1088/1475-7516/2019/10/071} {J. Cosmol. Astropart. Phys.\
  \bibinfo {volume} {10}\bibfield  {year} {\bibinfo  {year} { (\textbf
  {2019})}\ }\bibfield  {pages} {\bibinfo  {pages} {071}},\ }\Eprint
  {https://arxiv.org/abs/1904.12878} {arXiv:1904.12878} \BibitemShut {NoStop}%
\bibitem [{\citenamefont {Fumagalli}\ {\it
  et~al.}(2020{\natexlab{b}})\citenamefont {Fumagalli}, \citenamefont
  {Renaux-Petel},\ and\ \citenamefont {Witkowski}}]{Fumagalli:2020nvq}%
  \BibitemOpen
  \bibfield  {author} {\bibinfo {author} {\bibfnamefont {J.}~\bibnamefont
  {Fumagalli}}, \bibinfo {author} {\bibfnamefont {S.}~\bibnamefont
  {Renaux-Petel}},\ and\ \bibinfo {author} {\bibfnamefont {L.~T.}\ \bibnamefont
  {Witkowski}},\ }\bibinfo {title} {{Oscillations in the stochastic
  gravitational wave background from sharp features and particle production
  during inflation}},\ \href@noop {} {\  (\bibinfo {year}
  {2020}{\natexlab{b}})},\ \Eprint {https://arxiv.org/abs/2012.02761}
  {arXiv:2012.02761} \BibitemShut {NoStop}%
\bibitem [{\citenamefont {Dom\`enech}\ {\it et~al.}(2020)\citenamefont
  {Dom\`enech}, \citenamefont {Pi},\ and\ \citenamefont
  {Sasaki}}]{Domenech:2020kqm}%
  \BibitemOpen
  \bibfield  {author} {\bibinfo {author} {\bibfnamefont {G.}~\bibnamefont
  {Dom\`enech}}, \bibinfo {author} {\bibfnamefont {S.}~\bibnamefont {Pi}},\
  and\ \bibinfo {author} {\bibfnamefont {M.}~\bibnamefont {Sasaki}},\ }\bibinfo
  {title} {{Induced gravitational waves as a probe of thermal history of the
  universe}},\ \href {https://doi.org/10.1088/1475-7516/2020/08/017} {J.
  Cosmol. Astropart. Phys.\ \bibinfo {volume} {08}\bibfield  {year} {\bibinfo
  {year} { (\textbf {2020})}\ }\bibfield  {pages} {\bibinfo  {pages} {017}},\
  }\Eprint {https://arxiv.org/abs/2005.12314} {arXiv:2005.12314} \BibitemShut
  {NoStop}%
\bibitem [{\citenamefont {Braglia}\ {\it et~al.}(2021)\citenamefont {Braglia},
  \citenamefont {Chen},\ and\ \citenamefont {Hazra}}]{Braglia:2020taf}%
  \BibitemOpen
  \bibfield  {author} {\bibinfo {author} {\bibfnamefont {M.}~\bibnamefont
  {Braglia}}, \bibinfo {author} {\bibfnamefont {X.}~\bibnamefont {Chen}},\ and\
  \bibinfo {author} {\bibfnamefont {D.~K.}\ \bibnamefont {Hazra}},\ }\bibinfo
  {title} {{Probing Primordial Features with the Stochastic Gravitational Wave
  Background}},\ \href {https://doi.org/10.1088/1475-7516/2021/03/005} {J.
  Cosmol. Astropart. Phys.\ \bibinfo {volume} {03}\bibfield  {year} {\bibinfo
  {year} { (\textbf {2021})}\ }\bibfield  {pages} {\bibinfo  {pages} {005}},\
  }\Eprint {https://arxiv.org/abs/2012.05821} {arXiv:2012.05821} \BibitemShut
  {NoStop}%
\bibitem [{\citenamefont {Fumagalli}\ {\it et~al.}(2021)\citenamefont
  {Fumagalli}, \citenamefont {Renaux-Petel},\ and\ \citenamefont
  {Witkowski}}]{Fumagalli:2021cel}%
  \BibitemOpen
  \bibfield  {author} {\bibinfo {author} {\bibfnamefont {J.}~\bibnamefont
  {Fumagalli}}, \bibinfo {author} {\bibfnamefont {S.}~\bibnamefont
  {Renaux-Petel}},\ and\ \bibinfo {author} {\bibfnamefont {L.~T.}\ \bibnamefont
  {Witkowski}},\ }\bibinfo {title} {{Resonant features in the stochastic
  gravitational wave background}},\ \href@noop {} {\  (\bibinfo {year}
  {2021})},\ \Eprint {https://arxiv.org/abs/2105.06481} {arXiv:2105.06481}
  \BibitemShut {NoStop}%
\bibitem [{\citenamefont {Gao}\ and\ \citenamefont {Yang}(2021)}]{Gao:2021dfi}%
  \BibitemOpen
  \bibfield  {author} {\bibinfo {author} {\bibfnamefont {T.-J.}\ \bibnamefont
  {Gao}}\ and\ \bibinfo {author} {\bibfnamefont {X.-Y.}\ \bibnamefont {Yang}},\
  }\bibinfo {title} {{Double peaks of gravitational wave spectrum induced from
  inflection point inflation}},\ \href
  {https://doi.org/10.1140/epjc/s10052-021-09269-4} {\bibfield  {journal}
  {\bibinfo  {journal} {Eur. Phys. J. C}\ }\textbf {\bibinfo {volume} {81}},\
  \bibinfo {pages} {494} (\bibinfo {year} {2021})},\ \Eprint
  {https://arxiv.org/abs/2101.07616} {arXiv:2101.07616} \BibitemShut {NoStop}%
\bibitem [{\citenamefont {Zheng}\ {\it et~al.}(2021)\citenamefont {Zheng},
  \citenamefont {Shi},\ and\ \citenamefont {Qiu}}]{Zheng:2021vda}%
  \BibitemOpen
  \bibfield  {author} {\bibinfo {author} {\bibfnamefont {R.}~\bibnamefont
  {Zheng}}, \bibinfo {author} {\bibfnamefont {J.}~\bibnamefont {Shi}},\ and\
  \bibinfo {author} {\bibfnamefont {T.}~\bibnamefont {Qiu}},\ }\bibinfo {title}
  {{On Primordial Black Holes generated from inflation with solo/multi-bumpy
  potential}},\ \href@noop {} {\  (\bibinfo {year} {2021})},\ \Eprint
  {https://arxiv.org/abs/2106.04303} {arXiv:2106.04303} \BibitemShut {NoStop}%
\bibitem [{\citenamefont {Maldacena}(2003)}]{Maldacena:2002vr}%
  \BibitemOpen
  \bibfield  {author} {\bibinfo {author} {\bibfnamefont {J.~M.}\ \bibnamefont
  {Maldacena}},\ }\bibinfo {title} {{Non-Gaussian features of primordial
  fluctuations in single field inflationary models}},\ \href
  {https://doi.org/10.1088/1126-6708/2003/05/013} {J. High Energ. Phys.\
  \bibinfo {volume} {05}\bibfield  {year} {\bibinfo  {year} { (\textbf
  {2003})}\ }\bibfield  {pages} {\bibinfo  {pages} {013}},\ }\Eprint
  {https://arxiv.org/abs/astro-ph/0210603} {arXiv:astro-ph/0210603}
  \BibitemShut {NoStop}%
\bibitem [{\citenamefont {Amaro-Seoane}\ {\it et~al.}(2017)\citenamefont
  {Amaro-Seoane} {\it et~al.}}]{Audley:2017drz}%
  \BibitemOpen
  \bibfield  {author} {\bibinfo {author} {\bibfnamefont {P.}~\bibnamefont
  {Amaro-Seoane}} {\it et~al.} (\bibinfo {collaboration} {LISA}),\ }\bibinfo
  {title} {{Laser Interferometer Space Antenna}},\ \href@noop {} {\  (\bibinfo
  {year} {2017})},\ \Eprint {https://arxiv.org/abs/1702.00786}
  {arXiv:1702.00786} \BibitemShut {NoStop}%
\bibitem [{\citenamefont {Danzmann}(1997)}]{Danzmann:1997hm}%
  \BibitemOpen
  \bibfield  {author} {\bibinfo {author} {\bibfnamefont {K.}~\bibnamefont
  {Danzmann}},\ }\bibinfo {title} {{LISA: An ESA cornerstone mission for a
  gravitational wave observatory}},\ \href
  {https://doi.org/10.1088/0264-9381/14/6/002} {\bibfield  {journal} {\bibinfo
  {journal} {Class. Quant. Grav.}\ }\textbf {\bibinfo {volume} {14}},\ \bibinfo
  {pages} {1399} (\bibinfo {year} {1997})}\BibitemShut {NoStop}%
\bibitem [{\citenamefont {Luo}\ {\it et~al.}(2016)\citenamefont {Luo} {\it
  et~al.}}]{Luo:2015ght}%
  \BibitemOpen
  \bibfield  {author} {\bibinfo {author} {\bibfnamefont {J.}~\bibnamefont
  {Luo}} {\it et~al.} (\bibinfo {collaboration} {TianQin}),\ }\bibinfo {title}
  {{TianQin: a space-borne gravitational wave detector}},\ \href
  {https://doi.org/10.1088/0264-9381/33/3/035010} {\bibfield  {journal}
  {\bibinfo  {journal} {Class. Quant. Grav.}\ }\textbf {\bibinfo {volume}
  {33}},\ \bibinfo {pages} {035010} (\bibinfo {year} {2016})},\ \Eprint
  {https://arxiv.org/abs/1512.02076} {arXiv:1512.02076} \BibitemShut {NoStop}%
\bibitem [{\citenamefont {Hu}\ and\ \citenamefont {Wu}(2017)}]{Hu:2017mde}%
  \BibitemOpen
  \bibfield  {author} {\bibinfo {author} {\bibfnamefont {W.-R.}\ \bibnamefont
  {Hu}}\ and\ \bibinfo {author} {\bibfnamefont {Y.-L.}\ \bibnamefont {Wu}},\
  }\bibinfo {title} {{The Taiji Program in Space for gravitational wave physics
  and the nature of gravity}},\ \href {https://doi.org/10.1093/nsr/nwx116}
  {\bibfield  {journal} {\bibinfo  {journal} {Natl. Sci. Rev.}\ }\textbf
  {\bibinfo {volume} {4}},\ \bibinfo {pages} {685} (\bibinfo {year}
  {2017})}\BibitemShut {NoStop}%
\bibitem [{\citenamefont {Moore}\ {\it et~al.}(2015)\citenamefont {Moore},
  \citenamefont {Cole},\ and\ \citenamefont {Berry}}]{Moore:2014lga}%
  \BibitemOpen
  \bibfield  {author} {\bibinfo {author} {\bibfnamefont {C.~J.}\ \bibnamefont
  {Moore}}, \bibinfo {author} {\bibfnamefont {R.~H.}\ \bibnamefont {Cole}},\
  and\ \bibinfo {author} {\bibfnamefont {C.~P.~L.}\ \bibnamefont {Berry}},\
  }\bibinfo {title} {{Gravitational-wave sensitivity curves}},\ \href
  {https://doi.org/10.1088/0264-9381/32/1/015014} {\bibfield  {journal}
  {\bibinfo  {journal} {Class. Quant. Grav.}\ }\textbf {\bibinfo {volume}
  {32}},\ \bibinfo {pages} {015014} (\bibinfo {year} {2015})},\ \Eprint
  {https://arxiv.org/abs/1408.0740} {arXiv:1408.0740} \BibitemShut {NoStop}%
\bibitem [{\citenamefont {Mukhanov}(1985)}]{Mukhanov:1985rz}%
  \BibitemOpen
  \bibfield  {author} {\bibinfo {author} {\bibfnamefont {V.~F.}\ \bibnamefont
  {Mukhanov}},\ }\bibinfo {title} {{Gravitational Instability of the Universe
  Filled with a Scalar Field}},\ \href@noop {} {\bibfield  {journal} {\bibinfo
  {journal} {JETP Lett.}\ }\textbf {\bibinfo {volume} {41}},\ \bibinfo {pages}
  {493} (\bibinfo {year} {1985})}\BibitemShut {NoStop}%
\bibitem [{\citenamefont {Sasaki}(1986)}]{Sasaki:1986hm}%
  \BibitemOpen
  \bibfield  {author} {\bibinfo {author} {\bibfnamefont {M.}~\bibnamefont
  {Sasaki}},\ }\bibinfo {title} {{Large Scale Quantum Fluctuations in the
  Inflationary Universe}},\ \href {https://doi.org/10.1143/PTP.76.1036}
  {\bibfield  {journal} {\bibinfo  {journal} {Prog. Theor. Phys.}\ }\textbf
  {\bibinfo {volume} {76}},\ \bibinfo {pages} {1036} (\bibinfo {year}
  {1986})}\BibitemShut {NoStop}%
\bibitem [{\citenamefont {Inomata}\ and\ \citenamefont
  {Nakama}(2019)}]{Inomata:2018epa}%
  \BibitemOpen
  \bibfield  {author} {\bibinfo {author} {\bibfnamefont {K.}~\bibnamefont
  {Inomata}}\ and\ \bibinfo {author} {\bibfnamefont {T.}~\bibnamefont
  {Nakama}},\ }\bibinfo {title} {{Gravitational waves induced by scalar
  perturbations as probes of the small-scale primordial spectrum}},\ \-href
  {https://doi.org/10.1103/PhysRevD.99.043511} {\bibfield  {journal} {\bibinfo
  {journal} {Phys. Rev. D}\ }\textbf {\bibinfo {volume} {99}},\ \bibinfo
  {pages} {043511} (\bibinfo {year} {2019})},\ \Eprint
  {https://arxiv.org/abs/1812.00674} {arXiv:1812.00674} \BibitemShut {NoStop}%
\bibitem [{\citenamefont {Inomata}\ {\it et~al.}(2016)\citenamefont {Inomata},
  \citenamefont {Kawasaki},\ and\ \citenamefont {Tada}}]{Inomata:2016uip}%
  \BibitemOpen
  \bibfield  {author} {\bibinfo {author} {\bibfnamefont {K.}~\bibnamefont
  {Inomata}}, \bibinfo {author} {\bibfnamefont {M.}~\bibnamefont {Kawasaki}},\
  and\ \bibinfo {author} {\bibfnamefont {Y.}~\bibnamefont {Tada}},\ }\bibinfo
  {title} {{Revisiting constraints on small scale perturbations from big-bang
  nucleosynthesis}},\ \href {https://doi.org/10.1103/PhysRevD.94.043527}
  {\bibfield  {journal} {\bibinfo  {journal} {Phys. Rev. D}\ }\textbf {\bibinfo
  {volume} {94}},\ \bibinfo {pages} {043527} (\bibinfo {year} {2016})},\
  \Eprint {https://arxiv.org/abs/1605.04646} {arXiv:1605.04646} \BibitemShut
  {NoStop}%
\bibitem [{\citenamefont {Fixsen}\ {\it et~al.}(1996)\citenamefont {Fixsen},
  \citenamefont {Cheng}, \citenamefont {Gales}, \citenamefont {Mather},
  \citenamefont {Shafer},\ and\ \citenamefont {Wright}}]{Fixsen:1996nj}%
  \BibitemOpen
  \bibfield  {author} {\bibinfo {author} {\bibfnamefont {D.~J.}\ \bibnamefont
  {Fixsen}}, \bibinfo {author} {\bibfnamefont {E.~S.}\ \bibnamefont {Cheng}},
  \bibinfo {author} {\bibfnamefont {J.~M.}\ \bibnamefont {Gales}}, \bibinfo
  {author} {\bibfnamefont {J.~C.}\ \bibnamefont {Mather}}, \bibinfo {author}
  {\bibfnamefont {R.~A.}\ \bibnamefont {Shafer}},\ and\ \bibinfo {author}
  {\bibfnamefont {E.~L.}\ \bibnamefont {Wright}},\ }\bibinfo {title} {{The
  Cosmic Microwave Background spectrum from the full COBE FIRAS data set}},\
  \href {https://doi.org/10.1086/178173} {\bibfield  {journal} {\bibinfo
  {journal} {Astrophys. J.}\ }\textbf {\bibinfo {volume} {473}},\ \bibinfo
  {pages} {576} (\bibinfo {year} {1996})},\ \Eprint
  {https://arxiv.org/abs/astro-ph/9605054} {arXiv:astro-ph/9605054}
  \BibitemShut {NoStop}%
\bibitem [{\citenamefont {Kinney}(2005)}]{Kinney:2005vj}%
  \BibitemOpen
  \bibfield  {author} {\bibinfo {author} {\bibfnamefont {W.~H.}\ \bibnamefont
  {Kinney}},\ }\bibinfo {title} {{Horizon crossing and inflation with large
  eta}},\ \href {https://doi.org/10.1103/PhysRevD.72.023515} {\bibfield
  {journal} {\bibinfo  {journal} {Phys. Rev. D}\ }\textbf {\bibinfo {volume}
  {72}},\ \bibinfo {pages} {023515} (\bibinfo {year} {2005})},\ \Eprint
  {https://arxiv.org/abs/gr-qc/0503017} {arXiv:gr-qc/0503017} \BibitemShut
  {NoStop}%
\bibitem [{\citenamefont {Dimopoulos}(2017)}]{Dimopoulos:2017ged}%
  \BibitemOpen
  \bibfield  {author} {\bibinfo {author} {\bibfnamefont {K.}~\bibnamefont
  {Dimopoulos}},\ }\bibinfo {title} {{Ultra slow-roll inflation demystified}},\
  \href {https://doi.org/10.1016/j.physletb.2017.10.066} {\bibfield  {journal}
  {\bibinfo  {journal} {Phys. Lett. B}\ }\textbf {\bibinfo {volume} {775}},\
  \bibinfo {pages} {262} (\bibinfo {year} {2017})},\ \Eprint
  {https://arxiv.org/abs/1707.05644} {arXiv:1707.05644} \BibitemShut {NoStop}%
\bibitem [{\citenamefont {Byrnes}\ {\it et~al.}(2010)\citenamefont {Byrnes},
  \citenamefont {Gerstenlauer}, \citenamefont {Nurmi}, \citenamefont
  {Tasinato},\ and\ \citenamefont {Wands}}]{Byrnes:2010ft}%
  \BibitemOpen
  \bibfield  {author} {\bibinfo {author} {\bibfnamefont {C.~T.}\ \bibnamefont
  {Byrnes}}, \bibinfo {author} {\bibfnamefont {M.}~\bibnamefont
  {Gerstenlauer}}, \bibinfo {author} {\bibfnamefont {S.}~\bibnamefont {Nurmi}},
  \bibinfo {author} {\bibfnamefont {G.}~\bibnamefont {Tasinato}},\ and\
  \bibinfo {author} {\bibfnamefont {D.}~\bibnamefont {Wands}},\ }\bibinfo
  {title} {{Scale-dependent non-Gaussianity probes inflationary physics}},\
  \href {https://doi.org/10.1088/1475-7516/2010/10/004} {J. Cosmol. Astropart.
  Phys.\ \bibinfo {volume} {10}\bibfield  {year} {\bibinfo  {year} { (\textbf
  {2010})}\ }\bibfield  {pages} {\bibinfo  {pages} {004}},\ }\Eprint
  {https://arxiv.org/abs/1007.4277} {arXiv:1007.4277} \BibitemShut {NoStop}%
\bibitem [{\citenamefont {Ade}\ {\it et~al.}(2016)\citenamefont {Ade} {\it
  et~al.}}]{Ade:2015ava}%
  \BibitemOpen
  \bibfield  {author} {\bibinfo {author} {\bibfnamefont {P.~A.~R.}\
  \bibnamefont {Ade}} {\it et~al.} (\bibinfo {collaboration} {Planck}),\
  }\bibinfo {title} {{Planck 2015 results. XVII. Constraints on primordial
  non-Gaussianity}},\ \href {https://doi.org/10.1051/0004-6361/201525836}
  {\bibfield  {journal} {\bibinfo  {journal} {Astron. Astrophys.}\ }\textbf
  {\bibinfo {volume} {594}},\ \bibinfo {pages} {A17} (\bibinfo {year}
  {2016})},\ \Eprint {https://arxiv.org/abs/1502.01592} {arXiv:1502.01592}
  \BibitemShut {NoStop}%
\bibitem [{\citenamefont {Hazra}\ {\it et~al.}(2013)\citenamefont {Hazra},
  \citenamefont {Sriramkumar},\ and\ \citenamefont {Martin}}]{Hazra:2012yn}%
  \BibitemOpen
  \bibfield  {author} {\bibinfo {author} {\bibfnamefont {D.~K.}\ \bibnamefont
  {Hazra}}, \bibinfo {author} {\bibfnamefont {L.}~\bibnamefont {Sriramkumar}},\
  and\ \bibinfo {author} {\bibfnamefont {J.}~\bibnamefont {Martin}},\ }\bibinfo
  {title} {{BINGO: A code for the efficient computation of the scalar
  bi-spectrum}},\ \href {https://doi.org/10.1088/1475-7516/2013/05/026} {J.
  Cosmol. Astropart. Phys.\ \bibinfo {volume} {05}\bibfield  {year} {\bibinfo
  {year} { (\textbf {2013})}\ }\bibfield  {pages} {\bibinfo  {pages} {026}},\
  }\Eprint {https://arxiv.org/abs/1201.0926} {arXiv:1201.0926} \BibitemShut
  {NoStop}%
\bibitem [{\citenamefont {Arroja}\ and\ \citenamefont
  {Tanaka}(2011)}]{Arroja:2011yj}%
  \BibitemOpen
  \bibfield  {author} {\bibinfo {author} {\bibfnamefont {F.}~\bibnamefont
  {Arroja}}\ and\ \bibinfo {author} {\bibfnamefont {T.}~\bibnamefont
  {Tanaka}},\ }\bibinfo {title} {{A note on the role of the boundary terms for
  the non-Gaussianity in general k-inflation}},\ \href
  {https://doi.org/10.1088/1475-7516/2011/05/005} {J. Cosmol. Astropart. Phys.\
  \bibinfo {volume} {05}\bibfield  {year} {\bibinfo  {year} { (\textbf
  {2011})}\ }\bibfield  {pages} {\bibinfo  {pages} {005}},\ }\Eprint
  {https://arxiv.org/abs/1103.1102} {arXiv:1103.1102} \BibitemShut {NoStop}%
\bibitem [{\citenamefont {Zhang}\ {\it et~al.}(2021)\citenamefont {Zhang},
  \citenamefont {Gong}, \citenamefont {Lin}, \citenamefont {Lu},\ and\
  \citenamefont {Yi}}]{Zhang:2020uek}%
  \BibitemOpen
  \bibfield  {author} {\bibinfo {author} {\bibfnamefont {F.}~\bibnamefont
  {Zhang}}, \bibinfo {author} {\bibfnamefont {Y.}~\bibnamefont {Gong}},
  \bibinfo {author} {\bibfnamefont {J.}~\bibnamefont {Lin}}, \bibinfo {author}
  {\bibfnamefont {Y.}~\bibnamefont {Lu}},\ and\ \bibinfo {author}
  {\bibfnamefont {Z.}~\bibnamefont {Yi}},\ }\bibinfo {title} {{Primordial
  non-Gaussianity from G-inflation}},\ \href
  {https://doi.org/10.1088/1475-7516/2021/04/045} {J. Cosmol. Astropart. Phys.\
  \bibinfo {volume} {04}\bibfield  {year} {\bibinfo  {year} { (\textbf
  {2021})}\ }\bibfield  {pages} {\bibinfo  {pages} {045}},\ }\Eprint
  {https://arxiv.org/abs/2012.06960} {arXiv:2012.06960} \BibitemShut {NoStop}%
\bibitem [{\citenamefont {Creminelli}\ {\it et~al.}(2007)\citenamefont
  {Creminelli}, \citenamefont {Senatore}, \citenamefont {Zaldarriaga},\ and\
  \citenamefont {Tegmark}}]{Creminelli:2006rz}%
  \BibitemOpen
  \bibfield  {author} {\bibinfo {author} {\bibfnamefont {P.}~\bibnamefont
  {Creminelli}}, \bibinfo {author} {\bibfnamefont {L.}~\bibnamefont
  {Senatore}}, \bibinfo {author} {\bibfnamefont {M.}~\bibnamefont
  {Zaldarriaga}},\ and\ \bibinfo {author} {\bibfnamefont {M.}~\bibnamefont
  {Tegmark}},\ }\bibinfo {title} {{Limits on f\_NL parameters from WMAP 3yr
  data}},\ \href {https://doi.org/10.1088/1475-7516/2007/03/005} {J. Cosmol.
  Astropart. Phys.\ \bibinfo {volume} {03}\bibfield  {year} {\bibinfo  {year} {
  (\textbf {2007})}\ }\bibfield  {pages} {\bibinfo  {pages} {005}},\ }\Eprint
  {https://arxiv.org/abs/astro-ph/0610600} {arXiv:astro-ph/0610600}
  \BibitemShut {NoStop}%
\bibitem [{\citenamefont {Creminelli}\ and\ \citenamefont
  {Zaldarriaga}(2004)}]{Creminelli:2004yq}%
  \BibitemOpen
  \bibfield  {author} {\bibinfo {author} {\bibfnamefont {P.}~\bibnamefont
  {Creminelli}}\ and\ \bibinfo {author} {\bibfnamefont {M.}~\bibnamefont
  {Zaldarriaga}},\ }\bibinfo {title} {{Single field consistency relation for
  the 3-point function}},\ \href
  {https://doi.org/10.1088/1475-7516/2004/10/006} {J. Cosmol. Astropart. Phys.\
  \bibinfo {volume} {10}\bibfield  {year} {\bibinfo  {year} { (\textbf
  {2004})}\ }\bibfield  {pages} {\bibinfo  {pages} {006}},\ }\Eprint
  {https://arxiv.org/abs/astro-ph/0407059} {arXiv:astro-ph/0407059}
  \BibitemShut {NoStop}%
\bibitem [{\citenamefont {Carr}(1975)}]{Carr:1975qj}%
  \BibitemOpen
  \bibfield  {author} {\bibinfo {author} {\bibfnamefont {B.~J.}\ \bibnamefont
  {Carr}},\ }\bibinfo {title} {{The Primordial black hole mass spectrum}},\
  \href {https://doi.org/10.1086/153853} {\bibfield  {journal} {\bibinfo
  {journal} {Astrophys. J.}\ }\textbf {\bibinfo {volume} {201}},\ \bibinfo
  {pages} {1} (\bibinfo {year} {1975})}\BibitemShut {NoStop}%
\bibitem [{\citenamefont {Carr}\ {\it et~al.}(2016)\citenamefont {Carr},
  \citenamefont {Kuhnel},\ and\ \citenamefont {Sandstad}}]{Carr:2016drx}%
  \BibitemOpen
  \bibfield  {author} {\bibinfo {author} {\bibfnamefont {B.}~\bibnamefont
  {Carr}}, \bibinfo {author} {\bibfnamefont {F.}~\bibnamefont {Kuhnel}},\ and\
  \bibinfo {author} {\bibfnamefont {M.}~\bibnamefont {Sandstad}},\ }\bibinfo
  {title} {{Primordial Black Holes as Dark Matter}},\ \href
  {https://doi.org/10.1103/PhysRevD.94.083504} {\bibfield  {journal} {\bibinfo
  {journal} {Phys. Rev. D}\ }\textbf {\bibinfo {volume} {94}},\ \bibinfo
  {pages} {083504} (\bibinfo {year} {2016})},\ \Eprint
  {https://arxiv.org/abs/1607.06077} {arXiv:1607.06077} \BibitemShut {NoStop}%
\bibitem [{\citenamefont {\"Ozsoy}\ {\it et~al.}(2018)\citenamefont {\"Ozsoy},
  \citenamefont {Parameswaran}, \citenamefont {Tasinato},\ and\ \citenamefont
  {Zavala}}]{Ozsoy:2018flq}%
  \BibitemOpen
  \bibfield  {author} {\bibinfo {author} {\bibfnamefont {O.}~\bibnamefont
  {\"Ozsoy}}, \bibinfo {author} {\bibfnamefont {S.}~\bibnamefont
  {Parameswaran}}, \bibinfo {author} {\bibfnamefont {G.}~\bibnamefont
  {Tasinato}},\ and\ \bibinfo {author} {\bibfnamefont {I.}~\bibnamefont
  {Zavala}},\ }\bibinfo {title} {{Mechanisms for Primordial Black Hole
  Production in String Theory}},\ \href
  {https://doi.org/10.1088/1475-7516/2018/07/005} {J. Cosmol. Astropart. Phys.\
  \bibinfo {volume} {07}\bibfield  {year} {\bibinfo  {year} { (\textbf
  {2018})}\ }\bibfield  {pages} {\bibinfo  {pages} {005}},\ }\Eprint
  {https://arxiv.org/abs/1803.07626} {arXiv:1803.07626} \BibitemShut {NoStop}%
\bibitem [{\citenamefont {Tada}\ and\ \citenamefont
  {Yokoyama}(2019)}]{Tada:2019amh}%
  \BibitemOpen
  \bibfield  {author} {\bibinfo {author} {\bibfnamefont {Y.}~\bibnamefont
  {Tada}}\ and\ \bibinfo {author} {\bibfnamefont {S.}~\bibnamefont
  {Yokoyama}},\ }\bibinfo {title} {{Primordial black hole tower: Dark matter,
  earth-mass, and LIGO black holes}},\ \href
  {https://doi.org/10.1103/PhysRevD.100.023537} {\bibfield  {journal} {\bibinfo
   {journal} {Phys. Rev. D}\ }\textbf {\bibinfo {volume} {100}},\ \bibinfo
  {pages} {023537} (\bibinfo {year} {2019})},\ \Eprint
  {https://arxiv.org/abs/1904.10298} {arXiv:1904.10298} \BibitemShut {NoStop}%
\bibitem [{\citenamefont {Musco}(2019)}]{Musco:2018rwt}%
  \BibitemOpen
  \bibfield  {author} {\bibinfo {author} {\bibfnamefont {I.}~\bibnamefont
  {Musco}},\ }\bibinfo {title} {{Threshold for primordial black holes:
  Dependence on the shape of the cosmological perturbations}},\ \href
  {https://doi.org/10.1103/PhysRevD.100.123524} {\bibfield  {journal} {\bibinfo
   {journal} {Phys. Rev. D}\ }\textbf {\bibinfo {volume} {100}},\ \bibinfo
  {pages} {123524} (\bibinfo {year} {2019})},\ \Eprint
  {https://arxiv.org/abs/1809.02127} {arXiv:1809.02127} \BibitemShut {NoStop}%
\bibitem [{\citenamefont {De~Luca}\ {\it et~al.}(2019)\citenamefont {De~Luca},
  \citenamefont {Franciolini}, \citenamefont {Kehagias}, \citenamefont
  {Peloso}, \citenamefont {Riotto},\ and\ \citenamefont
  {\"Unal}}]{DeLuca:2019qsy}%
  \BibitemOpen
  \bibfield  {author} {\bibinfo {author} {\bibfnamefont {V.}~\bibnamefont
  {De~Luca}}, \bibinfo {author} {\bibfnamefont {G.}~\bibnamefont
  {Franciolini}}, \bibinfo {author} {\bibfnamefont {A.}~\bibnamefont
  {Kehagias}}, \bibinfo {author} {\bibfnamefont {M.}~\bibnamefont {Peloso}},
  \bibinfo {author} {\bibfnamefont {A.}~\bibnamefont {Riotto}},\ and\ \bibinfo
  {author} {\bibfnamefont {C.}~\bibnamefont {\"Unal}},\ }\bibinfo {title} {{The
  Ineludible non-Gaussianity of the Primordial Black Hole Abundance}},\ \href
  {https://doi.org/10.1088/1475-7516/2019/07/048} {J. Cosmol. Astropart. Phys.\
  \bibinfo {volume} {07}\bibfield  {year} {\bibinfo  {year} { (\textbf
  {2019})}\ }\bibfield  {pages} {\bibinfo  {pages} {048}},\ }\Eprint
  {https://arxiv.org/abs/1904.00970} {arXiv:1904.00970} \BibitemShut {NoStop}%
\bibitem [{\citenamefont {Aghanim}\ {\it et~al.}(2020)\citenamefont {Aghanim}
  {\it et~al.}}]{Aghanim:2018eyx}%
  \BibitemOpen
  \bibfield  {author} {\bibinfo {author} {\bibfnamefont {N.}~\bibnamefont
  {Aghanim}} {\it et~al.} (\bibinfo {collaboration} {Planck}),\ }\bibinfo
  {title} {{Planck 2018 results. VI. Cosmological parameters}},\ \href
  {https://doi.org/10.1051/0004-6361/201833910} {\bibfield  {journal} {\bibinfo
   {journal} {Astron. Astrophys.}\ }\textbf {\bibinfo {volume} {641}},\
  \bibinfo {pages} {A6} (\bibinfo {year} {2020})},\ \Eprint
  {https://arxiv.org/abs/1807.06209} {arXiv:1807.06209} \BibitemShut {NoStop}%
\bibitem [{\citenamefont {Harada}\ {\it et~al.}(2013)\citenamefont {Harada},
  \citenamefont {Yoo},\ and\ \citenamefont {Kohri}}]{Harada:2013epa}%
  \BibitemOpen
  \bibfield  {author} {\bibinfo {author} {\bibfnamefont {T.}~\bibnamefont
  {Harada}}, \bibinfo {author} {\bibfnamefont {C.-M.}\ \bibnamefont {Yoo}},\
  and\ \bibinfo {author} {\bibfnamefont {K.}~\bibnamefont {Kohri}},\ }\bibinfo
  {title} {{Threshold of primordial black hole formation}},\ \href
  {https://doi.org/10.1103/PhysRevD.88.084051} {\bibfield  {journal} {\bibinfo
  {journal} {Phys. Rev. D}\ }\textbf {\bibinfo {volume} {88}},\ \bibinfo
  {pages} {084051} (\bibinfo {year} {2013})},\ \bibinfo {note} {[Erratum:
  Phys.Rev.D 89, 029903 (2014)]},\ \Eprint {https://arxiv.org/abs/1309.4201}
  {arXiv:1309.4201} \BibitemShut {NoStop}%
\bibitem [{\citenamefont {Escriv\`a}\ {\it et~al.}(2020)\citenamefont
  {Escriv\`a}, \citenamefont {Germani},\ and\ \citenamefont
  {Sheth}}]{Escriva:2019phb}%
  \BibitemOpen
  \bibfield  {author} {\bibinfo {author} {\bibfnamefont {A.}~\bibnamefont
  {Escriv\`a}}, \bibinfo {author} {\bibfnamefont {C.}~\bibnamefont {Germani}},\
  and\ \bibinfo {author} {\bibfnamefont {R.~K.}\ \bibnamefont {Sheth}},\
  }\bibinfo {title} {{Universal threshold for primordial black hole
  formation}},\ \href {https://doi.org/10.1103/PhysRevD.101.044022} {\bibfield
  {journal} {\bibinfo  {journal} {Phys. Rev. D}\ }\textbf {\bibinfo {volume}
  {101}},\ \bibinfo {pages} {044022} (\bibinfo {year} {2020})},\ \Eprint
  {https://arxiv.org/abs/1907.13311} {arXiv:1907.13311} \BibitemShut {NoStop}%
\bibitem [{\citenamefont {Musco}\ {\it et~al.}(2021)\citenamefont {Musco},
  \citenamefont {De~Luca}, \citenamefont {Franciolini},\ and\ \citenamefont
  {Riotto}}]{Musco:2020jjb}%
  \BibitemOpen
  \bibfield  {author} {\bibinfo {author} {\bibfnamefont {I.}~\bibnamefont
  {Musco}}, \bibinfo {author} {\bibfnamefont {V.}~\bibnamefont {De~Luca}},
  \bibinfo {author} {\bibfnamefont {G.}~\bibnamefont {Franciolini}},\ and\
  \bibinfo {author} {\bibfnamefont {A.}~\bibnamefont {Riotto}},\ }\bibinfo
  {title} {{Threshold for primordial black holes. II. A simple analytic
  prescription}},\ \href {https://doi.org/10.1103/PhysRevD.103.063538}
  {\bibfield  {journal} {\bibinfo  {journal} {Phys. Rev. D}\ }\textbf {\bibinfo
  {volume} {103}},\ \bibinfo {pages} {063538} (\bibinfo {year} {2021})},\
  \Eprint {https://arxiv.org/abs/2011.03014} {arXiv:2011.03014} \BibitemShut
  {NoStop}%
\bibitem [{\citenamefont {Carr}\ {\it et~al.}(2010)\citenamefont {Carr},
  \citenamefont {Kohri}, \citenamefont {Sendouda},\ and\ \citenamefont
  {Yokoyama}}]{Carr:2009jm}%
  \BibitemOpen
  \bibfield  {author} {\bibinfo {author} {\bibfnamefont {B.~J.}\ \bibnamefont
  {Carr}}, \bibinfo {author} {\bibfnamefont {K.}~\bibnamefont {Kohri}},
  \bibinfo {author} {\bibfnamefont {Y.}~\bibnamefont {Sendouda}},\ and\
  \bibinfo {author} {\bibfnamefont {J.}~\bibnamefont {Yokoyama}},\ }\bibinfo
  {title} {{New cosmological constraints on primordial black holes}},\ \href
  {https://doi.org/10.1103/PhysRevD.81.104019} {\bibfield  {journal} {\bibinfo
  {journal} {Phys. Rev. D}\ }\textbf {\bibinfo {volume} {81}},\ \bibinfo
  {pages} {104019} (\bibinfo {year} {2010})},\ \Eprint
  {https://arxiv.org/abs/0912.5297} {arXiv:0912.5297} \BibitemShut {NoStop}%
\bibitem [{\citenamefont {Laha}(2019)}]{Laha:2019ssq}%
  \BibitemOpen
  \bibfield  {author} {\bibinfo {author} {\bibfnamefont {R.}~\bibnamefont
  {Laha}},\ }\bibinfo {title} {{Primordial Black Holes as a Dark Matter
  Candidate Are Severely Constrained by the Galactic Center 511 keV $\gamma$
  -Ray Line}},\ \href {https://doi.org/10.1103/PhysRevLett.123.251101}
  {\bibfield  {journal} {\bibinfo  {journal} {Phys. Rev. Lett.}\ }\textbf
  {\bibinfo {volume} {123}},\ \bibinfo {pages} {251101} (\bibinfo {year}
  {2019})},\ \Eprint {https://arxiv.org/abs/1906.09994} {arXiv:1906.09994}
  \BibitemShut {NoStop}%
\bibitem [{\citenamefont {Dasgupta}\ {\it et~al.}(2020)\citenamefont
  {Dasgupta}, \citenamefont {Laha},\ and\ \citenamefont
  {Ray}}]{Dasgupta:2019cae}%
  \BibitemOpen
  \bibfield  {author} {\bibinfo {author} {\bibfnamefont {B.}~\bibnamefont
  {Dasgupta}}, \bibinfo {author} {\bibfnamefont {R.}~\bibnamefont {Laha}},\
  and\ \bibinfo {author} {\bibfnamefont {A.}~\bibnamefont {Ray}},\ }\bibinfo
  {title} {{Neutrino and positron constraints on spinning primordial black hole
  dark matter}},\ \href {https://doi.org/10.1103/PhysRevLett.125.101101}
  {\bibfield  {journal} {\bibinfo  {journal} {Phys. Rev. Lett.}\ }\textbf
  {\bibinfo {volume} {125}},\ \bibinfo {pages} {101101} (\bibinfo {year}
  {2020})},\ \Eprint {https://arxiv.org/abs/1912.01014} {arXiv:1912.01014}
  \BibitemShut {NoStop}%
\bibitem [{\citenamefont {Niikura}\ {\it et~al.}(2019)\citenamefont {Niikura}
  {\it et~al.}}]{Niikura:2017zjd}%
  \BibitemOpen
  \bibfield  {author} {\bibinfo {author} {\bibfnamefont {H.}~\bibnamefont
  {Niikura}} {\it et~al.},\ }\bibinfo {title} {{Microlensing constraints on
  primordial black holes with Subaru/HSC Andromeda observations}},\ \href
  {https://doi.org/10.1038/s41550-019-0723-1} {\bibfield  {journal} {\bibinfo
  {journal} {Nature Astron.}\ }\textbf {\bibinfo {volume} {3}},\ \bibinfo
  {pages} {524} (\bibinfo {year} {2019})},\ \Eprint
  {https://arxiv.org/abs/1701.02151} {arXiv:1701.02151} \BibitemShut {NoStop}%
\bibitem [{\citenamefont {Griest}\ {\it et~al.}(2013)\citenamefont {Griest},
  \citenamefont {Cieplak},\ and\ \citenamefont {Lehner}}]{Griest:2013esa}%
  \BibitemOpen
  \bibfield  {author} {\bibinfo {author} {\bibfnamefont {K.}~\bibnamefont
  {Griest}}, \bibinfo {author} {\bibfnamefont {A.~M.}\ \bibnamefont
  {Cieplak}},\ and\ \bibinfo {author} {\bibfnamefont {M.~J.}\ \bibnamefont
  {Lehner}},\ }\bibinfo {title} {{New Limits on Primordial Black Hole Dark
  Matter from an Analysis of Kepler Source Microlensing Data}},\ \href
  {https://doi.org/10.1103/PhysRevLett.111.181302} {\bibfield  {journal}
  {\bibinfo  {journal} {Phys. Rev. Lett.}\ }\textbf {\bibinfo {volume} {111}},\
  \bibinfo {pages} {181302} (\bibinfo {year} {2013})}\BibitemShut {NoStop}%
\bibitem [{\citenamefont {Tisserand}\ {\it et~al.}(2007)\citenamefont
  {Tisserand} {\it et~al.}}]{Tisserand:2006zx}%
  \BibitemOpen
  \bibfield  {author} {\bibinfo {author} {\bibfnamefont {P.}~\bibnamefont
  {Tisserand}} {\it et~al.} (\bibinfo {collaboration} {EROS-2}),\ }\bibinfo
  {title} {{Limits on the Macho Content of the Galactic Halo from the EROS-2
  Survey of the Magellanic Clouds}},\ \href
  {https://doi.org/10.1051/0004-6361:20066017} {\bibfield  {journal} {\bibinfo
  {journal} {Astron. Astrophys.}\ }\textbf {\bibinfo {volume} {469}},\ \bibinfo
  {pages} {387} (\bibinfo {year} {2007})},\ \Eprint
  {https://arxiv.org/abs/astro-ph/0607207} {arXiv:astro-ph/0607207}
  \BibitemShut {NoStop}%
\bibitem [{\citenamefont {Ali-Ha\"\i{}moud}\ {\it et~al.}(2017)\citenamefont
  {Ali-Ha\"\i{}moud}, \citenamefont {Kovetz},\ and\ \citenamefont
  {Kamionkowski}}]{Ali-Haimoud:2017rtz}%
  \BibitemOpen
  \bibfield  {author} {\bibinfo {author} {\bibfnamefont {Y.}~\bibnamefont
  {Ali-Ha\"\i{}moud}}, \bibinfo {author} {\bibfnamefont {E.~D.}\ \bibnamefont
  {Kovetz}},\ and\ \bibinfo {author} {\bibfnamefont {M.}~\bibnamefont
  {Kamionkowski}},\ }\bibinfo {title} {{Merger rate of primordial black-hole
  binaries}},\ \href {https://doi.org/10.1103/PhysRevD.96.123523} {\bibfield
  {journal} {\bibinfo  {journal} {Phys. Rev. D}\ }\textbf {\bibinfo {volume}
  {96}},\ \bibinfo {pages} {123523} (\bibinfo {year} {2017})},\ \Eprint
  {https://arxiv.org/abs/1709.06576} {arXiv:1709.06576} \BibitemShut {NoStop}%
\bibitem [{\citenamefont {Raidal}\ {\it et~al.}(2017)\citenamefont {Raidal},
  \citenamefont {Vaskonen},\ and\ \citenamefont {Veerm\"ae}}]{Raidal:2017mfl}%
  \BibitemOpen
  \bibfield  {author} {\bibinfo {author} {\bibfnamefont {M.}~\bibnamefont
  {Raidal}}, \bibinfo {author} {\bibfnamefont {V.}~\bibnamefont {Vaskonen}},\
  and\ \bibinfo {author} {\bibfnamefont {H.}~\bibnamefont {Veerm\"ae}},\
  }\bibinfo {title} {{Gravitational Waves from Primordial Black Hole
  Mergers}},\ \href {https://doi.org/10.1088/1475-7516/2017/09/037} {J. Cosmol.
  Astropart. Phys.\ \bibinfo {volume} {09}\bibfield  {year} {\bibinfo  {year} {
  (\textbf {2017})}\ }\bibfield  {pages} {\bibinfo  {pages} {037}},\ }\Eprint
  {https://arxiv.org/abs/1707.01480} {arXiv:1707.01480} \BibitemShut {NoStop}%
\bibitem [{\citenamefont {Ali-Ha\"\i{}moud}\ and\ \citenamefont
  {Kamionkowski}(2017)}]{Ali-Haimoud:2016mbv}%
  \BibitemOpen
  \bibfield  {author} {\bibinfo {author} {\bibfnamefont {Y.}~\bibnamefont
  {Ali-Ha\"\i{}moud}}\ and\ \bibinfo {author} {\bibfnamefont {M.}~\bibnamefont
  {Kamionkowski}},\ }\bibinfo {title} {{Cosmic microwave background limits on
  accreting primordial black holes}},\ \href
  {https://doi.org/10.1103/PhysRevD.95.043534} {\bibfield  {journal} {\bibinfo
  {journal} {Phys. Rev. D}\ }\textbf {\bibinfo {volume} {95}},\ \bibinfo
  {pages} {043534} (\bibinfo {year} {2017})},\ \Eprint
  {https://arxiv.org/abs/1612.05644} {arXiv:1612.05644} \BibitemShut {NoStop}%
\bibitem [{\citenamefont {Poulin}\ {\it et~al.}(2017)\citenamefont {Poulin},
  \citenamefont {Serpico}, \citenamefont {Calore}, \citenamefont {Clesse},\
  and\ \citenamefont {Kohri}}]{Poulin:2017bwe}%
  \BibitemOpen
  \bibfield  {author} {\bibinfo {author} {\bibfnamefont {V.}~\bibnamefont
  {Poulin}}, \bibinfo {author} {\bibfnamefont {P.~D.}\ \bibnamefont {Serpico}},
  \bibinfo {author} {\bibfnamefont {F.}~\bibnamefont {Calore}}, \bibinfo
  {author} {\bibfnamefont {S.}~\bibnamefont {Clesse}},\ and\ \bibinfo {author}
  {\bibfnamefont {K.}~\bibnamefont {Kohri}},\ }\bibinfo {title} {{CMB bounds on
  disk-accreting massive primordial black holes}},\ \href
  {https://doi.org/10.1103/PhysRevD.96.083524} {\bibfield  {journal} {\bibinfo
  {journal} {Phys. Rev. D}\ }\textbf {\bibinfo {volume} {96}},\ \bibinfo
  {pages} {083524} (\bibinfo {year} {2017})},\ \Eprint
  {https://arxiv.org/abs/1707.04206} {arXiv:1707.04206} \BibitemShut {NoStop}%
\bibitem [{\citenamefont {Wang}\ {\it et~al.}(2019)\citenamefont {Wang},
  \citenamefont {Terada},\ and\ \citenamefont {Kohri}}]{Wang:2019kaf}%
  \BibitemOpen
  \bibfield  {author} {\bibinfo {author} {\bibfnamefont {S.}~\bibnamefont
  {Wang}}, \bibinfo {author} {\bibfnamefont {T.}~\bibnamefont {Terada}},\ and\
  \bibinfo {author} {\bibfnamefont {K.}~\bibnamefont {Kohri}},\ }\bibinfo
  {title} {{Prospective constraints on the primordial black hole abundance from
  the stochastic gravitational-wave backgrounds produced by coalescing events
  and curvature perturbations}},\ \href
  {https://doi.org/10.1103/PhysRevD.99.103531} {\bibfield  {journal} {\bibinfo
  {journal} {Phys. Rev. D}\ }\textbf {\bibinfo {volume} {99}},\ \bibinfo
  {pages} {103531} (\bibinfo {year} {2019})},\ \bibinfo {note} {[Erratum:
  Phys.Rev.D 101, 069901 (2020)]},\ \Eprint {https://arxiv.org/abs/1903.05924}
  {arXiv:1903.05924} \BibitemShut {NoStop}%
\bibitem [{\citenamefont {Franciolini}\ {\it et~al.}(2018)\citenamefont
  {Franciolini}, \citenamefont {Kehagias}, \citenamefont {Matarrese},\ and\
  \citenamefont {Riotto}}]{Franciolini:2018vbk}%
  \BibitemOpen
  \bibfield  {author} {\bibinfo {author} {\bibfnamefont {G.}~\bibnamefont
  {Franciolini}}, \bibinfo {author} {\bibfnamefont {A.}~\bibnamefont
  {Kehagias}}, \bibinfo {author} {\bibfnamefont {S.}~\bibnamefont
  {Matarrese}},\ and\ \bibinfo {author} {\bibfnamefont {A.}~\bibnamefont
  {Riotto}},\ }\bibinfo {title} {{Primordial Black Holes from Inflation and
  non-Gaussianity}},\ \href {https://doi.org/10.1088/1475-7516/2018/03/016} {J.
  Cosmol. Astropart. Phys.\ \bibinfo {volume} {03}\bibfield  {year} {\bibinfo
  {year} { (\textbf {2018})}\ }\bibfield  {pages} {\bibinfo  {pages} {016}},\
  }\Eprint {https://arxiv.org/abs/1801.09415} {arXiv:1801.09415} \BibitemShut
  {NoStop}%
\bibitem [{\citenamefont {Kehagias}\ {\it et~al.}(2019)\citenamefont
  {Kehagias}, \citenamefont {Musco},\ and\ \citenamefont
  {Riotto}}]{Kehagias:2019eil}%
  \BibitemOpen
  \bibfield  {author} {\bibinfo {author} {\bibfnamefont {A.}~\bibnamefont
  {Kehagias}}, \bibinfo {author} {\bibfnamefont {I.}~\bibnamefont {Musco}},\
  and\ \bibinfo {author} {\bibfnamefont {A.}~\bibnamefont {Riotto}},\ }\bibinfo
  {title} {{Non-Gaussian Formation of Primordial Black Holes: Effects on the
  Threshold}},\ \href {https://doi.org/10.1088/1475-7516/2019/12/029} {J.
  Cosmol. Astropart. Phys.\ \bibinfo {volume} {12}\bibfield  {year} {\bibinfo
  {year} { (\textbf {2019})}\ }\bibfield  {pages} {\bibinfo  {pages} {029}},\
  }\Eprint {https://arxiv.org/abs/1906.07135} {arXiv:1906.07135} \BibitemShut
  {NoStop}%
\bibitem [{\citenamefont {Atal}\ and\ \citenamefont
  {Germani}(2019)}]{Atal:2018neu}%
  \BibitemOpen
  \bibfield  {author} {\bibinfo {author} {\bibfnamefont {V.}~\bibnamefont
  {Atal}}\ and\ \bibinfo {author} {\bibfnamefont {C.}~\bibnamefont {Germani}},\
  }\bibinfo {title} {{The role of non-gaussianities in Primordial Black Hole
  formation}},\ \href {https://doi.org/10.1016/j.dark.2019.100275} {\bibfield
  {journal} {\bibinfo  {journal} {Phys. Dark Univ.}\ }\textbf {\bibinfo
  {volume} {24}},\ \bibinfo {pages} {100275} (\bibinfo {year} {2019})},\
  \Eprint {https://arxiv.org/abs/1811.07857} {arXiv:1811.07857} \BibitemShut
  {NoStop}%
\bibitem [{\citenamefont {Riccardi}\ {\it et~al.}(2021)\citenamefont
  {Riccardi}, \citenamefont {Taoso},\ and\ \citenamefont
  {Urbano}}]{Riccardi:2021rlf}%
  \BibitemOpen
  \bibfield  {author} {\bibinfo {author} {\bibfnamefont {F.}~\bibnamefont
  {Riccardi}}, \bibinfo {author} {\bibfnamefont {M.}~\bibnamefont {Taoso}},\
  and\ \bibinfo {author} {\bibfnamefont {A.}~\bibnamefont {Urbano}},\ }\bibinfo
  {title} {{Solving peak theory in the presence of local non-gaussianities}},\
  \href@noop {} {\  (\bibinfo {year} {2021})},\ \Eprint
  {https://arxiv.org/abs/2102.04084} {arXiv:2102.04084} \BibitemShut {NoStop}%
\bibitem [{\citenamefont {Lu}\ {\it et~al.}(2020)\citenamefont {Lu},
  \citenamefont {Ali}, \citenamefont {Gong}, \citenamefont {Lin},\ and\
  \citenamefont {Zhang}}]{Lu:2020diy}%
  \BibitemOpen
  \bibfield  {author} {\bibinfo {author} {\bibfnamefont {Y.}~\bibnamefont
  {Lu}}, \bibinfo {author} {\bibfnamefont {A.}~\bibnamefont {Ali}}, \bibinfo
  {author} {\bibfnamefont {Y.}~\bibnamefont {Gong}}, \bibinfo {author}
  {\bibfnamefont {J.}~\bibnamefont {Lin}},\ and\ \bibinfo {author}
  {\bibfnamefont {F.}~\bibnamefont {Zhang}},\ }\bibinfo {title} {{Gauge
  transformation of scalar induced gravitational waves}},\ \href
  {https://doi.org/10.1103/PhysRevD.102.083503} {\bibfield  {journal}
  {\bibinfo  {journal} {Phys. Rev. D}\ }\textbf {\bibinfo {volume} {102}},\
  \bibinfo {pages} {083503} (\bibinfo {year} {2020})},\ \Eprint
  {https://arxiv.org/abs/2006.03450} {arXiv:2006.03450} \BibitemShut {NoStop}%
\bibitem [{\citenamefont {Ali}\ {\it et~al.}(2021)\citenamefont {Ali},
  \citenamefont {Gong},\ and\ \citenamefont {Lu}}]{Ali:2020sfw}%
  \BibitemOpen
  \bibfield  {author} {\bibinfo {author} {\bibfnamefont {A.}~\bibnamefont
  {Ali}}, \bibinfo {author} {\bibfnamefont {Y.}~\bibnamefont {Gong}},\ and\
  \bibinfo {author} {\bibfnamefont {Y.}~\bibnamefont {Lu}},\ }\bibinfo {title}
  {{Gauge transformation of scalar induced tensor perturbation during matter
  domination}},\ \href {https://doi.org/10.1103/PhysRevD.103.043516} {\bibfield
   {journal} {\bibinfo  {journal} {Phys. Rev. D}\ }\textbf {\bibinfo {volume}
  {103}},\ \bibinfo {pages} {043516} (\bibinfo {year} {2021})},\ \Eprint
  {https://arxiv.org/abs/2009.11081} {arXiv:2009.11081} \BibitemShut {NoStop}%
\bibitem [{\citenamefont {Chang}\ {\it et~al.}(2020{\natexlab{a}})\citenamefont
  {Chang}, \citenamefont {Wang},\ and\ \citenamefont {Zhu}}]{Chang:2020tji}%
  \BibitemOpen
  \bibfield  {author} {\bibinfo {author} {\bibfnamefont {Z.}~\bibnamefont
  {Chang}}, \bibinfo {author} {\bibfnamefont {S.}~\bibnamefont {Wang}},\ and\
  \bibinfo {author} {\bibfnamefont {Q.-H.}\ \bibnamefont {Zhu}},\ }\bibinfo
  {title} {{Note on gauge invariance of second order cosmological
  perturbations}}\ \href {https://doi.org/10.1088/1674-1137/ac0c74}
  {10.1088/1674-1137/ac0c74} (\bibinfo {year} {2020}{\natexlab{a}}),\ \Eprint
  {https://arxiv.org/abs/2009.11025} {arXiv:2009.11025} \BibitemShut {NoStop}%
\bibitem [{\citenamefont {Chang}\ {\it et~al.}(2020{\natexlab{b}})\citenamefont
  {Chang}, \citenamefont {Wang},\ and\ \citenamefont {Zhu}}]{Chang:2020iji}%
  \BibitemOpen
  \bibfield  {author} {\bibinfo {author} {\bibfnamefont {Z.}~\bibnamefont
  {Chang}}, \bibinfo {author} {\bibfnamefont {S.}~\bibnamefont {Wang}},\ and\
  \bibinfo {author} {\bibfnamefont {Q.-H.}\ \bibnamefont {Zhu}},\ }\bibinfo
  {title} {{Gauge Invariant Second Order Gravitational Waves}},\ \href@noop {}
  {\  (\bibinfo {year} {2020}{\natexlab{b}})},\ \Eprint
  {https://arxiv.org/abs/2009.11994} {arXiv:2009.11994} \BibitemShut {NoStop}%
\bibitem [{\citenamefont {Chang}\ {\it et~al.}(2020{\natexlab{c}})\citenamefont
  {Chang}, \citenamefont {Wang},\ and\ \citenamefont {Zhu}}]{Chang:2020mky}%
  \BibitemOpen
  \bibfield  {author} {\bibinfo {author} {\bibfnamefont {Z.}~\bibnamefont
  {Chang}}, \bibinfo {author} {\bibfnamefont {S.}~\bibnamefont {Wang}},\ and\
  \bibinfo {author} {\bibfnamefont {Q.-H.}\ \bibnamefont {Zhu}},\ }\bibinfo
  {title} {{On the Gauge Invariance of Scalar Induced Gravitational Waves:
  Gauge Fixings Considered}},\ \href@noop {} {\  (\bibinfo {year}
  {2020}{\natexlab{c}})},\ \Eprint {https://arxiv.org/abs/2010.01487}
  {arXiv:2010.01487} \BibitemShut {NoStop}%
\bibitem [{\citenamefont {Dom\`enech}\ and\ \citenamefont
  {Sasaki}(2021)}]{Domenech:2020xin}%
  \BibitemOpen
  \bibfield  {author} {\bibinfo {author} {\bibfnamefont {G.}~\bibnamefont
  {Dom\`enech}}\ and\ \bibinfo {author} {\bibfnamefont {M.}~\bibnamefont
  {Sasaki}},\ }\bibinfo {title} {{Approximate gauge independence of the induced
  gravitational wave spectrum}},\ \href
  {https://doi.org/10.1103/PhysRevD.103.063531} {\bibfield  {journal} {\bibinfo
   {journal} {Phys. Rev. D}\ }\textbf {\bibinfo {volume} {103}},\ \bibinfo
  {pages} {063531} (\bibinfo {year} {2021})},\ \Eprint
  {https://arxiv.org/abs/2012.14016} {arXiv:2012.14016} \BibitemShut {NoStop}%
\bibitem [{\citenamefont {Inomata}\ and\ \citenamefont
  {Terada}(2020)}]{Inomata:2019yww}%
  \BibitemOpen
  \bibfield  {author} {\bibinfo {author} {\bibfnamefont {K.}~\bibnamefont
  {Inomata}}\ and\ \bibinfo {author} {\bibfnamefont {T.}~\bibnamefont
  {Terada}},\ }\bibinfo {title} {{Gauge Independence of Induced Gravitational
  Waves}},\ \href {https://doi.org/10.1103/PhysRevD.101.023523} {\bibfield
  {journal} {\bibinfo  {journal} {Phys. Rev. D}\ }\textbf {\bibinfo {volume}
  {101}},\ \bibinfo {pages} {023523} (\bibinfo {year} {2020})},\ \Eprint
  {https://arxiv.org/abs/1912.00785} {arXiv:1912.00785} \BibitemShut {NoStop}%
\bibitem [{\citenamefont {Tomikawa}\ and\ \citenamefont
  {Kobayashi}(2020)}]{Tomikawa:2019tvi}%
  \BibitemOpen
  \bibfield  {author} {\bibinfo {author} {\bibfnamefont {K.}~\bibnamefont
  {Tomikawa}}\ and\ \bibinfo {author} {\bibfnamefont {T.}~\bibnamefont
  {Kobayashi}},\ }\bibinfo {title} {{Gauge dependence of gravitational waves
  generated at second order from scalar perturbations}},\ \href
  {https://doi.org/10.1103/PhysRevD.101.083529} {\bibfield  {journal} {\bibinfo
   {journal} {Phys. Rev. D}\ }\textbf {\bibinfo {volume} {101}},\ \bibinfo
  {pages} {083529} (\bibinfo {year} {2020})},\ \Eprint
  {https://arxiv.org/abs/1910.01880} {arXiv:1910.01880} \BibitemShut {NoStop}%
\bibitem [{\citenamefont {Yuan}\ {\it et~al.}(2020)\citenamefont {Yuan},
  \citenamefont {Chen},\ and\ \citenamefont {Huang}}]{Yuan:2019fwv}%
  \BibitemOpen
  \bibfield  {author} {\bibinfo {author} {\bibfnamefont {C.}~\bibnamefont
  {Yuan}}, \bibinfo {author} {\bibfnamefont {Z.-C.}\ \bibnamefont {Chen}},\
  and\ \bibinfo {author} {\bibfnamefont {Q.-G.}\ \bibnamefont {Huang}},\
  }\bibinfo {title} {{Scalar induced gravitational waves in different
  gauges}},\ \href {https://doi.org/10.1103/PhysRevD.101.063018} {\bibfield
  {journal} {\bibinfo  {journal} {Phys. Rev. D}\ }\textbf {\bibinfo {volume}
  {101}},\ \bibinfo {pages} {063018} (\bibinfo {year} {2020})},\ \Eprint
  {https://arxiv.org/abs/1912.00885} {arXiv:1912.00885} \BibitemShut {NoStop}%
\bibitem [{\citenamefont {De~Luca}\ {\it et~al.}(2020)\citenamefont {De~Luca},
  \citenamefont {Franciolini}, \citenamefont {Kehagias},\ and\ \citenamefont
  {Riotto}}]{DeLuca:2019ufz}%
  \BibitemOpen
  \bibfield  {author} {\bibinfo {author} {\bibfnamefont {V.}~\bibnamefont
  {De~Luca}}, \bibinfo {author} {\bibfnamefont {G.}~\bibnamefont
  {Franciolini}}, \bibinfo {author} {\bibfnamefont {A.}~\bibnamefont
  {Kehagias}},\ and\ \bibinfo {author} {\bibfnamefont {A.}~\bibnamefont
  {Riotto}},\ }\bibinfo {title} {{On the Gauge Invariance of Cosmological
  Gravitational Waves}},\ \href {https://doi.org/10.1088/1475-7516/2020/03/014}
  {J. Cosmol. Astropart. Phys.\ \bibinfo {volume} {03}\bibfield  {year}
  {\bibinfo  {year} { (\textbf {2020})}\ }\bibfield  {pages} {\bibinfo  {pages}
  {014}},\ }\Eprint {https://arxiv.org/abs/1911.09689} {arXiv:1911.09689}
  \BibitemShut {NoStop}%
\bibitem [{\citenamefont {Verde}\ {\it et~al.}(2000)\citenamefont {Verde},
  \citenamefont {Wang}, \citenamefont {Heavens},\ and\ \citenamefont
  {Kamionkowski}}]{Verde:1999ij}%
  \BibitemOpen
  \bibfield  {author} {\bibinfo {author} {\bibfnamefont {L.}~\bibnamefont
  {Verde}}, \bibinfo {author} {\bibfnamefont {L.-M.}\ \bibnamefont {Wang}},
  \bibinfo {author} {\bibfnamefont {A.}~\bibnamefont {Heavens}},\ and\ \bibinfo
  {author} {\bibfnamefont {M.}~\bibnamefont {Kamionkowski}},\ }\bibinfo {title}
  {{Large scale structure, the cosmic microwave background, and primordial
  non-gaussianity}},\ \href {https://doi.org/10.1046/j.1365-8711.2000.03191.x}
  {\bibfield  {journal} {\bibinfo  {journal} {Mon. Not. Roy. Astron. Soc.}\
  }\textbf {\bibinfo {volume} {313}},\ \bibinfo {pages} {L141} (\bibinfo {year}
  {2000})},\ \Eprint {https://arxiv.org/abs/astro-ph/9906301}
  {arXiv:astro-ph/9906301} \BibitemShut {NoStop}%
\bibitem [{\citenamefont {Komatsu}\ and\ \citenamefont
  {Spergel}(2001)}]{Komatsu:2001rj}%
  \BibitemOpen
  \bibfield  {author} {\bibinfo {author} {\bibfnamefont {E.}~\bibnamefont
  {Komatsu}}\ and\ \bibinfo {author} {\bibfnamefont {D.~N.}\ \bibnamefont
  {Spergel}},\ }\bibinfo {title} {{Acoustic signatures in the primary microwave
  background bispectrum}},\ \href {https://doi.org/10.1103/PhysRevD.63.063002}
  {\bibfield  {journal} {\bibinfo  {journal} {Phys. Rev. D}\ }\textbf {\bibinfo
  {volume} {63}},\ \bibinfo {pages} {063002} (\bibinfo {year} {2001})},\
  \Eprint {https://arxiv.org/abs/astro-ph/0005036} {arXiv:astro-ph/0005036}
  \BibitemShut {NoStop}%
\bibitem [{\citenamefont {Chen}\ {\it et~al.}(2007)\citenamefont {Chen},
  \citenamefont {Huang}, \citenamefont {Kachru},\ and\ \citenamefont
  {Shiu}}]{Chen:2006nt}%
  \BibitemOpen
  \bibfield  {author} {\bibinfo {author} {\bibfnamefont {X.}~\bibnamefont
  {Chen}}, \bibinfo {author} {\bibfnamefont {M.-x.}\ \bibnamefont {Huang}},
  \bibinfo {author} {\bibfnamefont {S.}~\bibnamefont {Kachru}},\ and\ \bibinfo
  {author} {\bibfnamefont {G.}~\bibnamefont {Shiu}},\ }\bibinfo {title}
  {{Observational signatures and non-Gaussianities of general single field
  inflation}},\ \href {https://doi.org/10.1088/1475-7516/2007/01/002} {J.
  Cosmol. Astropart. Phys.\ \bibinfo {volume} {01}\bibfield  {year} {\bibinfo
  {year} { (\textbf {2007})}\ }\bibfield  {pages} {\bibinfo  {pages} {002}},\
  }\Eprint {https://arxiv.org/abs/hep-th/0605045} {arXiv:hep-th/0605045}
  \BibitemShut {NoStop}%
\bibitem [{\citenamefont {Ferdman}\ {\it et~al.}(2010)\citenamefont {Ferdman}
  {\it et~al.}}]{Ferdman:2010xq}%
  \BibitemOpen
  \bibfield  {author} {\bibinfo {author} {\bibfnamefont {R.~D.}\ \bibnamefont
  {Ferdman}} {\it et~al.},\ }\bibinfo {title} {{The European Pulsar Timing
  Array: current efforts and a LEAP toward the future}},\ \href
  {https://doi.org/10.1088/0264-9381/27/8/084014} {\bibfield  {journal}
  {\bibinfo  {journal} {Class. Quant. Grav.}\ }\textbf {\bibinfo {volume}
  {27}},\ \bibinfo {pages} {084014} (\bibinfo {year} {2010})},\ \Eprint
  {https://arxiv.org/abs/1003.3405} {arXiv:1003.3405} \BibitemShut {NoStop}%
\bibitem [{\citenamefont {Hobbs}\ {\it et~al.}(2010)\citenamefont {Hobbs} {\it
  et~al.}}]{Hobbs:2009yy}%
  \BibitemOpen
  \bibfield  {author} {\bibinfo {author} {\bibfnamefont {G.}~\bibnamefont
  {Hobbs}} {\it et~al.},\ }\bibinfo {title} {{The international pulsar timing
  array project: using pulsars as a gravitational wave detector}},\ \href
  {https://doi.org/10.1088/0264-9381/27/8/084013} {\bibfield  {journal}
  {\bibinfo  {journal} {Class. Quant. Grav.}\ }\textbf {\bibinfo {volume}
  {27}},\ \bibinfo {pages} {084013} (\bibinfo {year} {2010})},\ \Eprint
  {https://arxiv.org/abs/0911.5206} {arXiv:0911.5206} \BibitemShut {NoStop}%
\bibitem [{\citenamefont {McLaughlin}(2013)}]{McLaughlin:2013ira}%
  \BibitemOpen
  \bibfield  {author} {\bibinfo {author} {\bibfnamefont {M.~A.}\ \bibnamefont
  {McLaughlin}},\ }\bibinfo {title} {{The North American Nanohertz Observatory
  for Gravitational Waves}},\ \href
  {https://doi.org/10.1088/0264-9381/30/22/224008} {\bibfield  {journal}
  {\bibinfo  {journal} {Class. Quant. Grav.}\ }\textbf {\bibinfo {volume}
  {30}},\ \bibinfo {pages} {224008} (\bibinfo {year} {2013})},\ \Eprint
  {https://arxiv.org/abs/1310.0758} {arXiv:1310.0758} \BibitemShut {NoStop}%
\bibitem [{\citenamefont {Hobbs}(2013)}]{Hobbs:2013aka}%
  \BibitemOpen
  \bibfield  {author} {\bibinfo {author} {\bibfnamefont {G.}~\bibnamefont
  {Hobbs}},\ }\bibinfo {title} {{The Parkes Pulsar Timing Array}},\ \href
  {https://doi.org/10.1088/0264-9381/30/22/224007} {\bibfield  {journal}
  {\bibinfo  {journal} {Class. Quant. Grav.}\ }\textbf {\bibinfo {volume}
  {30}},\ \bibinfo {pages} {224007} (\bibinfo {year} {2013})},\ \Eprint
  {https://arxiv.org/abs/1307.2629} {arXiv:1307.2629} \BibitemShut {NoStop}%
\bibitem [{\citenamefont {Harry}(2010)}]{Harry:2010zz}%
  \BibitemOpen
  \bibfield  {author} {\bibinfo {author} {\bibfnamefont {G.~M.}\ \bibnamefont
  {Harry}} (\bibinfo {collaboration} {LIGO Scientific}),\ }\bibinfo {title}
  {{Advanced LIGO: The next generation of gravitational wave detectors}},\
  \href {https://doi.org/10.1088/0264-9381/27/8/084006} {\bibfield  {journal}
  {\bibinfo  {journal} {Class. Quant. Grav.}\ }\textbf {\bibinfo {volume}
  {27}},\ \bibinfo {pages} {084006} (\bibinfo {year} {2010})}\BibitemShut
  {NoStop}%
\bibitem [{\citenamefont {Aasi}\ {\it et~al.}(2015)\citenamefont {Aasi} {\it
  et~al.}}]{TheLIGOScientific:2014jea}%
  \BibitemOpen
  \bibfield  {author} {\bibinfo {author} {\bibfnamefont {J.}~\bibnamefont
  {Aasi}} {\it et~al.} (\bibinfo {collaboration} {LIGO Scientific}),\ }\bibinfo
  {title} {{Advanced LIGO}},\ \href
  {https://doi.org/10.1088/0264-9381/32/7/074001} {\bibfield  {journal}
  {\bibinfo  {journal} {Class. Quant. Grav.}\ }\textbf {\bibinfo {volume}
  {32}},\ \bibinfo {pages} {074001} (\bibinfo {year} {2015})},\ \Eprint
  {https://arxiv.org/abs/1411.4547} {arXiv:1411.4547} \BibitemShut {NoStop}%
\end{thebibliography}
%

\end{CJK*}
\end{document}